\documentclass{article}

\usepackage{PRIMEarxiv}

\usepackage[utf8]{inputenc} 
\usepackage[T1]{fontenc}    
\usepackage{hyperref}       
\usepackage{url}            
\usepackage{booktabs}       
\usepackage{amsfonts}       
\usepackage{nicefrac}       
\usepackage{microtype}      
\usepackage{lipsum}
\usepackage{fancyhdr}       
\usepackage{graphicx}       
\usepackage{pifont}
\usepackage{amsmath}
\usepackage{xcolor,multirow,multicol}
\usepackage{hyperref}
\usepackage{rotating}
\graphicspath{{media/}}     

\newcommand\update[1]{\textcolor{purple}{#1}}

\title{Large Language Model Inference Acceleration: \\ A Comprehensive Hardware Perspective}

\author{
  Jinhao Li\thanks{Email: kimholee@sjtu.edu.cn} \\
  Shanghai Jiao Tong University \\
  \& SII\thanks{SII: Shanghai Innovation Institute} \\
  \And
  Jiaming Xu \\
  Shanghai Jiao Tong University \\
  \& Infinigence-AI  \\
  \And
  Shan Huang \\
  Shanghai Jiao Tong University \\
  \& SII \\
  \And
  ~~~~Yonghua Chen \\
  ~~~~~~~~~~~~~Infinigence-AI~~~~~~~~  \\
  \And
  ~~~Wen Li \\
  ~~~~~~~~~~~~~~~~~Infinigence-AI~~~~~~~~~~~~~  \\
  \And
  Jun Liu \\
  Shanghai Jiao Tong University \\
  \And
  Yaoxiu Lian \\
  Shanghai Jiao Tong University \\
  \And
  Jiayi Pan \\
  Shanghai Jiao Tong University \\
  \And
  Li Ding \\
  Shanghai Jiao Tong University \\
  \And
  Hao Zhou \\
  Shanghai Jiao Tong University \\
  \And
  Yu Wang \\
  ~~~~~~~~Tsinghua University~~~~~~~~ \\
  \And
  Guohao Dai\thanks{Corresponding author: daiguohao@sjtu.edu.cn} \\
  Shanghai Jiao Tong University \\
  \& Infinigence-AI \& SII \\
}

\begin{document}
\maketitle

\begin{abstract}
Large Language Models (LLMs) have demonstrated remarkable capabilities across various fields, from natural language understanding to text generation.
Compared to non-generative LLMs like BERT and DeBERTa, generative LLMs like GPT series and Llama series are currently the main focus due to their superior algorithmic performance.
The advancements in generative LLMs are closely intertwined with the development of hardware capabilities. 
Various hardware platforms exhibit distinct hardware characteristics, which can help improve LLM inference performance.
Therefore, this paper comprehensively surveys efficient generative LLM inference on different hardware platforms.
First, we provide an overview of the algorithm architecture of mainstream generative LLMs and delve into the inference process. 
Then, we summarize different optimization methods for different platforms such as CPU, GPU, FPGA, ASIC, and PIM/NDP, and provide inference results for generative LLMs.
Furthermore, we perform a qualitative and quantitative comparison of inference performance with batch sizes 1 and 8 on different hardware platforms by considering hardware power consumption, absolute inference speed (\textbf{tokens/s}), and energy efficiency (\textbf{tokens/J}).
We compare the performance of the same optimization methods across different hardware platforms, the performance across different hardware platforms, and the performance of different methods on the same hardware platform. 
This provides a systematic and comprehensive summary of existing inference acceleration work by integrating software optimization methods and hardware platforms.
We point out that the development of edge intelligence has gained significant momentum, driven by the increasing capability of LLMs and the increasing demands of edge applications. 
And three trends (multimodality, inference-time compute, and higher inference energy efficiency) are promising to redefine the capabilities of edge artificial intelligence systems.
Our project available at \url{https://dai.sjtu.edu.cn/project.html}
\setcounter{footnote}{0}
\footnote{Github: https://github.com/Kimho666/LLM\_Hardware\_Survey}.

\end{abstract}

\keywords{Generative Large Language Model \and Hardware \and CPU \and GPU \and FPGA \and ASIC \and PIM \and NDP}

\section{Introduction}



\begin{figure}[!t]
  \centering
  \includegraphics[width=0.99\textwidth]{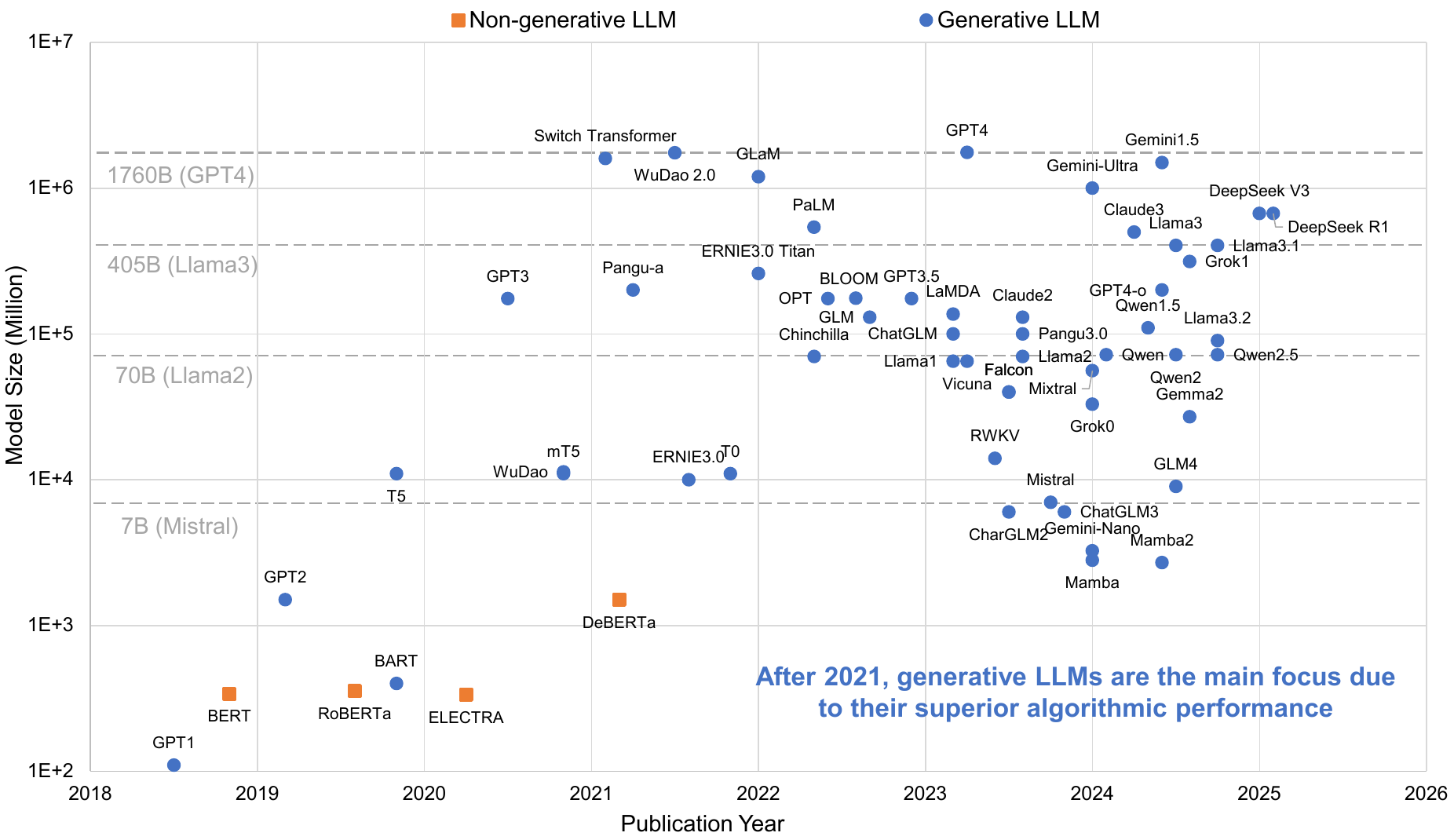}
  \vspace{-10pt}
  \caption{Typical LLM model size in the past six years.}
  \vspace{-15pt}
  \label{fig:gllm}
\end{figure}

Large Language Models (LLMs) have become cornerstones of modern artificial intelligence, demonstrating remarkable capabilities across a spectrum of fields, from natural language understanding to text generation~\cite{mo2024large,wu2024large,li2024pre,shu2024rewritelm,thakur2024verigen}.
LLMs can be categorized into two primary types: generative LLMs and non-generative LLMs.
Non-generative LLMs, such as BERT~\cite{devlin2018bert}, RoBERTa~\cite{liu2019roberta}, ELECTRA~\cite{clark2020electra}, and DeBERTa~\cite{he2020deberta}, are designed to classify and make predictions based on input text. 
These models typically range in size from millions of parameters, allowing them to excel in tasks that require discernment and nuanced understanding. 
BERT, introduced in 2018, only has 340 million parameters.
RoBERTa, introduced in 2019, slightly increases to 355 million parameters.
And DeBERTa, released in 2021, increases to 1.5 billion parameters.
Generative LLMs, like GPT series~\cite{radford2018improving,radford2019language,brown2020language,achiam2023gpt}, T5~\cite{raffel2020exploring}, OPT~\cite{zhang2022opt}, BLOOM~\cite{le2023bloom}, and Llama series~\cite{llama,llama2,meta2024llama3}, have taken language generation to new heights. 
The model size increase of generative LLMs~\cite{radford2018improving,radford2019language,lewis2019bart, brown2020language, raffel2020exploring, du2021glm, xue2020mt5, fedus2021switch, zeng2021pangu, sun2021ernie, sanh2021multitask, du2022glam, wang2021ernie, chowdhery2022palm, zhang2022opt, le2023bloom, hoffmann2022training, achiam2023gpt, thoppilan2022lamda, anthropic2024claude, chatglm, chatglm2, chatglm3, llama, llama2, zheng2023judging, almazrouei2023falcon, jiang2024mixtral, peng2023rwkv, xai2024grok, jiang2023mistral, team2023gemini, mamba, mamba2, team2024gemma, openai2024gpt4o, bai2023qwen, qwen2.5, yang2024qwen2, qwen1.5, meta2024llama3, meta2024llama31, meta2024llama32, reid2024gemini, deepmind2024gemini, anthropic2024claude3, openai2024gpt35turbo, grok2024} are particularly notable in the past 6 years, as shown in Figure~\ref{fig:gllm}. 
In 2018, GPT1 has only 110 million parameters, which grows to 1.5 billion in GPT2 in 2019. 
GPT3, launched in 2020, grows to 175B parameters dramatically, and GPT3.5 maintains the same size.
After 2022, the model size maintains to several hundreds and thousands of billions like GPT4, Llama3, and Grok1~\cite{grok1}.
The evolution of LLMs is characterized by an exponential growth in model parameters, which enhances their performance and versatility.
Compared to non-generative LLMs, generative LLMs are currently the primary focus of research and development in the field of LLMs for their superior algorithmic performance.
In recent years, after reaching the trillion-parameter scale in 2022, the parameter size of generative LLMs has stopped growing at an exponential rate. Two main reasons can explain this phenomenon: (1) As the amount of computation increases, the demand for computing power also rises significantly. The slow growth of hardware capabilities, particularly the slowing down of Moore's Law~\cite{schaller1997moore}, limits the improvement of single-chip computing power.
(2) Researchers have found that model performance is not solely dependent on the number of parameters, but also on the quantity and quality of training data~\cite{kaplan2020scaling}. By providing more qualified training tokens, the algorithmic performance can be further improved~\cite{llama2,meta2024llama3}.
At the same time, the model size of generative LLMs have shifted from "increasing" to "remaining stable" or even "shrinking".
More models with fewer parameters are released, which are more suitable for deployment on edge devices.

The advancements in generative LLMs are closely intertwined with the development of hardware capabilities. 
Due to the continuation of Moore's Law, from 2018 to 2022, GPU manufacturing processes have progressed from 12nm to 3nm, and the floating-point performance of single GPU die has increased from 130 TFLOPS to 989 TFLOPS.
During model training, GPUs are used predominantly due to the user-friendliness of the CUDA programming stack~\cite{cook2012cuda} and the high scalability of GPU chips (\textit{e.g.} NVLink~\cite{foley2017ultra}). 
During inference, various hardware options like CPU, GPU, FPGA, and ASIC exhibit distinct hardware characteristics, which can help improving LLM inference performance. 
CPUs offer high programmability with a computing power of approximately 4 to 70 TOPS and with power consumption around from 4W to >200W. 
Modern CPUs (including some System-on-Chips, SoCs) enhance AI performance by integrating domain-specific architecture (DSA) units. These include Apple’s Neural Engine in the M2 Ultra~\cite{Apple_M2_Ultra}, Qualcomm’s NPU in the Snapdragon 8 Gen3~\cite{Snapdragon8Gen3}, and Intel’s AVX/AMX ISA extensions~\cite{4th_Gen_Intel_Xeon}.
GPUs excel in parallelism and computing power, delivering between $\sim$70 to >1000 TOPS and featuring an impressive memory bandwidth of up to 1555 GB/s. 
On one hand, GPUs integrate a large number of SIMD cores and Tensor Cores in NVIDIA V100/A100/H100~\cite{nvv100,nva100,nvh100} or Matrix Cores in AMD Instinct MI100/MI200/MI300 series~\cite{amdcdna1,amdcdna2,amdcdna3} to enhance computing powers.
On the other hand, GPUs support lower precision computations, such as INT8, FP8 and INT4~\cite{nva100,nvh100}, which allows for more multiplication units to be packed into a given chip area.
Nevertheless, their power consumption is significantly higher, ranging from $\sim$20W to >700W. 
FPGAs offer substantial parallelism and optimization capabilities, with computing performance between 50 to 100 TOPS. They are also more power-efficient, consuming about 75 to 100W. 
AMD's Xilinx offers Zynq~\cite{zynq}, Virtex~\cite{virtex}, and Versal~\cite{versal} FPGA series for edge-side and cloud-side computing. 
The Zynq series, like ZCU102, combine ARM processors with FPGA programmable logic, has a power consumption of 10W-30W. 
The Virtex series, such as the VU13P, are designed for high-performance computing and data centers, providing up to 6.2 TFLOPS of computing power, with 3,840 DSP units and power consumption ranging from 30W to 70W. 
The Versal series, such as the VCK5000, integrate AI Engines~\cite{versal_aie} and efficient parallel computation units, offering >50 TOPS of computing power and 4,000 DSP units, with power consumption of 225W.
ASICs like Groq LPU~\cite{groq_lpu} and Cerebras WSE-3~\cite{lie2024wafer} are often designed for specific applications with the customized architecture, offering higher computational efficiency and better energy efficiency.

Here, we list and compare the existing LLM inference surveys in Table~\ref{tab:survey}.
Previous surveys~\cite{miao2023towards,zhu2023survey,qu2024mobile,park2024comprehensive,zhou2024survey} primarily summarize various software optimization methods like quantization, sparsity, fast decoding for generative LLMs from an algorithm perspective.
However, they do not consider that different optimization methods exhibit different inference performance across different hardware platforms, and similarly, they also lack a fair and quantitative comparison.
Surveys~\cite{kachris2024survey,koilia2024hardware} focus on accelerating transformer-based LLMs, including non-generative LLMs like BERT and a small number of generative LLMs like GPT, but merely list the work done on different hardware platforms. And they lack a summary and abstraction of the optimization methods used by different accelerators.
Additionally, it only provides a relative comparison of speedup and energy efficiency with different baselines, which is unfair. 
Like \cite{kachris2024survey,koilia2024hardware}, surveys~\cite{wolters2024memory,kang2024survey} mainly focus on non-generative LLMs on one or two specific hardware platforms. 
Our survey focuses solely on generative LLMs, summarizing various software optimization methods in conjunction with multiple hardware platforms, including CPUs, GPUs, FPGAs, ASICs, and PIM/NDPs. 
What matters most for generative LLMs is the absolute inference speed and the inference energy efficiency.
Therefore, for the first time, we use the number of tokens generated per second (tokens per second, tokens/s) and the number of tokens generated per joule (tokens per joule, tokens/J) to evaluate LLM acceleration. 
Besides these two metrics, we also conduct the comparison on \textbf{(1) the performance of the same optimization methods across different hardware platforms}, \textbf{(2) the performance across different hardware platforms}, and \textbf{(3) the performance of different methods on the same hardware platform}. 
This can provide the systematic and comprehensive summary of existing inference accelerations with software optimization methods and hardware platforms.

\begin{table}[htbp]
\footnotesize
\centering
\caption{Comparison of existing LLM surveys}
\begin{tabular}{|c|c|c|ccccc|c|}
\hline
\multicolumn{1}{|c|}{\multirow{2}{*}{Survey}} & \multicolumn{1}{c|}{Generative} & \multicolumn{1}{c|}{Software} & \multicolumn{5}{c|}{Hardware Platforms}                                                                             & \multicolumn{1}{c|}{Quantitative} \\ \cline{4-8}
\multicolumn{1}{|c|}{}                        & \multicolumn{1}{c|}{LLM}                                & \multicolumn{1}{c|}{Optimization}                                       & \multicolumn{1}{l|}{CPU} & \multicolumn{1}{l|}{GPU} & \multicolumn{1}{l|}{FPGA} & \multicolumn{1}{l|}{ASIC} & PIM/NDP & \multicolumn{1}{c|}{Comparison}                                         \\ \hline
\cite{miao2023towards,zhu2023survey,qu2024mobile,park2024comprehensive,zhou2024survey} &  \color{green}\ding{51}  & \color{green}\ding{51} & \multicolumn{1}{c|}{\color{red}\ding{55}}    & \multicolumn{1}{c|}{\color{red}\ding{55}}    & \multicolumn{1}{c|}{\color{red}\ding{55}}    & \multicolumn{1}{c|}{\color{red}\ding{55}}    & \multicolumn{1}{c|}{\color{red}\ding{55}}    & \multicolumn{1}{c|}{\color{red}\ding{55}}    \\ \hline
\cite{kachris2024survey,koilia2024hardware}  &  \color{red}\ding{55}  & \color{red}\ding{55} & \multicolumn{1}{c|}{\color{green}\ding{51}}    & \multicolumn{1}{c|}{\color{green}\ding{51}}    & \multicolumn{1}{c|}{\color{green}\ding{51}}    & \multicolumn{1}{c|}{\color{green}\ding{51}}    & \multicolumn{1}{c|}{\color{green}\ding{51}}    & \multicolumn{1}{c|}{\color{green}\ding{51}}    \\ \hline
\cite{wolters2024memory}  &  \color{red}\ding{55}  & \color{red}\ding{55} & \multicolumn{1}{c|}{\color{red}\ding{55}}    & \multicolumn{1}{c|}{\color{red}\ding{55}}    & \multicolumn{1}{c|}{\color{red}\ding{55}}    & \multicolumn{1}{c|}{\color{red}\ding{55}}    & \multicolumn{1}{c|}{\color{green}\ding{51}}    & \multicolumn{1}{c|}{\color{green}\ding{51}}    \\ \hline
\cite{kang2024survey}   &                            \color{red}\ding{55}  & \color{green}\ding{51} & \multicolumn{1}{c|}{\color{red}\ding{55}}    & \multicolumn{1}{c|}{\color{red}\ding{55}}    & \multicolumn{1}{c|}{\color{green}\ding{51}}  & \multicolumn{1}{c|}{\color{green}\ding{51}}    & \multicolumn{1}{c|}{\color{red}\ding{55}}    & \multicolumn{1}{c|}{\color{red}\ding{55}}    \\ \hline
\textbf{Ours}  &  \color{green}\ding{51}  & \color{green}\ding{51} & \multicolumn{1}{c|}{\color{green}\ding{51}}    & \multicolumn{1}{c|}{\color{green}\ding{51}}    & \multicolumn{1}{c|}{\color{green}\ding{51}}    & \multicolumn{1}{c|}{\color{green}\ding{51}}    & \multicolumn{1}{c|}{\color{green}\ding{51}}    & \multicolumn{1}{c|}{\color{green}\ding{51}}    \\ \hline
\end{tabular}
\label{tab:survey}
\end{table}

    

This survey aims to systematically summarize the optimization methods for generative LLMs across different hardware platforms.
Section~\ref{sec:gllm} delves into the inference process of LLMs, providing an overview of the architecture and functioning of mainstream generative LLMs. 
Section~\ref{platforms} first summarizes the different optimization methods on various platforms such as CPU, GPU, FPGA, ASIC, and PIM/NDP in tabular form, and then provides a detailed description of each method and related works. Additionally, for each method, we also perform a qualified and quantitative comparison to show the difference among the hardware platforms.
Section~\ref{sec:comparison} performs a qualitative and quantitative comparison of inference performance with batch sizes 1 and 8 on different hardware platforms.
Furthermore, in section~\ref{sec:discussion}, we point out that three trends (multimodality, inference-time compute, and higher inference energy efficiency) are promising to redefine the capabilities of edge artificial intelligence systems. 
Section~\ref{sec:conclusion} summarizes the work of this survey.

\section{Generative LLM Architecture}\label{sec:gllm}
\subsection{Overview}
\begin{figure*}[!t]
  \centering
  \includegraphics[width=0.95\textwidth]{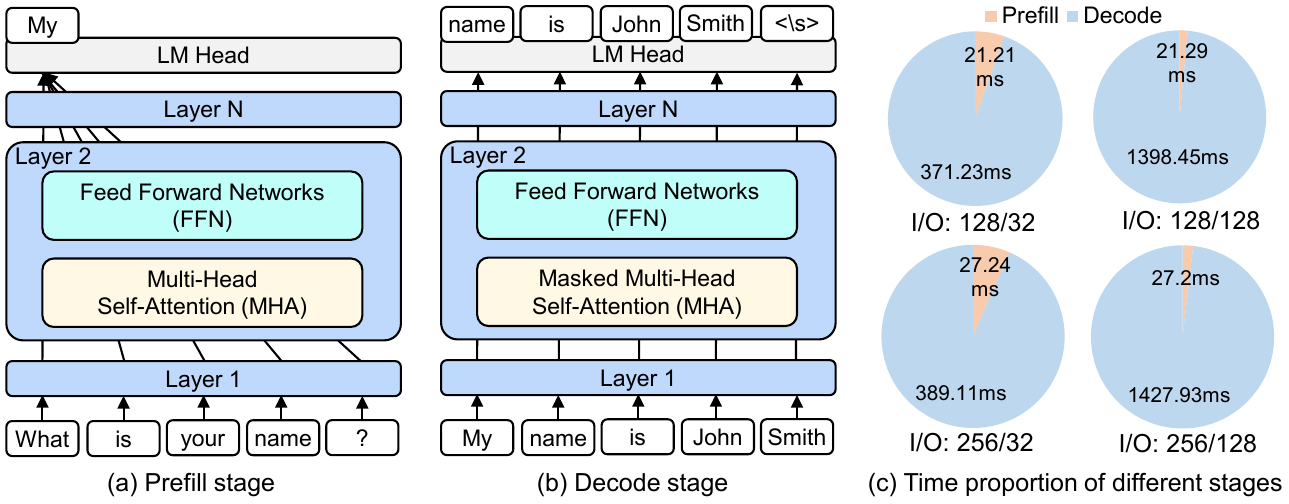}
  \vspace{-10pt}
  \caption{LLM inference includes prefill and decode stages. During inference in daily scenarios (input tokens: 128$\leq$I$\leq$256, output tokens: O$\geq$32), the time of decode stage is dominant.}
  \vspace{-10pt}
  \label{fig:inference_breakdown}
\end{figure*}

The most common generative LLM is based on the transformer structure due to its abilities for capturing long-term dependencies~\cite{vaswani2017attention}. 
The inference of generative LLM consists of two stages, the prefill stage and the decode stage, as shown in Figure~\ref{fig:inference_breakdown}.
In the prefill stage, the input text is converted into embeddings and input into the LLM all at once. Each transformer layer performs operations, and the intermediate results (Key and Value, KV cache) are saved for reuse during the decode stage. 
After completing calculating the last layer and LM head, the first token (`My') is generated. 
Then, in the decode stage, the LLM generates each token output autoregressively and updates the KV cache each time until generating the stop string (`</\/s>').
For daily application scenarios where the number of input tokens is from 128 to 256 and output tokens larger than 32, we obtain that the time proportion of the decode stage exceeds 80\% by profiling Llama2-7B on single NVIDIA A100 GPU, as shown in Figure~\ref{fig:inference_breakdown}(c).

In the prefill stage, the attention computation has quadratical computational and storage complexity related with input text length. 
To address this issue, many hardware-efficient LLMs are designed for more efficient processing of longer length. 
Therefore, we summarize these LLMs in sub-section~\ref{eff_gllm}.
And for the decode stage, we summarize hardware optimization methods for different platforms such as CPU, GPU, FPGA, ASIC, and PIM/NDP, and provide inference results for generative LLMs in latter sub-sections.

\subsection{Efficient Generative LLM}\label{eff_gllm}

\subsubsection{Attention-based LLM}

In LLMs, the most common type of attention is self-attention, where each token in the input sequence attends to every other token, capturing both local and global dependencies. This is achieved by computing three vectors for each token: Query ($Q$), Key ($K$), and Value ($V$). The attention scores are calculated as the scaled dot product of $Q$ and $K$, followed by a softmax operation to determine the weight of each token in the sequence. These weights are then used to compute a weighted sum of the $V$ vectors, producing the attention output.

\begin{equation}\label{eq:attention}
  O=Softmax(\frac{QK^T}{\sqrt{d_k}})V
\end{equation}

Many methods focus on simplifying the attention mechanism.
Transformer-XL~\cite{transformer-xl} adopts a segment-level recurrence mechanism and a positional encoding scheme to learn dependencies beyond a fixed length without disrupting temporal coherence. 
Linear Transformer~\cite{linear-transformer} represents self-attention as a linear dot product of kernel feature maps and alters the computation order by leveraging the associativity of matrix multiplication. This modification reduces the complexity from $O(L^2)$ to $O(L)$, where $L$ is the context length, significantly accelerating the computation of autoregressive Transformers. 
Another efficient structure is the Attention-Free Transformer (AFT)~\cite{attention-free}. Unlike vanilla transformers, which first compute the query-key product, AFT combines the key and value with a set of learned positional biases before performing element-wise multiplication with the query. As a result, the memory complexity of AFT is linear with respect to both the context size and feature dimensions, enabling support for larger input lengths and model sizes.
Based on AFT, the Receptance Weighted Key Value (RWKV)~\cite{peng2023rwkv} combines the efficient parallel training capabilities of Transformers with the efficient inference of RNNs. It leverages linear attention mechanisms and allows the model to be expressed as either a transformer or an RNN.
It also enables parallel computation during training while maintaining constant computational and memory complexity during inference. 
DiJiang~\cite{dijiang} introduces a novel frequency-domain kernelization method based on the Discrete Cosine Transform (DCT). It points out that improving attention mechanisms often requires extensive retraining, which is impractical for large language models with vast numbers of parameters. This approach enables the conversion of a pre-trained standard Transformer into a model with linear complexity and low training costs, utilizing a weighted quasi-Monte Carlo method for sampling. 
Extensive experiments demonstrate that this method achieves performance comparable to the vanilla transformer while significantly reducing training costs and substantially increasing inference speed. 

\subsubsection{SSM-based LLM}

State Space Model (SSM) defines a linear mapping from an input $x$ to output $y$ through a hidden state $h$, where $A$ is the state matrix, $B$ is the input matrix, and $C$ is the output matrix.
In generative LLMs, SSM can understand and compress the input text into hidden states, and then generate output text based on these states.

\begin{equation}\label{eq:cont_ssm}
\begin{split}
  h &= A \times h+B \times x \\
   y &= C \times h
\end{split}
\end{equation}

The Structured State Space Sequence Model (S4)~\cite{s4} involves conditioning the matrix $A$ with low-rank corrections, enabling it to be stably diagonalized, and simplifying the SSM to computations that involve an in-depth exploration of the Cauchy kernel. 
It offers significantly higher computational efficiency compared to previous methods while retaining its theoretical advantages.
The Gated State Space Model (GSS)~\cite{gss} is
built on the effectiveness of gated activation functions. GSS demonstrates significantly faster training speeds on TPUs compared to S4, and it competes effectively with several Transformer-based LLMs. 
Hyena~\cite{hyena} addresses that existing sub-quadratic methods based on low-rank and sparse approximations need to be combined with dense attention layers to match the performance of transformers. 
Therefore, it introduces a sub-quadratic direct replacement for attention, constructed using interleaved implicit parameterized long convolutions and data-controlled gating. 
Hyena can improve accuracy by over 50 points compared to operators relying on state space models and other implicit and explicit methods. 
Due to the inability to perform content-based reasoning for linear attention, gated convolution, recurrent models, and S4, Mamba~\cite{mamba} makes the SSM parameters a function of the input and enables the model to selectively propagate or forget information along the sequence length dimension based on the current token. 
Additionally, a hardware-aware parallel algorithm was designed for the recurrent mode, enhancing computational efficiency.
DenseSSM~\cite{densemamba} enhance Mamba by selectively integrating shallow layer hidden states into deeper layers. Despite the dense connections, DenseSSM maintains both training parallelism and inference efficiency. 
Mamba2~\cite{mamba2} demonstrates that transformer and SSM model families are closely related through various decompositions of a well-studied class of structured quasi-separable matrices. Mamba2 also introduces the State Space Duality (SSD) framework, with its core layer being an improved version of the selective SSM used in Mamba and offering a 2-8$\times$ speedup. 

\subsubsection{Hybrid LLM}

Some other LLMs integrate attention-based and SSM-based LLMs, leveraging the complete information extraction ability of attention and the information compression capability of SSM to enhance the performance for long inputs.
The Block-State-Transformer (BST)~\cite{block-state-transformer} integrates an SSM sublayer for long-range contextualization with a block-transformer sublayer for short-term sequence representation. This architecture combines the strengths of SSMs and block attention, and explores three distinct, fully parallelizable variants. 
Griffin~\cite{griffin} combines gated linear recurrence with local attention, featuring the Hawk layer (a type of RNN with gated linear recurrence). 
Jamba~\cite{jamba} interleaves blocks of transformer and Mamba layers, harnessing the strengths of both model families. In some of these layers, mixture of expert (MoE) is added to increase model capacity while keeping the number of active parameters manageable. 
Unlike BST, Infini-Transformer~\cite{infini-attention} combines masked local attention and long-term linear attention within a single Transformer block.
This Infini-Attention mechanism incorporates compressed memory into the original attention mechanism within the constraints of limited memory and computational resources. 
MEGALODON~\cite{megalodon} is a neural architecture designed for efficient sequence modeling with infinite context length. It builds on the MEGA architecture (exponential moving average with gated attention) and introduces complex exponential moving average (CEMA), timestep normalization layer, normalized attention mechanism, and a pre-norm configuration with two-hop residuals to enhance its capability and stability. 

\section{Optimizations on Hardware Platforms}\label{platforms}
In this section, we provide an overview of the hardware platforms and various optimization techniques used in LLM inference. As shown in Table~\ref{tab:all_methods}, the hardware platforms include CPU, GPU, FPGA, ASIC, and PIM/NDP, while the optimization methods include quantization, sparsity, fast decoding, operator optimization, heterogeneous cooperation, and homogeneous cooperation. 
In the following sections, we will provide a detailed explanation of the principles of each optimization method and related works, followed by a qualified and quantitative comparison.

\begin{sidewaystable}[!htbp]
\vspace{-10pt}
\scriptsize
    \centering
    \vspace{-10pt}
    \caption{Existing LLM Inference Optimizations across Different Hardware Platforms}
    \vspace{5pt}
    \begin{tabular}  {|p{1.5cm}|p{3cm}|p{4.5cm}|p{3.5cm}|p{3cm}|p{3.5cm}|}
    \hline
        \textbf{Methods} & \textbf{CPU} & \textbf{GPU} & \textbf{FPGA} & \textbf{ASIC} & \textbf{PIM/NDP} \\ \hline
        \textbf{Quantization} & Shen et al.~\cite{shen2023efficient}, T-MAC~\cite{wei2024t}, Snapdragon 8 Gen3~\cite{Snapdragon8Gen3}, llama.cpp~\cite{llamacpp}, NoMAD-Attention~\cite{zhang2024nomad}, \update{Zhou et al.~\cite{zhou2024all}, Jayanth et al.~\cite{jayanth2024towards}, Gope et al.~\cite{gope2024highly}, Yu et al.~\cite{yu2024dynamic}, HeteroLLM~\cite{chen2025heterollm}, DECA~\cite{gerogiannis2025deca}} & GPTQ~\cite{frantar2022gptq}, AWQ~\cite{lin2024awq}, SpQR~\cite{dettmers2023spqr}, SqueezeLLM~\cite{kim2023squeezellm}, LLM-MQ~\cite{li2023llm}, APTQ~\cite{guan2024aptq}, Li et al.~\cite{li2024fast}, LUT-GEMM~\cite{park2022lut}, FLUTE~\cite{guo2024fast}, FP6-LLM~\cite{xia2024fp6}, LLM.int8~\cite{dettmers2022gpt3}, SmoothQuant~\cite{xiao2023smoothquant}, QUIK~\cite{ashkboos2023towards}, Atom~\cite{zhao2024atom}, LLM-FP4~\cite{liu2023llm}, 
        \update{GQSA~\cite{gqsa}, Su et al.~\cite{su2025accurate}, Ma et al.~\cite{ma2025efficient}}
        & FlexRun~\cite{hur2023fast}, HLSTransform~\cite{he2024hlstransform}, SECDA-LLM~\cite{haris2024designing}, Chen et al.~\cite{chen2024understanding}, FlightLLM~\cite{zeng2024flightllm}, EdgeLLM~\cite{huang2025edgellm}, \update{On-Device Qwen2.5~\cite{xiang2025device}, LightMamba~\cite{wei2025lightmamba}, TeLLMe~\cite{qiao2025tellme}, TerEffic~\cite{yin2025tereffic}, Li et al.~\cite{li2025pushing}, MEADOW~\cite{moitra2025meadow}, TTD~\cite{huang2025tensor}, LlamaF~\cite{xu2024llamaf}, LNS-LLM~\cite{haghi2024bridging}, LoopLynx~\cite{zheng2025looplynx}, AccLLM~\cite{liang2025accllm}} & FIGNA~\cite{jang2024figna}, MECLA~\cite{qin2024mecla}, OliVe~\cite{guo2023olive}, Li et al.~\cite{li2024quantization}, Tender~\cite{10609625}, \update{Cai et al.~\cite{cai2025adaptive}, FineQ~\cite{xie2025fineq}, BitMoD~\cite{chen2025bitmod}, OFQ-LLM~\cite{wang2025ofq}, M-ANT~\cite{hu2025m}, FlexiBit~\cite{tahmasebi2024flexibit}, Anda~\cite{fang2025anda}, AirGun~\cite{kim2024airgun}, Ecco~\cite{cheng2025ecco}, OwL-P~\cite{lee2025integer}, FGMP~\cite{hooper2025fgmp}, LightRot~\cite{kim2025lightrot}} & Guo et al.~\cite{guo2024towards}, TransPIM~\cite{zhou2022transpim}, Sharda et al.~\cite{sharda2024accelerator}, \update{SoftmAP~\cite{rakka2025softmap}, MVDRAM~\cite{kubo2025mvdram}, PIM-LLM~\cite{malekar2025pim}, ROMA~\cite{wang2025roma}, Lue et al.~\cite{lue2024prospects}} \\ \hline
        \textbf{Sparsity} & Turbo Sparse~\cite{song2024turbo}, ProSparse~\cite{song2024prosparse}, \update{SparAMX~\cite{abouelhamayed2025sparamx}, DECA~\cite{gerogiannis2025deca}} & LLM-pruner~\cite{ma2023llm}, SparseGPT~\cite{frantar2023sparsegpt}, Wanda~\cite{sun2023wanda}, E-Sparse~\cite{li2023sparse}, Flash-LLM~\cite{xia2023flash}, Agarwalla et al.~\cite{agarwalla2024enabling}, DejaVu~\cite{liu2023deja}, Sparse Transformer~\cite{child2019generating}, Bigbird~\cite{zaheer2020big}, StreamingLLM~\cite{xiao2023streamingllm}, Longformer~\cite{beltagy2020longformer}, Adaptively Sparse Attention~\cite{correia2019adaptively}, Reformer~\cite{kitaev2020reformer}, Sparse Flash Attention~\cite{pagliardini2023faster}, Sparse Sinkhorn Attention~\cite{tay2020sparse}, H$_2$O~\cite{zhang2024h2o}, \update{shihab et al.~\cite{shihab2025efficient}, SpInfer~\cite{spinfer}, SoLA~\cite{sola}, R-Sparse~\cite{r-sparse}} & FlightLLM~\cite{zeng2024flightllm}, EdgeLLM~\cite{huang2025edgellm}, \update{ChatOPU~\cite{zhao2024chatopu}, Zhang et al.~\cite{zhang2024fine}, AccLLM~\cite{liang2025accllm}} & Spatten~\cite{wang2021spatten}, TF-MVP~\cite{yoo2023tf}, SOFA~\cite{wang2024sofa} & LauWS~\cite{li2024lauws}, HARDSEA~\cite{liu2023hardsea}, Sharda et al.~\cite{sharda2024accelerator}, \update{FlexCiM~\cite{ramachandran2025accelerating}} \\ \hline
        \textbf{Fast Decoding} & \update{Dovetail~\cite{zhang2024dovetail}, ML-SpecQD~\cite{georganas2025ml}} & LLMA~\cite{yang2023inference}, Speculative decoding~\cite{stern2018blockwise}, Lookahead~\cite{lookahead}, Medusa~\cite{medusa}, EAGLE~\cite{eagle1,eagle2}, Ouroboros~\cite{zhao2024ouroboros}, Sequoia~\cite{chen2024sequoia}, Draft\&Verify~\cite{zhang2023draft}, Kangaroo~\cite{liu2024kangaroo}, LayerSkip~\cite{layerskip}, Adainfer~\cite{adainfer}, RAEE~\cite{RAEE}, MOD~\cite{MOD}, \update{SpecEE~\cite{specee}, AMUSD~\cite{amusd}, PipeSpec~\cite{pipespec}, Adaptix~\cite{adaptix}, SPIN~\cite{spin}, PARD~\cite{pard},
        Judge Decoding~\cite{judge_decoding}, Falcon~\cite{falcon}} & ~ & C-Transformer~\cite{kim202420} & SpecPIM~\cite{li2024specpim} \\ \hline
    \textbf{Operator Optimization} & \update{V-Seek~\cite{rodrigo2025vseekacceleratingllmreasoning}, FlexInfer~\cite{flexinfer}}  & FlashAttention~\cite{flashattention, flashattention2}, FlashDecoding~\cite{flashdecoding}, FlashDecoding++~\cite{hong2024flashdecoding}, DeepSpeed~\cite{deepspeed}, vLLM~\cite{kwon2023vllm}, OpenPPL~\cite{OpenPPL}, cuBLAS~\cite{cuBLAS}, TensorRT-LLM~\cite{tensorrt-llm}, CUTLASS~\cite{CUTLASS}, ByteTransformer~\cite{bytetransformer}, 
        \update{SpInfer~\cite{spinfer}, FlashFormer~\cite{flashformer}, Dong et al.~\cite{dong2025accelerating}} & \update{MEADOW~\cite{moitra2025meadow}, HAAN~\cite{peng2025haan}}  & LPU~\cite{moon2024lpu}, Groq LPU~\cite{groq_lpu}, ConSmax~\cite{liu2024consmax}, MARCA~\cite{li2024marca}, TCP~\cite{kim2024tcp}, Habana Gaudi~\cite{habana_gaudi},  Gaudi2~\cite{habana_gaudi2}, Gaudi3~\cite{kaplan2024intel}, Cerebras WSE-3~\cite{lie2024wafer}, \update{PICACHU~\cite{qin2025picachu}, WaferLLM~\cite{he2025waferllm}} & PIMnast~\cite{ibrahim2024balanced}, AttentionLego~\cite{cong2024attentionlego}, PIM-GPT~\cite{wu2024pim}, SAL-PIM~\cite{han2024sal}, PipePIM~\cite{jeong2024pipepim}, \update{SoftmAP~\cite{rakka2025softmap}} \\ \hline
        \textbf{Heterogeneous Cooperation} & Kim et al.~\cite{kim2024exploiting}, PowerInfer~\cite{song2023powerinfer}, PowerInfer-2~\cite{xue2024powerinfer}, \update{Dovetail~\cite{zhang2024dovetail}, Twinpilots~\cite{yu2024twinpilots}, Vellaisamy et al.~\cite{vellaisamy2025characterizing}, PRIMA.CPP~\cite{li2025prima}, HeteroLLM~\cite{chen2025heterollm}} & \update{TightLLM~\cite{tightllm}} & \update{GLITCHES~\cite{yang2024glitches}, HPU~\cite{rhee2025hpu}} & ~ & NeuPIMs~\cite{heo2024neupims}, IANUS~\cite{seo2024ianus}, MoNDE~\cite{kim2024monde}, Sharda et al.~\cite{sharda2024accelerator}, AttAcc~\cite{choi2023unleashing,park2024attacc}, Kang et al.~\cite{kang2023era}, Kim et al.~\cite{kim2024breakthrough}, H3D-Transformer~\cite{luo2024h3d}, CXL-PNM~\cite{park2024lpddr}, 3D-HI~\cite{sharma2023heterogeneous}, SK Hynix AiMX/AiMX-xPU~\cite{AiMX,kim2024sk}, Cambricon-LLM~\cite{yu2024cambricon}, \update{PAISE~\cite{lee2025paise}, PIM-AI~\cite{ortega2024pim}, Pyramid~\cite{yan2025pyramid}, PIM-LLM~\cite{malekar2025pim}, Lue et al.~\cite{lue2024prospects}, Hermes~\cite{liu2025make}} \\ \hline
        \textbf{Homogeneous Cooperation} & He et al.~\cite{he2024distributed,he2024inference} & ~ & DFX~\cite{hong2022dfx}, \update{LoopLynx~\cite{zheng2025looplynx}} & \update{Chen et al.~\cite{chen2024agile}} & ~ \\ \hline
    \end{tabular}
    \label{tab:all_methods}
\end{sidewaystable}

\subsection{Quantization}

\subsubsection{Overview}

Quantization converts the model's weights and activations from high-precision formats (32-bit floating-point numbers) to low-precision formats (such as 4-bit integers). This process aims to reduce the model's storage requirements and computational costs while maintaining its accuracy. 
From the perspective of data format, quantization includes uniform and non-uniform quantization. 
Uniform quantization is a method where the value range is divided into several equal intervals. In uniform quantization, the entire range of values is partitioned into equally sized intervals, with each interval mapped to a discrete representation value. These discrete values are typically represented using fewer bits (e.g., 8 bits). The advantages of uniform quantization include its simplicity and high computational efficiency. However, it may not effectively capture the data distribution characteristics, especially when the data distribution is uneven, potentially leading to significant information loss.
Non-uniform quantization, on the other hand, uses intervals of varying sizes based on the actual data distribution. It divides the data range into different-sized intervals, for example, using smaller intervals in regions where the data distribution is dense and larger intervals where the distribution is sparse. This approach can better preserve the details and features of the data, thus improving the model's accuracy. Non-uniform quantization typically requires additional computation and storage to manage the quantization intervals, but it provides higher precision and effectiveness in quantization.

Granularity in quantization is crucial for determining model performance and efficiency. The granularity are group-wise, channel-wise, and tensor-wise.
Group-wise granularity is a coarser approach where multiple channels or layers are quantized with the same parameters. This means that within a group, all channels or layers use identical quantization settings. The advantage of group-level granularity is its simplicity and relatively low computational and storage overhead. However, it may not capture the individual characteristics of each channel or layer as effectively, potentially resulting in some compromise in model performance.
Channel-wise granularity involves quantizing each channel individually within the model. Each channel can have its own quantization parameters, allowing for more precise adjustments according to the weight distribution and activation characteristics of each channel. This granularity offers a balance between precision and flexibility, though it increases the complexity of implementation and computation.
Tensor-wise granularity is the most detailed approach, where each tensor (such as weight tensors or activation tensors) is quantized separately. This means that each tensor has its own quantization parameters, enabling the highest degree of adaptation to the specific characteristics of each tensor and providing the best precision. However, this level of granularity comes with the highest computational and storage costs and is the most complex to implement.
There are two main quantization methods: weight-only quantization and weight-activation quantization.

\begin{figure*}[!thbp]
  \centering
  \includegraphics[width=0.98\textwidth]{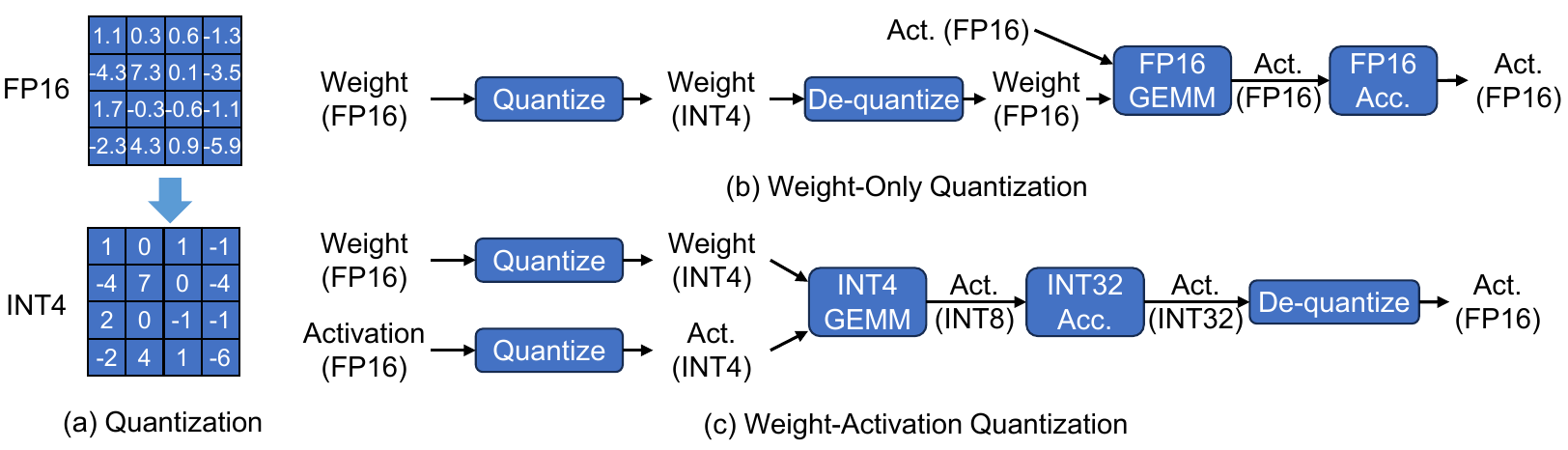}
  \vspace{-10pt}
  \caption{Two main quantization methods: weight-only quantization and weight-activation quantization.}
  \vspace{-5pt}
  \label{fig:quantization}
\end{figure*}

\textbf{Weight-Only Quantization.}
Weight-only quantization involves converting the model's weight parameters from high-precision formats (like 32-bit floating-point numbers) to low-precision formats (such as 8-bit integers). This process typically includes discretizing the weights by mapping them to a finite set of discrete values and then representing these values with fewer bits (e.g., 8 bits). This approach significantly reduces storage requirements and accelerates computation. Weight-only quantization can be implemented using methods such as uniform quantization, which divides the weight range into equal intervals, or non-uniform quantization, which adjusts the intervals based on the distribution of weights to better preserve model accuracy. Matrix decomposition quantization is a specialized method where a large matrix is approximated by the product of several smaller matrices. This technique reduces the computational and storage requirements by representing a large matrix with multiple smaller matrices, which can be stored and processed in lower precision formats. This method is particularly beneficial for managing extremely large models, as it helps lower computational complexity and storage overhead.

\textbf{Weight-Activation Quantization.}
Weight-activation quantization extends the concept of weight-only quantization to include the activations generated during model inference. In this method, both the weights and the activations at each layer are quantized to lower precision formats. This reduces memory bandwidth requirements and enhances inference speed. The challenge with weight-activation quantization is to manage the trade-off between quantization errors and model accuracy. Techniques such as dynamic range quantization or specific quantization schemes are used to balance precision and computational efficiency.
Table~\ref{tab:quant} shows the usage of two quantization methods across different hardware platforms. 

\textbf{KV Cache Quantization}
Su et al.~\cite{su2025accurate} reveals that a small subset of unusual tokens exhibit unique characteristics that deviate from the pattern that keys are distributed by channel and the Values are distributed by token, which can substantially impact quantization accuracy. So they develop a outlier token tracing method to identify these tokens accurately during the decoding process and exclude them from quantization as outlier tokens. They achieve around 480 tokens/s for Llama2-7B model on a single NVIDIA A100 GPU.

\begin{table}[htbp]
    \centering
    \caption{\update{Quantization on CPU, GPU, FPGA, ASIC, and PIM/NDP}}
    \begin{tabular} {|c|c|c|c|}
    
        \hline
        Hardware & Weight-Only Quantization & Weight-Activation Quantization & \update{KV Cache Quantization} \\
        \hline
        CPU & \ding{51} & \ding{55} & \ding{55} \\
        \hline
        GPU & \ding{51} & \ding{51} & \ding{51}\\
        \hline
        FPGA & \ding{55} & \ding{51} & \ding{55} \\ 
        \hline
        ASIC & \ding{55} & \ding{51} & \ding{55} \\ 
        \hline
        PIM/NDP & \ding{55} & \ding{51} & \ding{55} \\ 
        \hline
    \end{tabular}
    \label{tab:quant}
\end{table}

\subsubsection{CPU}
\textbf{Weight-Only Quantization.}
The optimization methods of quantization on CPUs mainly focus on weight-only quantization.
Shen et al.~\cite{shen2023efficient} leverage Intel Neural Compressor to automate the INT4 quantization process with negligible accuracy loss, supporting various quantization recipes such as GPTQ~\cite{frantar2022gptq}, AWQ~\cite{lin2024awq} and TEQ~\cite{cheng2023teq}. 
They further develop a tensor library tailored for CPUs, which supports mainstream instruction sets like AVX2, AVX512, AVX512\_VNNI, and AMX. 
By providing INT4 dequantization kernels on x86 CPUs, the experimental results on mainstream LLMs including Llama2, Llama and GPT-NeoX shows the latency of token generation is ranging from 12.5 tokens/s to 50 tokens/s for models with parameters ranging from 6B to 20B, by using a single socket of 4th Generation Intel Xeon Scalable Processors~\cite{4th_Gen_Intel_Xeon}. 
Qualcomm Snapdragon 8 Gen3 SoC~\cite{Snapdragon8Gen3} utilizes its proprietary Hexagon 700 AI processor and quantization techniques to support the efficient LLM execution. 
For Llama2-7B with 4-bit quantization, it achieves about 15 tokens/s.
Some open-source repositories like llama.cpp~\cite{llamacpp} are designed for efficient LLM inference across diverse hardware platforms including CPUs, GPUs and ASICs. 
For Llama2-7B with 4-bit quantization, llama.cpp achieves 6 tokens/s with a single core and 32 tokens/s with eight cores on Apple M2-Ultra processors.
For Llama2-7B with 2-bit quantization, llama.cpp achieves 4 tokens/s with a single core and 21 tokens/s with eight cores on M2-Ultra.

Due to the overheads of weight dequantization from integer to floating, 
T-MAC~\cite{wei2024t} leverages lookup tables (LUTs) for efficient low-bit LLM inference on edge CPUs, circumventing the need for dequantization and mixed precision matrix multiplication.
For Llama2-7B with 4-bit quantization, T-MAC achieves 10 tokens/s with a single core and 38 tokens/s with eight cores on Apple M2-Ultra processors~\cite{Apple_M2_Ultra}, and 3 tokens/s on Raspberry Pi 5~\cite{Raspberry_Pi_5} integrated with ARM Cortex-A76.
For Llama2-7B with 2-bit quantization, T-MAC achieves 17 tokens/s with a single core and 50 tokens/s with eight cores on M2-Ultra, and 6 tokens/s on Raspberry Pi 5.
Furthermore, due to the vast quantities of expensive Multiply-Add (MAD) matrix operations in the attention computations, NoMAD-Attention~\cite{zhang2024nomad} design an efficient attention algorithm that replaces MAD operations with in-register lookups.
Through hardware-aware algorithmic designs, NoMAD-Attention achieves the computation of attention scores using repeated fast accesses to SIMD registers despite their highly limited sizes. 
Empirical evaluations demonstrate that for CodeLlama-7B with 4-bit quantization, NoMAD-Attention achieves 9 tokens/s with short context (e.g. 128) and 4 tokens/s with long context (e.g. 16k) on 2 Intel Xeon E5-2695 V3 14-core CPUs. 

\update{
Zhou et al.~\cite{zhou2024all} apply different bitwidth to compress LLMs and deploy them on Intel Xeon CPU Max 9468.
Experiments show that for OPT-13B and OPT-30B, it achieves 33 tokens/s and 15.625 tokens/s with batchsize 1.
Gope et al.~\cite{gope2024highly} present a group-wise non-uniform codebook-based quantization method and highly optimized kernels for ultra-low-precision quantization. 
Experiments show that for Llama3-8B with 4-bit quantization, it achieves 45.5 tokens/s with batchsize 1 and 184.8 tokens/s with batchsize 8 on Arm Graviton3 CPU (64 cores).
Jayanth et al.~\cite{jayanth2024towards} deploy INT4 LLM on Intel Core Ultra 7 165H Processor and achieve 3.94 tokens/s and 15.6 tokens/s for Llama3-8B and TinyLlama-1B.
Yu et al.~\cite{yu2024dynamic} deploy INT4 LLM and llama.cpp on Intel Ultra-125H and Core i9-12900K and achieve 15.625 tokens/s and 6.393 tokens/s for Llama2-7B.
DECA~\cite{gerogiannis2025deca} first develops an analytical performance model with a 3D visual representation that provides insights into how memory resources, vector units, and hardware matrix engines interact to deliver compressed (quantization and sparsity) GeMM.
Then, DECA designs a new near-core ML-model decompression accelerator. DECA offloads tile de-sparsification and dequantization from the CPU, producing ready-to-use tiles for in-core GeMM engines. 
In a simulated 56-core Xeon 4 server with HBM, for Llama2-70B, DECA achieves 14-25 tokens/s with MXFP4 and BF8 with 5\%-20\% sparsity (bs=1) and 194-281 tokens/s (bs=16).
}

\subsubsection{GPU}
\textbf{Weight-Only Quantization.}
GPTQ~\cite{frantar2022gptq} is an one-shot weight quantization method based on approximate second-order information and error compensation, that is both highly-accurate and highly-efficient. 
It can quantize GPT models with 175 billion parameters in approximately four GPU hours, reducing the bitwidth down to 3-bit or 4-bit per weight, with negligible accuracy degradation relative to the uncompressed baseline. 
Experimental results show that the average time of per token of 3-bit OPT-175B model obtained via GPTQ running on a single A100 (80GB) is 14.1 tokens/s, which is about 3.25$\times$ faster than the FP16 version (running on 5 GPUs).
On more accessible GPUs, such as the NVIDIA A6000 (48GB), the average time of per token is 7.7 tokens/s (running on 2 GPUs), which is about 4.53$\times$ faster than the FP16 version (running on 8 GPUs).
AWQ~\cite{lin2024awq} is based on the observation that protecting 1\% of salient weights whose activations are extremely large can greatly reduce quantization error. 
It first searches for the optimal per-channel scaling and then multiplies the salient weights with the per-channel scalings.
It also reduces the bitwidth down to 3 or 4 bits per weight.
Experimental results with INT4 implementation show that for Llama-2-7B, it improves the inference speed from 52 tokens/s to 194 tokens/s on RTX 4090 desktop GPU (3.73$\times$ speedup).
For Llama-2-13B, the inference speed is 110 tokens/s on RTX 4090 desktop GPU.
On the laptop RTX 4070 GPU (8GB), it is able to run Llama-2-13B models at 33 tokens/s, while the FP16 implementation cannot fit 7B models. 

To further reduce the accuracy loss for smaller models in the 1-10B parameter range, SpQR~\cite{dettmers2023spqr} works by identifying and isolating outlier weights, which cause particularly-large quantization errors, and storing them in higher precision like half data type (16-bit), while compressing all other weights to 3-4 bits, and achieves relative accuracy losses of less than 1\% in perplexity for highly-accurate LLaMA and Falcon LLMs. 
Experimental results show that SpQR with 3-bit and 16-bit quantization achieves 57 tokens/s, 44 tokens/s, 22 tokens/s and 12 tokens/s on A100 GPU, respectively.
Unlike SpQR, SqueezeLLM~\cite{kim2023squeezellm} proposes a sensitivity-based non-uniform quantization method, which searches for the optimal bit precision assignment based on second-order information.
It also applies dense and sparse decomposition that stores outliers and sensitive weight values in an efficient sparse format. 
Experimental results show that SqueezeLLM with 3bit and 16-bit quantization achieves 63.5 tokens/s, 49.2 tokens/s, 29.1 tokens/s and 14.5 tokens/s on A6000 GPU, respectively.
LLM-MQ~\cite{li2023llm} proposes sensitivity-based precision allocation to assign the proper bitwidth for each layer within the given budget for weight memory based on their first-order information and quantization error. 
It also develops an efficient CUDA core kernels to accelerate LLMs by fusing the dequantization and general matrix-vector multiplication (GEMV). 
LLM-MQ deploys INT4 quantized Llama2-7B model on NVIDIA T4 GPU achieves up to 1.6$\times$ end-to-end speedup compared to the pytorch FP16 baseline. 
APTQ~\cite{guan2024aptq} proposes an attention-aware 2/4-bit mixed-precision quantization for LLMs, which considers not only the second-order information of each layer's weights, but also, for the first time, the nonlinear effect of attention outputs on the entire model.
Li et al.~\cite{li2024fast} are the first to propose an intra-weight mixed-precision quantization for LLMs to further reduce accuracy loss under 3-bit. 
By applying 2/4-bit mixed-precision quantization with memory alignment and exclusive 2-bit sparse outlier reservation with minimum speed degradation, it achieves 2.91-bit for each weight considering all scales/zeros for different models with negligible loss. 
Additionally, they design an asynchronous dequantization and fuse the dequantization and GEMV kernels during inference.
For Llama2-7B, it achieves 45.2 tokens/s on RTX 3090 GPU and 34.0 tokens/s on RTX 2080 GPU.

LUT-GEMM~\cite{park2022lut} proposes an efficient LUT-based GPU kernel for quantized matrix multiplication, which not only eliminates the resource-intensive dequantization process but also reduces computational costs compared to previous kernels for weight-only quantization. 
The impact of LUT-GEMM is facilitated by implementing high compression ratios through low-bit quantization and efficient LUT-based operations. 
For Llama-7B with 4-bit quantization, it achieves 163.9 tokens/s on A100 GPU, achieving a remarkable 1.64$\times$ token generation latency improvement compared to the pytorch FP16 baseline.
FLUTE~\cite{guo2024fast} is a flexible lookup table engine for LUT-quantized LLMs, which uses offline restructuring of the quantized weight matrix to minimize bit manipulations associated with unpacking, and vectorization and duplication of the lookup table to mitigate shared memory bandwidth constraints. 
For Llama3-8B with 4-bit quantization, it achieves 91.3-99.8 tokens/s and 113.7-121.7 tokens/s on NVIDIA A6000 and A100 GPUs, respectively.
For Llama3-8B with 3-bit quantization, it achieves 91.9-110.0 tokens/s and 117.7-135.5 tokens/s on NVIDIA A6000 and A100 GPUs, respectively.

To effectively reduce the size of LLMs and preserve the model accuracy, FP6-LLM~\cite{xia2024fp6} proposes FP6 quantization on GPUs with TC-FPx, the first full-stack GPU kernel design scheme with unified Tensor Core support of float-point weights for various quantization bit-width.
It solves the unfriendly memory access of model weights with irregular bit-width and high runtime overhead of weight de-quantization. 
Experimental results shows that for Llama2-13B with FP6 quantization, it achieves about 55 tokens/s on NVIDIA A100 GPU.

\update{
GQSA~\cite{gqsa}(Group Quantization and Sparse Acceleration) introduces a novel compression technique, 
integrating quantization and sparsification in
a tightly coupled manner, leveraging GPU-friendly structured group sparsity and quantization for efficient acceleration. In the setting of INT4 and 50\% sparsity, it achieves 343.43 tokens/s and 228.95 tokens/s on NVIDIA A100 for Llama2-7B and Llama2-13B models, respectively. 
Ma et al.~\cite{ma2025efficient} introduce a novel bipolar-INT data format that facilitates parallel computing and supports symmetric quantization, and implement an arbitrary precision matrix multiplication scheme that decomposes and recovers matrices at the bit level. Moreover, they design a data recovery-oriented memory management system to minimize memory access latency. For Llama2-7B, they achieve 51.94 tokens/s on a single RTX 3090 GPU.
}


\textbf{Weight-Activation Quantization.}
In addition to hardware units that support FP16 computations, NVIDIA GPUs also provide hardware units that support INT4, INT8, and FP8 computations. 
The number of these computation units can be 2$\times$ and 4$\times$ greater than FP16 on each chip. 
Compared to weight-only quantization, weight-activation quantization can utilize INT4, INT8, and FP8 computations, thereby maximizing the peak computational performance of the GPU. 
Since the prefill phase in LLM inference is compute-bound, weight-activation quantization can significantly enhance performance during this stage.
LLM.int8~\cite{dettmers2022gpt3} uses vector-wise quantization with separate normalization constants for each inner product in the matrix multiplication, to quantize most of the features.
For the outliers, it isolates the outlier feature dimensions into a 16-bit matrix multiplication while still more than 99.9\% of values are multiplied in 8-bit. 
For BLOOM-176B model, LLM.int8 achieves 4.05 tokens/s, 30.3 tokens/s and 109.77 tokens/s for batch size 1, 8 and 32, respectively, on 3 A100 GPUs in decode phase.
The inference speed is slightly slower but close to 16-bit inference with less GPU consumption.
SmoothQuant~\cite{xiao2023smoothquant} enables 8-bit weight and 8-bit activation (W8A8) quantization for LLMs. 
Based on the fact that weights are easy to quantize while activations are not, SmoothQuant smooths the activation outliers by offline migrating the quantization difficulty from activations to weights with a mathematically equivalent transformation. 
SmoothQuant enables an INT8 quantization of both weights and activations for all the matrix multiplications in LLMs. 
SmoothQuant achieves up to 1.56$\times$ speedup and 2$\times$ memory reduction for LLMs with negligible loss in accuracy.
QUIK~\cite{ashkboos2023towards} is for the first time, that the majority of inference computations for LLMs can be performed with both weights and activations being cast to 4 bits. 
QUIK compresses most of the weights and activations to 4-bit, while keeping some outlier weights and activations in higher-precision. 
It also provides GPU kernels matching the QUIK format with highly-efficient layer-wise runtimes, which lead to practical end-to-end throughput improvements of up to 3.4$\times$ relative to FP16 execution in prefill phase.

Prevalent quantization schemes (e.g., W8A8) cannot fully leverage the capabilities of modern GPUs, such as 4-bit integer operators, resulting in sub-optimal performance.
To maximize the throughput, Atom~\cite{zhao2024atom} significantly boosts serving throughput by using low-bit operators and considerably reduces memory consumption via low-bit quantization. 
It attains high accuracy by applying a novel mixed-precision and fine-grained quantization process. 
For single batch inference, Atom can achieve about 30 tokens/s for on a NVIDIA RTX 4090 GPU.
Atom improves end-to-end throughput by up to 7.73$\times$ compared to the FP16 and by 2.53$\times$ compared to INT8 quantization, while maintaining the same latency target.

Compared to integer quantization, floating-point (FP) quantization can better handle long-tail or bell-shaped distributions, and it has emerged as a default choice in many hardware platforms. 
LLM-FP4~\cite{liu2023llm} quantizes both weights and activations in LLMs down to 4-bit floating-point values (W4A4) with negligible accuracy loss.
Due to the lack of PF4 computing unit in GPUs, its decoding speed maybe slower than FP16 baseline.

\subsubsection{FPGA}
\textbf{Weight-Activation Quantization.}
FlexRun~\cite{hur2023fast} uses 8-bit quantization (W8A8), conducts an in-depth design space exploration to find the best accelerator architecture for a target LLM model, and automatically reconfigures the accelerator based on the exploration results.
With the implementation on Intel Stratix 10 GX and MX FPGAs, FlexRun outperforms the current state-of-the-art FPGA-based accelerator by  1.15$\times$–1.50$\times$ for GPT2, respectively. 
Compared to Nvidia’s V100 GPU, FlexRun achieves 2.69$\times$ higher performance on average for various GPT2 models.
HLSTransform~\cite{he2024hlstransform} uses HLS to design a FPGA accelerator and synthesis combined with pipelining, memory unrolling, and memory partitioning and transfer optimizations, with the addition of 8-bit integer quantization (W8A8).
On a tiny model with 110 million parameters, HLSTransform achieves 57.11 tokens/s on Xilinx Virtex UltraScale+ VU9P FPGA.
SECDA-LLM~\cite{haris2024designing} utilizes quantization (W3A8) and designs an efficient FPGA-based LLM accelerators for the llama.cpp inference framework. 
By deploying on the PYNQ-Z1 board, it achieves 0.588 tokens/s for the TinyLlama model (1.1B).
Chen et al.~\cite{chen2024understanding} investigate the feasibility and potential of model-speciic spatial acceleration for LLM inference on FPGAs. 
They introduce a comprehensive analytical model to estimate the LLM inference performance of FPGA accelerator with W4A8 quantization, and provide a library of high-level synthesis (HLS) kernels that are composable and reusable.
For Llama2-7B, during prefilling phase, they can achieves about 213 tokens/s, 43 tokens/s, and 320 tokens/s on Xilinx Alveo U280, VCK5000, and VHK158 FPGAs, respectively.
During decode stage, they can achieves about 200 tokens/s, 40 tokens/s, and 333 tokens/s on Xilinx Alveo U280, VCK5000, and VHK158 FPGAs, respectively.
\update{
On-Device Qwen2.5~\cite{xiang2025device} leverages AWQ with FPGA-accelerated execution pipelines and achieves 5.1 tokens/s for Qwen-0.5B on Xilinx Kria KV260 edge platform.
TeLLMe~\cite{qiao2025tellme} is the first ternary LLM accelerator for low-power FPGAs (e.g., AMD KV260) that fully supports both prefill and autoregressive decoding using 1.58-bit weights and 8-bit activations. 
Under a 7W power budget, it delivers up to 9 tokens/s throughput over 1024-token contexts and prefill latencies of 0.55-1.15s for 64-128 token prompts.
LightMamba~\cite{wei2025lightmamba} proposes an W4A4 and W8A8 FPGA-friendly post-training quantization algorithm that features rotation-assisted quantization and power-of-two SSM quantization to reduce the majority of computation to 4-bit. It can achieve 93 tokens/s, 7.21 tokens/s and 3.61 tokens/s on Xilinx Alveo U280 FPGA and Versal VCK190 FPGA for Mamba2-2.7B.
TerEffic~\cite{yin2025tereffic} is aslo tailored for ternary-quantized LLM inference.
Experimental results demonstrate that the HBM-assisted architecture processes 290 tokens/s and 727 tokens/s for Llama-7B and 2.7B parameter models while consuming only 46W on AMD Alveo U280.
Li et al.~\cite{li2025pushing} deploy a 4-bit LLaMA2-7B model, achieving a decoding speed of 4.9 token/s on Zynq-based KV260 with 6.57W.
MEADOW~\cite{moitra2025meadow} reduces the off-chip memory access for LLMs with a novel token-parallel head-sequential (TPHS) dataflow and mixed-precision quantization with weight packing. 
It achieves 0.25-2 tokens/s for OPT-1.1B on the low power Xilinx ZCU102 FPGA platform that consumes less than 10W. 
TTD~\cite{huang2025tensor} develops a tensor-train decomposition (TTD) for LLMs with a further hardware implementation on FPGA. The compressed LLMs are further implemented on FPGA hardware within a highly efficient group vector systolic array (GVSA) architecture, which has DSP-shared parallel vector PEs for TTD inference, as well as optimized data communication in the accelerator. 
Experimental results show that it achieves 69.7 tokens/s and 65.8 tokens/s for ChatGLM3-6B and Llama2-7B on AMD Alveo V80 FPGA, respectively.
LlamaF~\cite{xu2024llamaf} leverages W8A8 quantization techniques, asynchronous computation, and a fully-pipelined accelerator to enhance efficiency. 
The evaluations show that for TinyLlama-1.1B model on a Xilinx ZCU102 platform, it can achieve 1.5 tokens/s throughput with 0.291 tokens/J energy efficiency.
LNS-LLM~\cite{haghi2024bridging} points out that LLM exhibits significant outliers when directly adopting the Logarithmic Number System (LNS) format and presents a data-format/architecture co-design to bright this gap. 
On the format side, it proposes a dynamic and asymmetric LNS format, and on the architecture side, realizes the dynamic LNS format in a systolic array.
Experimental results show that on an Alveo U280 FPGA as a prototype, LNS-LLM can effectively handle the outliers and resolve the mismatch between LNS and LLM.
}

\subsubsection{ASIC}
\textbf{Weight-Only Quantization.}
Despite the memory footprint reduction achieved by weight-only quantization, the actual computing performance is not really improved due to dequantization from integer to float. 
FIGNA~\cite{jang2024figna} proposes dedicated FP-INT arithmetic units designed specifically for FP-INT MAC operations and integrates them on the accelerator. 
FIGNA with FP16-INT4 provides 3.2768 TOPS computing power and 26.58W power consumption by considering all memory access in 28nm at 100MHz.
Estimated result shows that for OPT-6.7B it can achieve 21.332 tokens/s in decode stage.
Different from normal quantization methods, MECLA~\cite{qin2024mecla} proposes a parameter-efficient scaling sub-matrix partition method (SSMP) to decompose large weight matrices into several tiny-scale source sub-matrices (SS) and derived sub-matrices (DS). 
For memory issues, SSMP avoids accessing the full weight matrix but only requires small SS and DS scaling scalars. 
For computation issues, the proposed accelerator fully exploits the intermediate data reuse of matrix multiplication via on-chip matrix regrouping, inner-product multiplication re-association, and outer-product partial sum reuse. 
Totally, it can reduce 83.6\% memory access and 72.2\% computation. 
For Llama2-7B and BLOOM-7B, compared to NVIDIA V100 GPU, MECLA achieves 6.74$\times$ and 5.91$\times$ inference speedup ($\sim$161 tokens/s and 141 tokens/s, respectively) under 91.84W (32 processors) power consumption.

\update{
FineQ~\cite{xie2025fineq} quantizes 3 weights into 8-bit with 2/3-bit mixed-precision and introduces an accelerator utilizing temporal coding that effectively supports the quantization algorithm while simplifying the multipliers in the systolic array. 
It can achieve up to 1.79× energy efficiency and reduces the area of the systolic array by 61.2\%. 
Anda~\cite{fang2025anda} proposes an adaptive data format with group-shared exponent bits and dynamic mantissa bit allocation. 
}

\textbf{Weight-Activation Quantization.}
Based on the key insight that outliers are important while the normal values next to them are not, OliVe~\cite{guo2023olive} adopts an outlier-victim pair (OVP) quantization and handles outlier values locally with low hardware overheads. 
This enables a memory-aligned W4A4/W8A8 quantization, which can be efficiently integrated to the existing hardware accelerators like systolic array and tensor core.
OliVe provides 0.71 TOPS computing power and 5.95W power consumption by considering all memory access under 22nm.
Estimated results shows that for Llama-7B it can achieve 9.95 tokens/s in decode stage.
Li et al.~\cite{li2024quantization} uniformly group weights and activations to ensure workload balance for hardware, and propose two approaches called channel sorting and channel selection to enhance the performance of quantization.
It provides 1.43 TOPS computing power and 2.36W power consumption by considering all memory access under 65nm.
Estimated results shows that for Llama-7B it can achieve 20.04 tokens/s in decode stage.
Tender~\cite{10609625} decomposes weight and activation matrices by groups with different size to smooth the impact of outliers. And the format of scale factors are powers of two apart, which avoids explicit dequantization and extension to the commodity tensor compute hardware.
It is 7.174W (1.60W+5.574W) power consumption by considering HBM2 memory access under 28nm.
Result shows that for OPT-6.7B, Tender achieves 1.33$\times$ speedup (53.33 tokens/s) than NVIDIA A100 GPU.

\update{
Cai et al.~\cite{cai2025adaptive} propose a grouped adaptive two-range quantization (ATRQ) with an in-group embedded identifier to encode outliers and develop a low-overhead ATRQ decoder and an outlier-bitsplit PE to reduce the hardware overhead associated with high-bit-width outliers.
Experimental results show that their design can achieve the similar speed (9.95 tokens/s) with OliVe for OPT-6.7B with $\sim$1.632W.
OFQ-LLM~\cite{wang2025ofq} is an outlier-flexing quantization accelrator for LLM inference. It can achieves the similar speed with OliVe~\cite{guo2023olive} with 1.64W for 7B models.
M-ANT~\cite{hu2025m} proposes an efficient real-time quantization mechanism and implements a specific processing element (PE) to efficiently support MANT and incorporate a real-time quantization unit. 
By integrating these components into a systolic array, MANT unifies the group-wise weight and KV cache quantization and addresses the associated challenges. 
Evaluation shows it can achieve, on average, 1.96$\times$ speedup and 1.54$\times$ energy reduction to OliVe accelerator (19.502 tokens/s, 7.57W).
Unlike previous bit-serial designs, which also provide flexibility but at the cost of performance due to its bit-wise temporal processing nature, FlexiBit’s architecture enables bit-parallel processing of any precision and format without compute unit under utilization~\cite{tahmasebi2024flexibit}. 
AirGun~\cite{kim2024airgun} applies hardware efficient per-tensor quantization for non-sensitive modules and per-input channel quantization for sensitive modules, and proposes an accelerator that fully utilizes the advantages of AirGun (12.935 tokens/s, 6.1964W). 
Ecco~\cite{cheng2025ecco} combines group-wise and nonuniform quantization with pre-defined shared k-means patterns and Huffman coding to exploit the inherent entropy characteristics of LLM cache data. 
It also introduces a parallel Huffman-based decoding process with a multi-stage pipeline design, reducing latency by two orders of magnitude and achieving throughput comparable to GPU L2 caches. 
OwL-P~\cite{lee2025integer} leverages outliers and shared exponents to facilitate the compression of both model weights and activations. 
FGMP~\cite{hooper2025fgmp} proposes a fine
grained FP4/FP8 mixed-precision quantization and a hardware implementation for performing dot products between two FP4 blocks, two FP8 blocks, or one FP4 and one FP8 block. 
LightRot~\cite{kim2025lightrot} proposes a lightweight rotation scheme and dedicated hardware accelerator designed for 4-bit LLM inference.
}

\subsubsection{PIM/NDP}
ReRAM-based analog PIM architectures perform integer MVMs using voltage, current, and conductance in the analog domain, limiting their application to the more accurate floating point (FP) data format. 
Guo et al.~\cite{guo2024towards} propose an ReRAM and 3D-SRAM-based hybrid PIM architecture with non-uniform data format, achieving FP-based algorithm accuracy, high device utilization, and high energy efficiency. 
At the software level, they first analyze the impact of quantization errors on the accuracy of attention-free LLMs. 
For the quantization error-insensitive MVM operations, they propose the PIM-oriented exponent-free non-uniform (PN) data format. 
The proposed PN format can be flexibly adjusted to fit the data distribution and approach the accuracy of the FP format using bit-slicing-based full INT operations. 
For the quantization error-sensitive EWM operations, they introduce the multiplication free approximated FP multiplications to reduce the additional hardware overhead for PIM. 
At the hardware level, they propose a hybrid PIM architecture, including an ReRAM analog PIM using shift-and-add for PN-based MVMs, and a 3D-SRAM digital PIM with high utilization for multiplication-free FP-based element-wise operations. 
Extensive experiments show that the proposed PIM architecture achieves up to 89$\times$ and 16$\times$ speedup with 2537$\times$ and 12$\times$ energy efficiency improvement compared with GPU and PIM-baseline, respectively.
TransPIM~\cite{zhou2022transpim} is a memory-based acceleration for Transformer using software and hardware co-design. 
In the software-level, TransPIM adopts a token-based dataflow to avoid the expensive inter-layer data movements introduced by previous layer-based dataflow. 
In the hardware-level, TransPIM introduces lightweight modifications in the conventional HBM architecture to support PIM-NMC hybrid processing and efficient data communication for accelerating Transformer-based models. 
TransPIM system uses the 8GB HBM as the memory with 2.15$mm^2$ area overhead and about 40.01W power consumption.
Experimental results show that for GPT2 models, TransPIM achieves at least 22.1$\times$ speedup than NVIDIA RTX 2080Ti GPU.
Other PIM/NDP accelerators like TransPIM~\cite{zhou2022transpim} and Sharda's method~\cite{sharda2024accelerator} also involve the quantization to further improve LLM inference.

\update{
Through on-the-fly vector encoding and horizontal matrix layout, MVDRAM~\cite{kubo2025mvdram} eliminates the overheads introduced by the Processing using DRAM's (PUD) fundamental limitation of column-to-column data movement. 
It can achieve 2-10 tokens/s and 7-20 tokens/J for Llama2-7B with 2-8-bit quantization.
}

\update{
ROMA~\cite{wang2025roma} is a QLoRA accelerator with a hybrid storage architecture that uses ROM for quantized base models and SRAM for LoRA weights and KV cache. 
ROMA can achieve 11600-24100 tokens/s for 2-bit Llama3-8B under 33.1W power consumption.
}

\begin{figure*}[!t]
  \centering
  \includegraphics[width=0.98\textwidth]{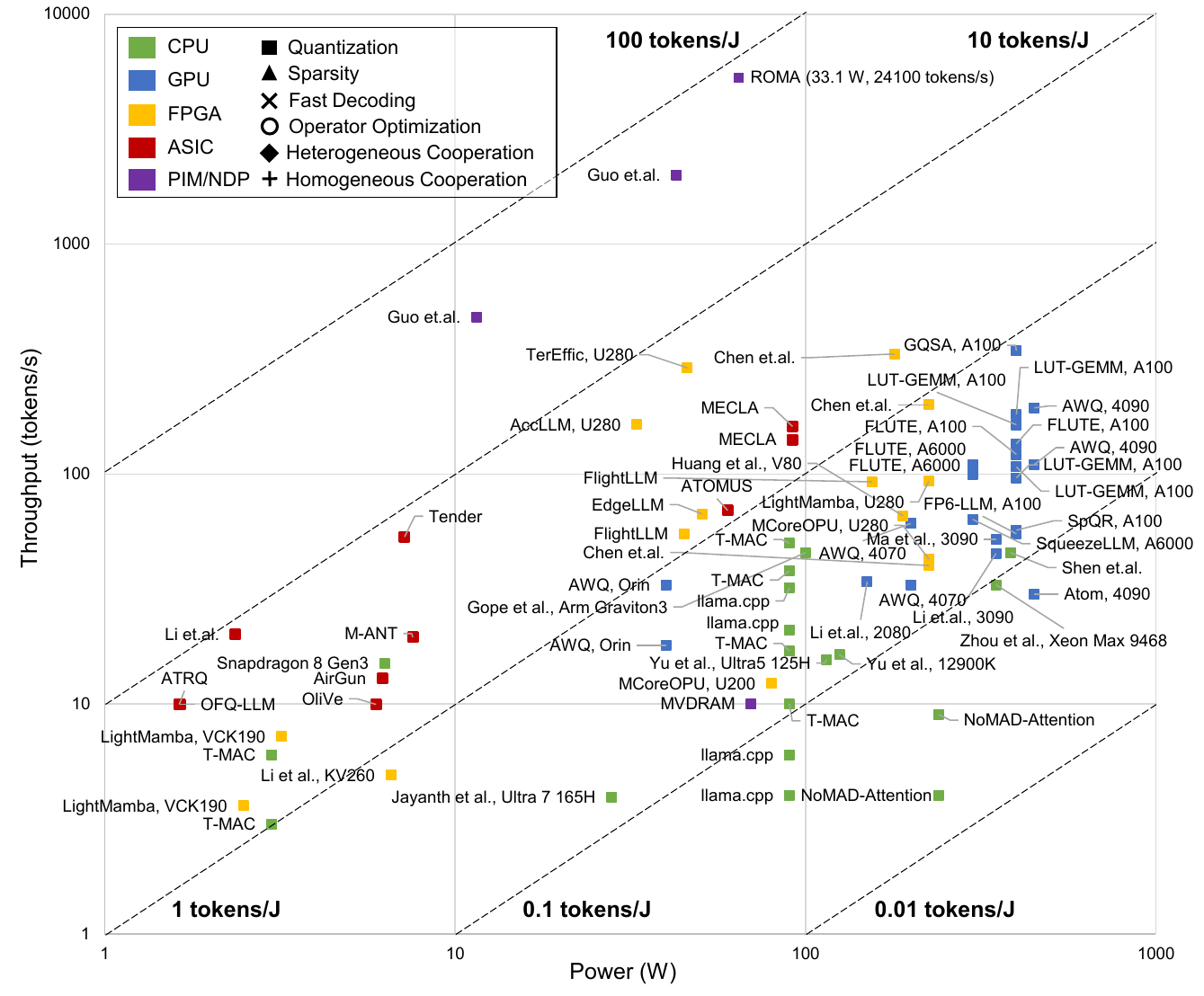}
  \vspace{-10pt}
  \caption{LLM ($\sim$ 7 billion parameters) decode stage throughput (batch size 1) vs power on different platforms with quantization.}
  \vspace{-10pt}
  \label{fig:sandain-quant}
\end{figure*}

\subsubsection{Quantitative Comparison}

\update{
We first compare the power consumption, inference speed and energy efficiency for different hardware platforms, in Figure~\ref{fig:sandain-quant}.
For quantization, power consumption ranges from 3W to 450W, with inference speeds between 3 tokens/s and 2700 tokens/s.
The energy efficiency ranges from 0.0057 tokens/J to 46.66 tokens/J.
T-MAC~\cite{wei2024t} (CPU) achieves the lowest power consumption with 3W and WaferLLM~\cite{he2025waferllm} (ASIC) achieves the highest throughput with batch size 1 (pre-silicon simulation result).
And Guo et al.~\cite{guo2024towards} also achieves the highest energy efficiency with 46.66 tokens/J.
\begin{itemize}
    \item 
    For CPUs, power consumption ranges from 3W to 385W, with inference speeds between 3 tokens/s and 50 tokens/s, located in the bottom part of the figure. The energy efficiency ranges from 0.0167 tokens/J to 2.38 token/J. Additionally, we observe edge CPUs (including CPU SoCs) with 3W to 9W power consumption exhibit higher energy efficiency (1 tokens/J to 2.38 tokens/J). 
    \item 
    For GPUs, power consumption ranges from 40W to 450W, with inference speeds between 18 tokens/s and 343.4 tokens/s, situated in the upper right part of the figure. The energy efficiency ranges from 0.067 tokens/J to 0.859 token/J. Compared to other hardware, GPUs can achieve higher absolute inference speeds due to their high computing power and high bandwidth. When quantization methods are used, the memory access bottlenecks in LLM inference are alleviated, further unlocking computing power.
    \item 
    For FPGAs, power consumption ranges from 2.490W to 225W, with inference speeds between 3.61 tokens/s and 333 tokens/s, also in the upper right part of the figure. The energy efficiency ranges from over 0.154 tokens/J to 6.30 tokens/J, which is higher than GPUs and server CPUs.
    \item 
    For ASICs, power consumption ranges from 1.632W to 91.84W, with inference speeds between 9.95 tokens/s and 161.1 tokens/s, found in the upper left section of the figure. The energy efficiency ranges from 1.16 tokens/J to over 8.49 tokens/J, outperforming CPU, GPU and FPGA hardware platforms.
    \item 
    For PIM/NDPs, power consumption ranges from 11.516W to 70W, with inference speeds between 10 tokens/s and 1998 tokens/s, found in the upper left section of the graph. The energy efficiency outperforms other hardware platforms.
\end{itemize}
}

Overall, both weight-only quantization and weight-activation quantization methods can enhance absolute inference speed and improve energy efficiency. 
Weight-only quantization reduces bandwidth requirements but introduces additional dequantization operations, which can increase hardware power consumption while improving absolute speed. 
On the other hand, weight-activation quantization reduces the hardware compute unit area and power consumption by using smaller-width computation units, leading to improved absolute speed while lowering overall hardware power consumption.

\subsection{Sparsity}
\subsubsection{Overview}
Sparsity reduces the number of non-zero elements and skip the multiplication and addition with zero to improve efficiency of computation and storage. 
Due to the presence of attention computations in standard transformer-based large models, sparsification methods include not only weight sparsity and activation sparsity but also attention sparsity.
\textbf{Weight sparsity} is primarily achieved through pruning methods, including global pruning, layer-wise pruning, and structured pruning, which reduce the size of weight matrices and leverage sparse matrix libraries for optimization. \textbf{Activation sparsity} focuses on reducing the computation of activation values by employing techniques such as activation pruning (e.g., threshold pruning) and dynamic sparsity, with hardware optimizations utilizing sparse data structures to enhance efficiency. \textbf{Attention sparsity} addresses the optimization of computations in self-attention mechanisms, employing methods like local attention, block-wise attention, and sparse attention matrices, which reduce computational load by limiting the calculation scope or using sparse matrix storage. These sparsity strategies help improve model inference efficiency, particularly when dealing with large-scale data and complex tasks.

\begin{figure*}[!t]
  \centering
  \includegraphics[width=0.8\textwidth]{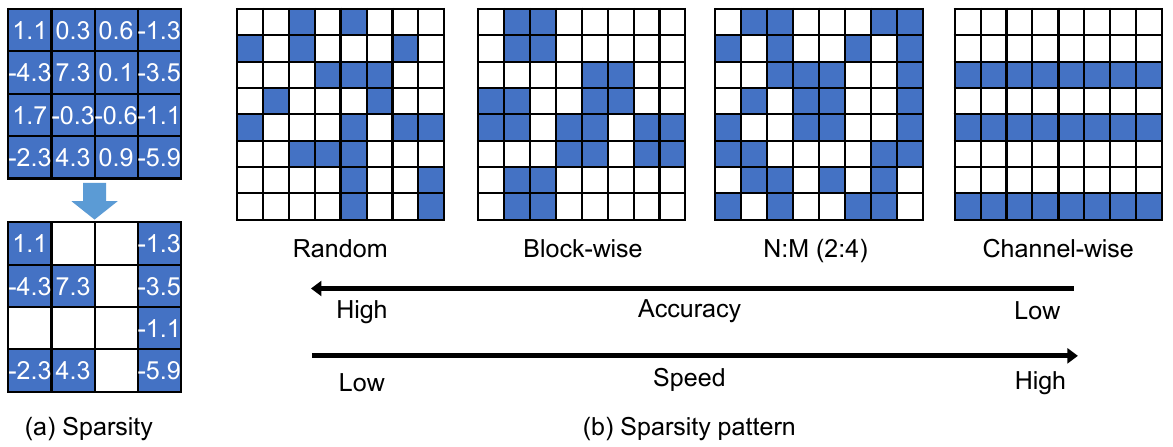}
  \vspace{-10pt}
  \caption{Sparsity and sparsity patterns.}
  \vspace{-10pt}
  \label{fig:sparsity}
\end{figure*}

Sparsity patterns can be categorized into random and structured sparsity as shown in Figure~\ref{fig:sparsity}. \textbf{Random pattern} involves a random distribution of zero elements within the matrix, achieving higher accuracy but potentially lower speed for computation. \textbf{Structured pattern} applies a specific pattern to the sparsity, improving computational efficiency by aligning with hardware optimizations. Within structured sparsity, common patterns include block-wise sparsity, N:M sparsity, channel-wise sparsity and some combinations of structured pattern sparsity. These structured patterns offer predictable and optimized computational benefits. 
Block-wise sparsity involves dividing the weight matrix into smaller blocks and applying sparsity within each block. 
N:M sparsity retains M non-zero elements out of every N elements, improving efficiency through hardware acceleration. NVIDIA's 2:4 sparse Tensor Core is a representative hardware unit for N:M sparsity, capable of achieving up to 2$\times$ computational acceleration.
Channel-wise sparsity aims to prune entire channels in a matrix, significantly reducing computation and storage needs.
Table~\ref{tab:sparsity} shows the usage of three sparsity methods across different hardware platforms. 

\begin{table}[htbp]
    \centering
    \caption{Sparsity on CPU, GPU, FPGA, ASIC, and PIM/NDP}
    \begin{tabular} {|c|c|c|c|}
    
        \hline
        Hardware & Weight Sparsity & Activation Sparsity & Attention Sparsity\\
        \hline
        CPU & \ding{55} & \ding{51} & \ding{55} \\
        \hline
        GPU & \ding{51} & \ding{55} & \ding{51} \\
        \hline
        FPGA & \ding{55} & \ding{55} & \ding{55} \\ 
        \hline
        ASIC & \ding{55} & \ding{55} & \ding{51} \\ 
        \hline
        PIM/NDP & \ding{55} & \ding{55} & \ding{55} \\ 
        \hline
    \end{tabular}
    \label{tab:sparsity}
\end{table}

\subsubsection{CPU}

\textbf{Activation Sparsity.}
Activation sparsity is determined by activation functions. 
Commonly using SwiGLU~\cite{shazeer2020glu} and GeGLU~\cite{hendrycks2016gaussian} exhibits limited sparsity for LLMs, but simply replacing these functions with ReLU fails to achieve sufficient sparsity. 
Turbo Sparse~\cite{song2024turbo} proposes the dReLU activation function to improve LLM activation sparsity, along with a high-quality training data mixture ratio to facilitate effective sparsity. 
By applying their sparsity method to the Mistral and Mixtral models, only 2.5 billion (35.7\%) and 4.3 billion (9.2\%) parameters are activated per inference iteration, respectively.
For Mistral-7B, Turbo Sparse achieves 8.71 tokens/s and 9.94 tokens/s on Intel i9-14900HX processor and Intel i7-12700K processor, respectively. 
For Mixtral-47B with 4-bit quantization, Turbo Sparse achieves 16.1 tokens/s, 11.98 tokens/s and 11.1 tokens/s on Intel i9-14900HX, Intel i7-12700K and SnapDragon 8 Gen3, respectively. 
ProSparse~\cite{song2024prosparse} also introduces activation function substitution, progressive sparsity regularization, and activation threshold shifting to help non-ReLU LLMs obtain high activation sparsity without performance degradation.
For Llama2-7B and Llama2-13B, ProSparse achieves high sparsity of 89.32\% and 88.80\%, and 16.3 tokens/s and 8.67 tokens/s, respectively, based on PowerInfer~\cite{song2023powerinfer} framework.

\textbf{Weight Sparsity.}
\update{
SparAMX~\cite{abouelhamayed2025sparamx} utilizes Advanced Matrix Extensions (AMX) support on the latest Intel CPUs with unstructured sparsity in linear layers to achieve a 1.42× reduction in end-to-end latency compared to the PyTorch implementation. It also provides a set of open-source customized sparse kernels to support all linear layers.
}

\subsubsection{GPU}
\textbf{Activation Sparsity.}
\update{
SoLA~\cite{sola} leverages Soft activation sparsity and Low-rank decomposition. SoLA
can identify and retain a minority of components significantly contributing to inference, while compressing the majority through low-rank decomposition, based on the activation pattern in the FFNs of LLMs. For Llama2-7B and 13B models, it achieves 76.69 tokens/s and 55.96 tokens/s on RTX 4090 GPU, respectively. 
R-Sparse~\cite{r-sparse} finds two key observations of FFNs: (i) the nonsparse components of the input can be regarded as a few bias terms, and (ii) the full computation can be effectively approximated by an appropriate combination of input channels and weight singular values. So it replaces the linear layers in LLMs with a rank-aware sparse inference method that leverages the sparsity of input channels and singular value components. For the Llama2-7B and Llama3-8B models, it achieves around 20 tokens/s and 18 tokens/s on a NVIDIA A6000 GPU, respectively. 
}

\textbf{Weight Sparsity.}
LLM-pruner~\cite{ma2023llm}, adopts structural pruning that selectively removes non-critical coupled structures based on gradient information, maximally preserving the majority of the LLM's functionality. 
To this end, the performance of pruned models can be efficiently recovered through tuning techniques, LoRA, in merely 3 hours, requiring only 50K data. 
We validate the LLM-Pruner on three LLMs, including Llama, Vicuna, and ChatGLM, and demonstrate that the compressed models still exhibit satisfactory capabilities in zero-shot classification and generation.

LLMs can be pruned to at least 50\% sparsity in one-shot, without any retraining, at minimal loss of accuracy. 
SparseGPT~\cite{frantar2023sparsegpt} requires a sophisticated weight update procedure in an iterative pruning process.
Wanda~\cite{sun2023wanda} prunes weights with the smallest magnitudes multiplied by the corresponding input activations, on a per-output basis. 
Notably, Wanda requires no retraining or weight update, where pruning process is faster. 
Besides unstructured pattern, these two methods generalizes to semi-structured N:M (2:4 and 4:8) patterns.
E-Sparse~\cite{li2023sparse} introduces entropy to quantify the information richness within each channel (intra-channel) of the input features, and adopts it to enhance the feature norms (crosschannel) as a metric to evaluate parameter importance. 
Furthermore, it proposes Channel Shuffle to reorder the information distribution in LLMs to obtain N:M Sparsity with less information loss.
2:4 sparsity as supported by NVIDIA GPUs of generation Ampere and newer theoretically offers 2$\times$ acceleration of matrix multiplications.  
In practical, 2:4 sparsity can achieve 1.54$\times$-1.79$\times$ speedup for MatMul, and end-to-end speedups are about 1.21$\times$-1.25$\times$ (due to some extra overheads from e.g. attention).

Based on the key observation that the bottleneck of LLM inference is the skinny matrix multiplications, Flash-LLM~\cite{xia2023flash} proposes a general Load-as-Sparse and Compute-as-Dense methodology for unstructured sparse matrix multiplication. 
Flash-LLM proposes a new sparse format called Tiled-CSL to relieve the memory bandwidth bottleneck and support the tile-by-tile SpMM execution with tensor cores.
For OPT-30B, Flash-LLM achieves 80\% sparsity with 1.44\% accuracy decrease and about 290 tokens/s, 500 tokens/s, 800 tokens/s, and 1187 tokens/s on single A100 GPU with batch sizes 8, 16, 32, and 64, respectively. 

Agarwalla et al.~\cite{agarwalla2024enabling} combine the SparseGPT one-shot pruning method and sparse pretraining to pretrain a high sparsity LLM.
They deploy model on GPU and CPU by utilizing Neural Magic’s DeepSparse engine and Neural Magic’s nm-vllm engine, respectively.
For Llama-7B, on NVIDIA A10 GPU, they achieve 44.4 tokens/s and 47.9 tokens/s with 50\% sparsity and 70\% sparsity, respectively.
On AMD EPYC 9R14 Processor, they achieve 4.4 tokens/s and 6.9 tokens/s with 50\% sparsity and 70\% sparsity, respectively.

Existing methods require costly retraining, forgo LLM’s in-context learning ability, or do not yield wall-clock time speedup on modern hardware. 
DejaVu~\cite{liu2023deja} predicts contextual sparsity on the fly given inputs to each layer, along with an asynchronous and hardware-aware implementation that speeds up LLM inference.
For OPT-175B model, DejaVu achieves up to 75\% sparsity and 50 tokens/s on 8 A100-80GB GPUs with batch size 1, which is over 2$\times$ and 6$\times$ faster than FasterTransformer and Hugging Face implementation, respectively.

For Mamba models, shihab et al.~\cite{shihab2025efficient} propose a novel unstructured pruning framework tailored for Mamba models that achieves up to 70\% parameter reduction while retaining over 95\% of the original performance. They integrate three key innovations: a gradient-aware magnitude pruning technique, an iterative pruning schedule, and a global pruning strategy. As a result, they achieve 2083.33 tokens/s and 1219.51 tokens/s on both Mamba-130M and Mamba-370M models on a single NVIDIA A100 GPU.

\textbf{Attention Sparsity.}
During the prefilling phase of LLM inference, attention computation complexity scales quadratically with input sequence length. 
Given limited GPU computing and memory resources, attention sparsification can reduce the number of attention values to accelerate prefilling phase.
For static sparsity, Sparse Transformer~\cite{child2019generating}, StreamingLLM~\cite{xiao2023streamingllm}, Bigbird~\cite{zaheer2020big}, and Longformer~\cite{beltagy2020longformer} use the manual combination of global and local patterns to replace the full attention patterns.
The local pattern captures the local context of each token within a fixed size or stride while the global pattern captures the relationship between the specific tokens to all other tokens.
For Llama2-7B and Llama2-13B models, StreamingLLM achieves 15.38-32.26 tokens/s and 9.43-20.83 tokens/s on single NVIDIA A6000 GPU, respectively.
For dynaimic sparsity, Adaptively Sparse Attention~\cite{correia2019adaptively} replaces softmax with $\alpha$-entmax, a differentiable generalization of softmax that allows low-scoring words to receive precisely zero weight and drops parts of the context that are no longer required for future generation.
Reformer~\cite{kitaev2020reformer} replaces dot-product attention by using locality-sensitive hashing, changing the complexity from $O(L^2)$ to $O(LlogL)$, where $L$ is the sequence length.
Sparse Flash Attention~\cite{pagliardini2023faster} extends FlashAttention~\cite{dao2022flashattention} GPU kernel and encompasses key/query dropping and hashing-based attention.
Sparse Sinkhorn Attention~\cite{tay2020sparse} adopts a learned sorting network to align keys with their relevant query buckets, ensuring that attention is computed only between the corresponding query-key pairs. 
H$_2$O~\cite{zhang2024h2o} observes that a small portion of tokens (called Heavy Hitters, H$_2$) contributes most of the value when computing attention scores.
H$_2$O introduces a dynamic attention sparsification method to adopt KV cache eviction policy that dynamically retains a balance of recent and H$_2$ tokens. 
For OPT-6.7B model, H$_2$O with 20\% H$_2$ achieves 30.4 tokens/s on single NVIDIA T4 GPU.

\subsubsection{FPGA}

\textbf{Weight Sparsity.}
FlightLLM~\cite{zeng2024flightllm} is the first real FPGA-based LLM accelerator which proposes a configurable sparse DSP chain to support different sparsity patterns with high computation efficiency. 
Then, it proposes an always-on-chip decode scheme to boost memory bandwidth with mixed-precision support.
Finally, it proposes a length adaptive compilation method to reduce the compilation overhead. 
For Llama2-7B model, FlightLLM achieves 55 tokens/s and 92.5 tokens/s with batch size 1 on the Xilinx Alveo U280 FPGA and Versal VHK158 FPGA, respectively.
EdgeLLM~\cite{huang2025edgellm} integrates 4-bit weight-only quantization and utilizes log-scale structural sparsity for weight parameters in the matrix multiplication operator.
For ChatGLM2-6B model, it achieves average 67 tokens/s with 55.07W power consumption on AMD Xilinx VCU128 FPGA.

\update{
Recent work mainly assumes structured model pruning and does not work well for unstructured sparsity. 
The reason behind is that it is difficult to exploit data reuse from the unstructured sparsity, leaving hardware underutilized. 
ChatOPU~\cite{zhao2024chatopu} proposes an FPGA-based overlay processor for LLMs, to support unstructured model pruning with better data reuse. 
It proposes a new diagonal dataflow on a systolic array to obtain efficient data reuse for both sparse and dense matrix multiplication. 
Then, it develops efficient encoding and decoding for the sparse parameters to save off-chip memory traffic and boosts the off-chip bandwidth utilization with pinned on-chip KVcache allocation and coalesced access throughout the LLM inference. 
Experimental results show that ChatOPU on Xilinx U200 FPGA can achieve 43.21 tokens/s and 166.26 tokens/s with 1 and 4 ChatOPU core for OPT-350M.
Zhang et al.~\cite{zhang2024fine} propose a sparse FPGA accelerator based on Fine-grained Structured Sparsity (FSS) and Channel Redistributed Fine-grained Structured Sparsity (CRFSS). 
For OPT-6.7B model, their approach achieves 51.744 tokens/s and 69.776 tokens/s with low-bit quantization on Alveo U280.
AccLLM~\cite{liang2025accllm} combines 2:4 semi-structured pruning, W2A8KV4 quantization, and sparse attention to enhance computational efficiency.
Extensive experimental results demonstrate for Llama2-7B model, it achieves 164 tokens/s with only 33W on Alveo U280.
}

\subsubsection{ASIC}
\textbf{Attention Sparsity.}
Spatten~\cite{wang2021spatten} leverages token sparsity, head sparsity, and quantization opportunities to reduce the attention computation and memory access.
It assesses the cumulative importance of each word by aggregating the attention matrix columns, subsequently pruning tokens with minimal cumulative significance from the input in subsequent layers.
It provides 2.88 TOPS computing power and 8.3W power consumption by considering all memory access under 40nm.
Experimental results shows that for GPT2-Medium, it can achieve about 35.86 tokens/s in decode stage.
TF-MVP~\cite{yoo2023tf} quantitatively analyzes sparsity patterns of pruned-transformer models with the cutting-edge fine-grained pruning scheme for the first time and presents the mixed-length vector pruning (MVP) procedure by utilizing this direction strength.
From hardware perspective, it introduces the TF-MVP architecture, a sparsity-aware cost-efficient accelerator design dedicated to the proposed pruned-transformer models.
Implemented in a 28nm CMOS technology at 400MHz, TF-MVP provides 0.835 TOPS with 1.721W on-chip power consumption for accelerating GPT-2 small model.
SOFA~\cite{wang2024sofa} predicts attention sparsity by using log-based add-only operations to avoid the significant overhead of prediction. 
Then, a distributed sorting and a sorted updating FlashAttention mechanism are proposed with a cross-stage coordinated tiling principle, which enables fine-grained and lightweight coordination among stages, helping optimize memory access and latency.  
SOFA provides 24.423 TOPS computing power and 3.4W power consumption under 28nm.
For Llama2-7B, compared to NVIDIA A100 GPU, it achieves 9.5$\times$ inference speedup in prefill stage.

\subsubsection{PIM/NDP}
\textbf{Weight Sparsity.}
LauWS~\cite{li2024lauws} proposes an unstructured sparsity method for NDP systems and is evaluated on a practical GDDR6-based bank NDP.
Not only the overall features of the total matrix but the features of the local region are preserved by ignoring non-feature values as much as possible, which is beneficial to the trade-off between high sparsity and least accuracy loss.
Compared to dense models, it achieves 1.23$\times$ and 1.24$\times$ speedup for GPT-2 small model and OPT-125M at 50\% sparsity, respectively.
Sharda et al.~\cite{sharda2024accelerator} propose to use the capacitorless 3D stackable DRAM to store much larger LLMs compared to conventional DRAM at higher density. 
To reduce the intermediate data size, they propose to use a layer-wise sparsity-quantization hybrid (LSQH) algorithm, which induces sparsity based on calculations performed using low-bit quantization to reduce both the energy consumption and the data storage requirements. 
Finally, a 3D heterogeneously integrated accelerator is designed by stacking a 3D DRAM with logic dies designed in the 3nm technology node at 1GHz. 
The evaluation shows that for Llama2-13B, it achieves 163k tokens/s in prefill stage with 193W power consumption. 

\update{
FlexCiM~\cite{ramachandran2025accelerating} presents a flexible layer-wise N:M sparsity accounting for the presence and distribution of outliers. 
FlexCiM enables support for diverse sparsity patterns by partitioning a digital CiM (DCiM) macro into smaller sub-macros, delivering up to 1.75× lower inference latency and 1.5× lower energy consumption compared to existing sparse accelerators.
}

\textbf{Attention Sparsity.}
HARDSEA~\cite{liu2023hardsea} proposes an attention sparsity method by predicting lightweight token relevance and design a hybrid analog-ReRAM and digital-SRAM in-memory computing accelerator. 
It employs ReRAM-CIM, whose precision is sensitive to circuit non-idealities, to take charge of token relevance prediction where only computing monotonicity is demanded. 
The SRAM-CIM, utilized for exact sparse attention computing, is reorganized as an on-memory-boundary computing scheme, thus adapting to irregular sparsity patterns. 
Experimental results show that HARDSEA prunes BERT and GPT-2 small model to 20\% sparsity without accuracy loss, achieving 5.8$\times$–6.7$\times$ speedup over NVIDIA RTX 3090 GPU. 

\begin{figure*}[!t]
  \centering
  \includegraphics[width=0.98\textwidth]{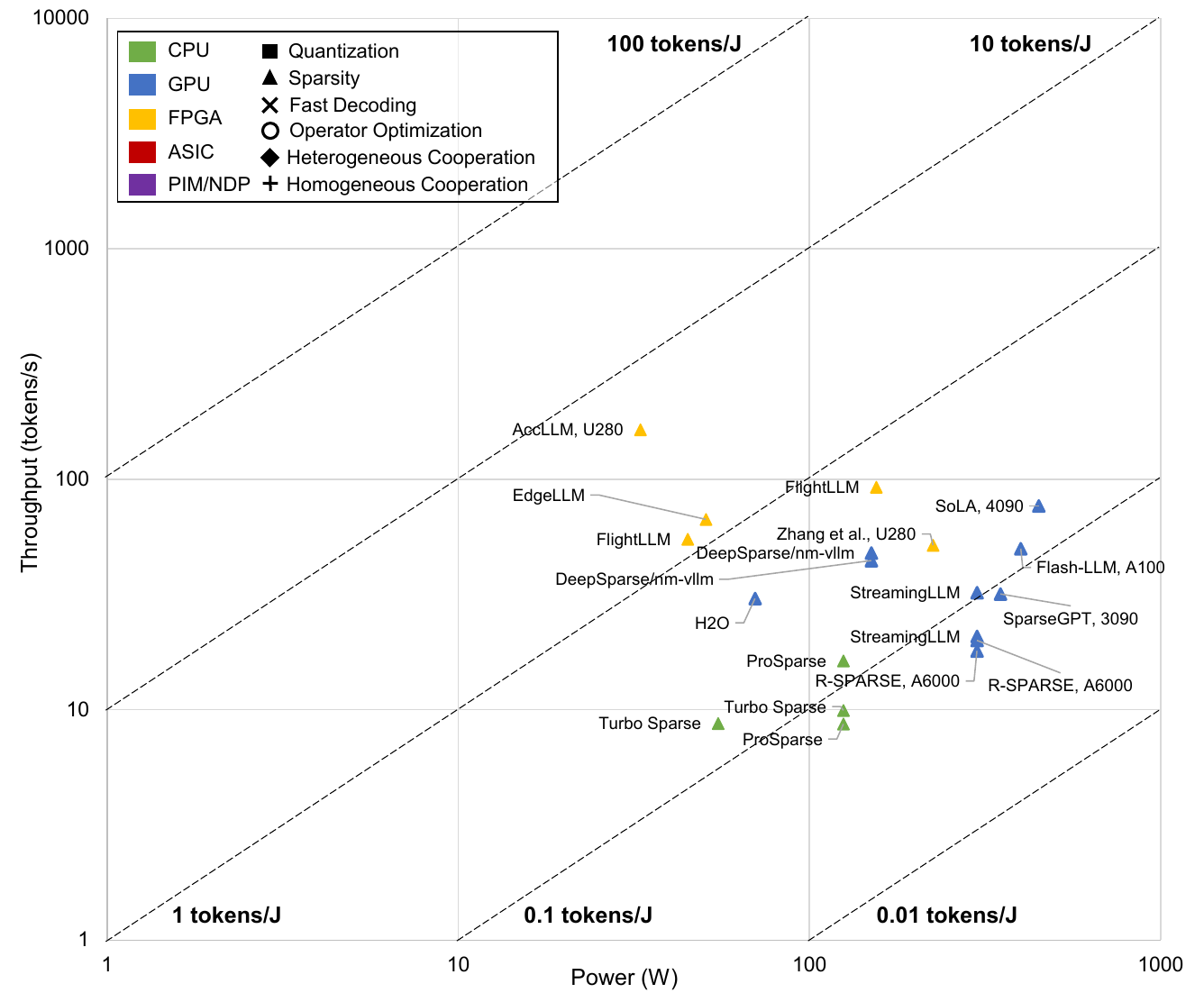}
  \vspace{-10pt}
  \caption{LLM ($\sim$ 7 billion parameters) decode stage throughput (batch size 1) vs power on different platforms with sparsity.}
  \vspace{-10pt}
  \label{fig:sandain-sparsity}
\end{figure*}

\subsubsection{Comparison}

\update{
In Figure~\ref{fig:sandain-sparsity}, for sparsity, power consumption ranges from 33W to 450W, with inference speeds between 8.67 tokens/s and 164 tokens/s.
The energy efficiency ranges from 0.06 tokens/J to 4.97 tokens/J.
AccLLM~\cite{liang2025accllm} (FPGA) achieves the lowest power consumption with 30.0167W and the highest throughput with batch size 1 (though it also applies quantization), which also achieves the highest energy efficiency with 4.97 tokens/J.
\begin{itemize}
    \item 
    For CPUs, power consumption ranges from 55W to 125W, with inference speeds between 8.67 tokens/s and 16.3 tokens/s, located in the bottom part of the figure. The energy efficiency ranges from 0.069 tokens/J to 0.158 token/J, which is much lower than FPGAs and ASICs. Currently, no edge-side CPUs have adopted sparsity methods to accelerate LLM inference. 
    \item 
    For GPUs, power consumption ranges from 70W to 450W, with inference speeds between 18 tokens/s and 66.69 tokens/s, situated in the middle right part of the figure. The energy efficiency ranges from 0.06 tokens/J to 0.434 token/J. Compared to CPUs, GPUs can achieve higher absolute inference speeds due to their high computing power and high bandwidth. However, the energy efficiency difference between CPUs is not significant.
    \item 
    For FPGAs, power consumption ranges from 45W to 225W, with inference speeds between 51.74 tokens/s and 164 tokens/s, also in the upper right part of the figure. The energy efficiency ranges from over 0.230 tokens/J to 4.97 tokens/J, which is much higher than GPUs and CPUs.
\end{itemize}
}

\subsection{Fast Decoding}
\subsubsection{Overview}

Traditional autoregressive decoding typically generates text token by token, choosing only the highest probability token at each step, known as greedy sampling. 
While this approach is simple and easy to implement, it may lead to a lack of diversity and creativity in the generated results. Another autoregressive method called nucleus sampling (or top-p sampling)~\cite{holtzman2019curious} considers multiple candidates during generation by setting a cumulative probability threshold p, allowing for sampling within a certain range. 
Although this method offers more diversity than greedy sampling, it still operates in a step-by-step generation manner. 
Currently, fast decoding techniques can be mainly divided into two categories: speculative decoding and skip layer.

\textbf{Speculative Decoding.} 
Speculative decoding is a technique for enhancing the generation efficiency of large language models (LLMs). Its core principle lies in using a draft model to quickly generate candidate outputs, which are then evaluated in depth by a main model, thereby accelerating the text generation process.
In the implementation of speculative decoding, a smaller draft model is first used to quickly generate multiple candidate words. This model can evaluate the context in a short time and propose various possible output options. Subsequently, the main model performs parallel evaluations of these candidates, calculating their probabilities or scores, and ultimately selects the candidate with the highest score for actual generation. Through this approach, speculative decoding combines speed and accuracy, significantly reducing the computation time while maintaining the quality of the generated text.
Common choices for draft models include: one option is to directly use a specific layer from the Transformer model, leveraging the existing architecture to maintain a certain level of feature extraction capability while accelerating inference speed; another option is to train a separate small model, which typically has fewer parameters and a simpler structure, focusing on rapidly generating candidate words and optimizing performance for specific tasks. Both methods have their advantages, and the choice can be made based on specific application needs, quality requirements, and computational resource constraints.

\textbf{Skip Layer.}
The working principle of skip layer technology is to dynamically and selectively skip certain layers during the model inference process, thereby reducing computational load and increasing generation speed. In practical implementation, the model evaluates the importance of each layer for the current task while processing input, and decides whether to execute the computation of specific layers based on preset heuristic rules or learned strategies.
In this method, the model typically consists of multiple layers, such as self-attention and feedforward layers in a Transformer. During inference, when the input features are relatively simple or the context complexity is low, the model can choose to skip the computation of certain intermediate layers. This choice may be based on real-time assessments, allowing the model to dynamically adjust its computation path according to different inputs.
To effectively implement skip layer, the model often requires optimization during the training phase, learning when to skip which layers. This can be achieved through methods such as policy gradients or reinforcement learning, enabling the model to adapt flexibly across various tasks.
By skipping unnecessary layers, skip layer technology can significantly accelerate inference speed while reducing computational resource consumption, making it particularly suitable for real-time systems and resource-constrained environments. Overall, this method enhances the efficiency and applicability of large language models by intelligently selecting the computation process.

\begin{figure*}[!t]
  \centering
  \includegraphics[width=0.6\textwidth]{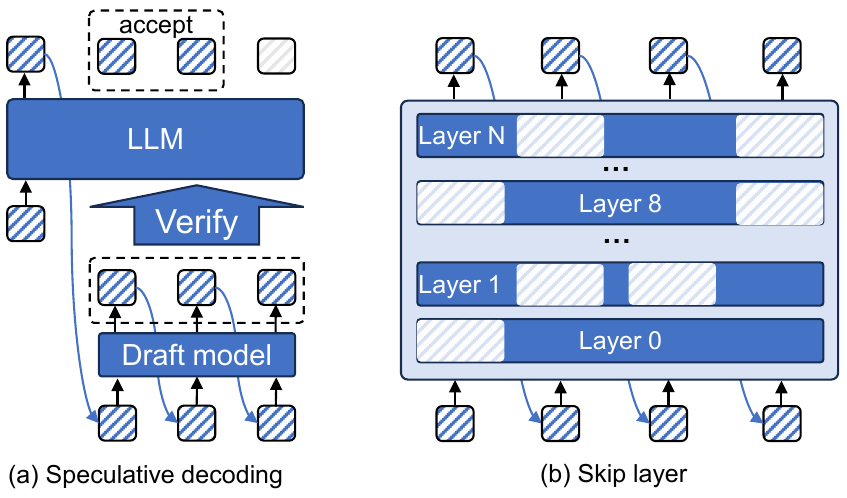}
  \vspace{-10pt}
  \caption{Fast decoding.}
  \vspace{-10pt}
  \label{fig:fastdec}
\end{figure*}

\begin{table}[htbp]
    \centering
    \caption{Fast decoding on CPU, GPU, FPGA, ASIC, and PIM/NDP}
    \begin{tabular} {|c|c|c|}
    
        \hline
        Hardware & Speculative Decoding & Skip Layer\\
        \hline
        CPU & \ding{51} & \ding{55} \\
        \hline
        GPU & \ding{51} & \ding{51} \\
        \hline
        FPGA & \ding{55} & \ding{55} \\ 
        \hline
        ASIC & \ding{51} & \ding{55} \\ 
        \hline
        PIM/NDP & \ding{51} & \ding{55} \\ 
        \hline
    \end{tabular}
    \label{tab:fastdecoding}
\end{table}

    

\subsubsection{CPU}
\update{
ML-SpecQD~\cite{georganas2025ml} uses MXFP4 models as drafts. 
And the MXFP4 draft token generation itself can be accelerated via speculative decoding by using yet another smaller draft. 
This multi-Level speculative decoding with MXFP4 quantized drafts achieves speedups up to 2.72x over the BF16 baseline.
}

\subsubsection{GPU}

\textbf{Speculative Decoding.}
Speculative decoding~\cite{stern2018blockwise} is proposed to overcome the inherently sequential process in the autoregressive decoding of LLM. The essential decoding mechanism is to make predictions (\textit{i.e.}, draft tokens) parallelly for multiple time steps and then select the longest prefix verified by a scoring model as the final output. 
Lookahead decoding~\cite{lookahead} adopts the \textit{Guess-and-Verify} paradigm as the whole decoding mechanism which generates the draft tokens by $n$-gram method and verifies the draft tokens during the forward at the same time. For Llama-7B and Llama-13B models with different datasets, it achieves 65.12-94.51 tokens/s and 56.01-90.05 tokens/s on a single NVIDIA A100-80GB GPU. To improve the acceptance rate while maintaining generation quality, many works focus on the scheme of generating draft tokens. 
Medusa~\cite{medusa} adds extra decoding heads to generate multiple subsequent tokens in parallel and uses a tree-based attention mechanism to construct multiple candidate continuations and verify them simultaneously in each decoding step. For Vicuna-7B and Vicuna-13B, it achieves 129.86 tokens/s and 98.54 tokens/s on a single NVIDIA A100-80GB GPU. 
EAGLE~\cite{eagle1,eagle2} selects a single transformer layer with the same configuration as LLM as the draft model to make predictions autoregressively and combines the feature and token embedding as the input of the draft model. During the verification phase on the target model, EAGLE chooses the tree-based attention mechanism similar to Lookahead and Medusa to ensure the correct relationship between the draft tokens. For Vicuna-7B, Vicuna-13B, Llama2-Chat-7B, and Llama-2-Chat-13B models, it achieves 139.95 tokens/s, 132.31 tokens/s, 133.98 tokens/s, and 156.80 tokens/s on a single NVIDIA A100-80GB GPU. Based on the basic paradigm of speculative decoding for prediction by draft models and verification by target models, some studies explore the optimizations on the modules in it. 
Ouroboros~\cite{zhao2024ouroboros} constructs a phrase candidate pool from the verification process of LLMs to provide candidates for the draft token generation of the draft model. Different from Medusa and EAGLE, Ouroboros uses the smaller LLMs (\textit{e.g.}, DeepSeek-7B) as the draft models for the target models (\textit{e.g.}, DeepSeek-34B). For Yi-34B, DeepSeek-34B and CodeLlama-34B models, it achieves 61.20 tokens/s, 41.00 tokens/s and 39.2 tokens/s on a single NVIDIA A100-80GB GPU. During the prediction phase, 
Sequoia~\cite{chen2024sequoia} introduces a dynamic programming algorithm and a hardware-aware tree optimizer to find the optimal tree structure based on the runtime features and the given hardware platform. During the verification phase, Sequoia uses a novel sampling and verification method that outperforms prior work across different decoding temperatures. For Llama2-7B and Llama2-13B models, it achieves 169.68 tokens/s and 149.20 tokens/s on a single NVIDIA A100-80GB GPU.

The studies mentioned above all require the help of an auxiliary model (\textit{e.g.}, a single transformer layer in EAGLE, several Medusa heads in Medusa, or a smaller LLM like Llama2-7B in Ouroboros) or the statistical methods (\textit{e.g.}, $n$-gram in Lookahead) to generate the predicted tokens and then the target model is utilized to verify the predicted tokens. Some other studies~\cite{zhang2023draft,liu2024kangaroo,layerskip} also explore the LLM inference acceleration without the need of auxiliary models, called self-speculative decoding. Draft\&Verify~\cite{zhang2023draft} generates draft tokens by selectively skipping certain intermediate layers of LLMs. Subsequently, the draft tokens will be verified in one forward pass. For the Llama2-13B model, it achieves 62.23 tokens/s on a single NVIDIA A100-40GB GPU. Kangaroo~\cite{liu2024kangaroo} adopts a fixed shallow sub-network of the LLM as the draft model, with the remaining layers serving as the target model. To enhance the representation ability of the draft model (\textit{i.e.}, the shallow sub-network), it trains an adapter module to follow the sub-network. For Vicuna-7B and Vicuna-13B models, it achieves 138.14 tokens/s and 105.89 tokens/s on a single NVIDIA A100-80GB GPU. LayerSkip~\cite{layerskip} proposes to exit at early layers and verify and correct with remaining layers of the model. During training, it applies layer dropout with low dropout rates for earlier layers and higher dropout rates for later layers, and adds an early exit loss to increase the accuracy of early exit at earlier layers. For the Llama2-13B model, it achieves 66.37 tokens/s on a single NVIDIA H100 GPU. 
LLMA~\cite{yang2023inference} is motivated by the observation that there are abundant identical text spans between the decoding result by an LLM and the reference that is available in many real-world scenarios (e.g., retrieved documents). 
LLMA first selects a text span from the reference and copies its tokens to the decoder and then efficiently checks the tokens’ appropriateness as the decoding result in parallel within one decoding step. 
For Llama-7B and Llama-13B models, it achieves 59.2 tokens/s and 41.1 tokens/s on a single NVIDIA V100 GPU, respectively.

\update{
AMUSD (Asynchronous Multi-device Speculative Decoding)~\cite{amusd} introduces a system that further accelerates generation by decoupling the draft and verify phases into a continuous, asynchronous approach. 
It enables both models to perform predictions independently on separate GPUs, instead of only one model(draft or verify) performing token generation at a time. For Llama3.1-8B model as a verify model and Llama3.2-1B model as a draft model, it achieves an average 71.02 tokens/s improvement over speculative decoding on two NVIDIA A100 GPUs with around 300W power consumption.
PipeSpec~\cite{pipespec} presents a framework that generalizes speculative decoding to k models arranged in a hierarchical pipeline, enabling asynchronous
execution with lightweight coordination for
prediction verification and rollback. For the Llama2-68M and 7B model combination, it achieves 79.18 tokens/s on a single NVIDIA A100 GPU, while for Llama3.1-1B, 8B, and 70B models, it achieves 20.51 tokens/s on two A100 GPUs.
Adaptix~\cite{adaptix} utilizes a tri-gram matrix-based LLM representation to dynamically approximate the output distribution of the LLM, allowing the model to adjust to changing token probabilities during the decoding process. It also implements a draft construction mechanism that effectively balances exploration and exploitation. For Llama2-7B and Llama-13B models, it achieves 85.62 tokens/s and 45.83 tokens/s on a NVIDIA A6000 GPU, respectively.
SPIN~\cite{spin} improves token speculation by using multiple heterogeneous SSMs(Small Speculative Models), with a learning-based algorithm for SSM selection that operates without prior knowledge of request difficulty. It also orchestrates speculation and verification phases by pipelining their executions on GPUs to achieve further acceleration. For Llama2-7B model as a verify model and Llama2-68M model as a draft model, it achieves 62.33 tokens/s on a NVIDIA V100 GPU.
PARD~\cite{pard}(PARallel Draft) enables low-cost adaptation of autoregressive draft models into parallel draft models. It predicts multiple future tokens in a single forward pass of the draft phase. For Llama3.1-8B model, it achieves 274.4 tokens/s on a single NVIDIA A100 GPU.
Judge Decoding~\cite{judge_decoding} noticed that even GPT-4o, as a draft model, cannot achieve high acceptance rates. So it investigated how its alignment between draft and target response leads to the rejection of objectively correct continuations, and then proposed an adapted verification scheme that judges responses in a versatile way. As a result, it achieves 141.8 tokens/s with Llama3.1-8B as draft model and Llama3.1-70B as a verify model on two NVIDIA H100 GPUs. To relieve the significant latency of the draft model in speculative decoding, Falcon~\cite{falcon} proposes the Coupled Sequential Glancing Distillation technique to fortify inter-token dependencies within the same block during draft model training to increase speculation accuracy. For the Llama2-Chat 7B model, it achieves 147.38 tokens/s on a single NVIDIA H800 GPU.
}

\textbf{Skip Layer.}
The skip layer method~\cite{adainfer,RAEE,MOD} is proposed based on the idea that not all layers of LLMs are necessary during inference. AdaInfer~\cite{adainfer} statistically analyzes the activated layers across tasks and proposes a simple algorithm to determine the inference termination moment based on the input instance adaptively. Due to the introduction of the overhead in each layer, which is not friendly to the decoding phase, AdaInfer only takes some Benchmarks for the Q\&A tasks and achieves 25.2 tokens/s for Llama2-7B on a single NVIDIA V100 GPU. Based on the basic dataflow of Adainfer, RAEE~\cite{RAEE} proposes to build the retrieval database to store the token information offline and leverages the information of the retrieved similar token by searching the pre-built retrieval database to guide the backbone model to exit at the layer. For the LM-BFF which is finetuned based on RoBERT-large-350M, RAEE achieves 26.23 tokens/s. The studies mentioned above both use continuous shallow layers for inference, called early exiting. MOD~\cite{MOD} decides whether to skip the current layer or not by pretraining the model to add a router in each layer like Mixture-of-Experts. This achieves a dynamic selection of partial layers for computation instead of forcing continuous layers. The model of MOD is not open source and no corresponding results on throughput are given in the paper.
\update{
SpecEE~\cite{specee} predicts whether it is early existing through a lightweight predictor by reducing the vocabulary space. And for this method, it designs the Two-level heuristic predictor scheduling and the Context-aware merged mapping for the predictor.  For Llama-7B and Llama-13B models, SpecEE + EAGLE achieves 124.66 tokens/s and 120.8 tokens/s on a single NVIDIA A100 GPU, respectively.
}

\subsubsection{ASIC}
\textbf{Speculative Decoding.}
C-Transformer~\cite{kim202420} adopts a big-little network, which is composed of the original GPT-2 big model and a 1/10$\times$ smaller model, and a reconfigurable homogeneous architecture to increase hardware utilization and energy efficiency.
During inference, only the little model computation is performed, and if the prediction probability of a specific token is over a predefined threshold, the big model computation is skipped leading to memory access reduction.
Then, workloads are divided into two domains: adder-based spike domain which is efficient for small input values, and multiplier-based non-spike domain which is efficient for large input values. 
It has 1.431W power consumption and is fabricated in 28nm CMOS technology. 
For GPT-2 large model, C-Transformer achieves 2146.75 tokens/s in prefill stage.

\begin{figure*}[!t]
  \centering
  \includegraphics[width=0.98\textwidth]{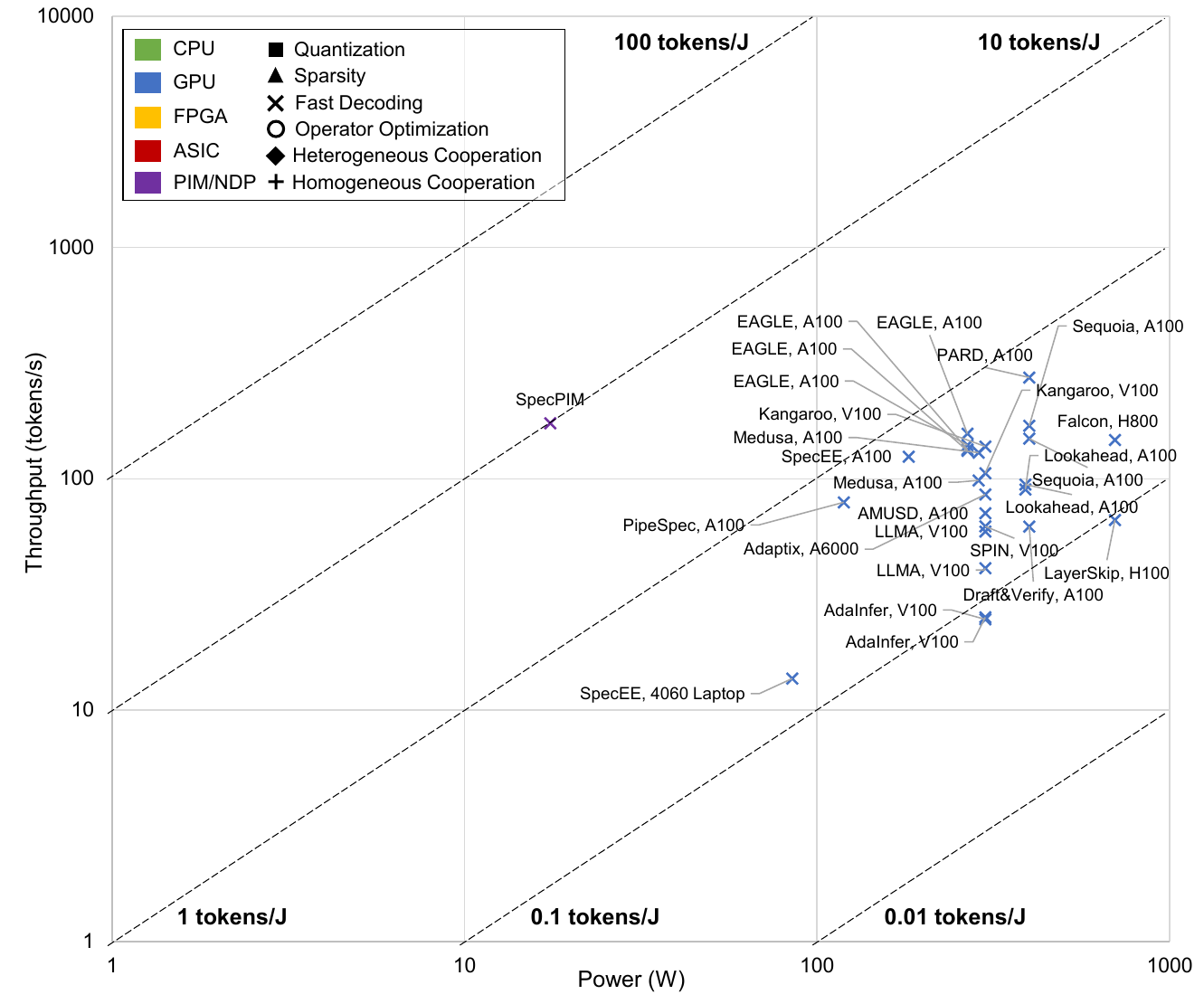}
  \vspace{-10pt}
  \caption{LLM ($\sim$ 7 billion parameters) decode stage throughput (batch size 1) vs power on different platforms with fast decoding.}
  \vspace{-10pt}
  \label{fig:sandian-fastdec}
\end{figure*}

\subsubsection{PIM/NDP}
\textbf{Speculative Decoding.}
SpecPIM~\cite{li2024specpim} aims to accelerate speculative inference on the PIM-enabled system by extensively exploring the heterogeneity brought by both the algorithm and the architecture. 
It constructs the architecture design space to satisfy each model's disparate resource demands and dedicates the dataflow design space to fully utilize the system's hardware resources. 
Based on the co-design space, it also proposes a design space exploration framework to provide the optimal design under different target scenarios. 
Compared with speculative inference on GPUs and existing PIM-based LLM accelerators, SpecPIM achieves 1.52$\times$/2.02$\times$ geomean speedup and 6.67$\times$/2.68$\times$ geomean higher energy efficiency.

\subsubsection{Comparison}

\update{
In Figure~\ref{fig:sandian-fastdec}, for fast decoding, power consumption ranges from 17.499W to 700W, with inference speeds between 13.7 tokens/s and 274.4 tokens/s.
The energy efficiency ranges from 0.082 tokens/J to 9.944 tokens/J.
SpecPIM~\cite{li2024specpim} (PIM/NDP) achieves the lowest power consumption with 17.499W, the highest throughput with batch size 1, and the highest energy efficiency.
\begin{itemize}
    \item 
    For GPUs, power consumption ranges from 85W to 700W, with inference speeds between 13.7 tokens/s and 274.4 tokens/s. The energy efficiency ranges from 0.082 tokens/J to 0.686 tokens/J. 
    \item 
    For PIM/NDPs, SpecPIM is the only one using speculative decoding, which outperforms GPU platforms.
\end{itemize}
}

\subsection{Operator Optimization}
\subsubsection{Overview}

Improving the execution efficiency of operators is crucial for LLM eras, which not only involves enhancing computational speed but also maximizing resource utilization. As the scale and complexity of models continue to increase, traditional operator execution methods become increasingly inefficient, prompting the need to explore various optimization strategies. 

The following four methods provide effective solutions for operator optimization, significantly enhancing the performance and responsiveness of models across different hardware platforms.
\textbf{Fusion.}
Operator fusion reduces the storage and transmission needs of intermediate data by merging multiple independent operators into a single entity. This approach not only lowers I/O overhead but also reduces redundant computations during the execution process, thereby improving overall efficiency. Operator fusion enables hardware to utilize cache and bandwidth more effectively, significantly boosting computational performance.
\textbf{Nonlinear Function Approximation.}
Common nonlinear activation functions in deep learning models often require specialized hardware support, which can occupy substantial chip resources. By employing linear approximations, we can achieve computations using less expensive hardware while maintaining algorithmic accuracy. This optimization strategy makes complex nonlinear calculations more efficient, making it suitable for resource-constrained environments.
\textbf{Coarse-grained Processing.}
When handling large-scale matrix operations, fine-grained computational units may lead to frequent resource scheduling and contention, introducing additional overhead. Coarse-grained processing simplifies the scheduling process by merging multiple small computational units into larger ones, reducing resource contention. This method effectively enhances computational coherence and efficiency, particularly in parallel computing environments.
\textbf{Storage Optimization.}
Storage optimization strategies focus on the arrangement of data in memory, aiming to minimize latency caused by memory access during computation. By strategically organizing data storage locations and access patterns, we can significantly improve data access efficiency and enhance overall computational performance. This optimization is closely related not only to hardware performance but also to algorithm design.

\begin{figure*}[!thbp]
  \centering
  \includegraphics[width=0.68\textwidth]{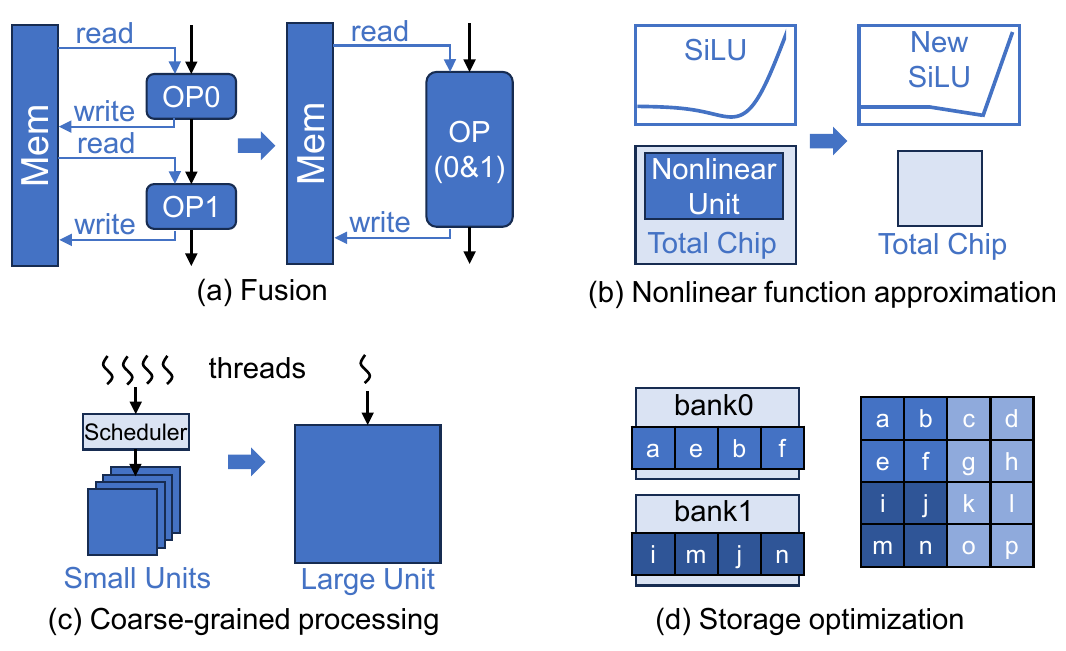}
  \vspace{-10pt}
  \caption{Operator optimizations.}
  \vspace{-10pt}
  \label{fig:operator}
\end{figure*}

Table~\ref{tab:operator} shows the usage of these four operator optimization methods across different hardware platforms. By comprehensively applying these four optimization methods, the efficiency of operator execution can be significantly improved, allowing deep learning models to handle complex tasks more effectively. Selecting and combining these optimization strategies appropriately for specific application scenarios and hardware platforms will be key to achieving high-performance computing.

\begin{table}[htbp]
\footnotesize
    \centering
    \caption{Operator optimization on CPU, GPU, FPGA, ASIC, and PIM/NDP}
    \begin{tabular} {|c|c|c|c|c|}
    
        \hline
        Hardware & Fusion & Nonlinear Function Approximation & Coarse-grained Processing & Storage Optimization \\
        \hline
        CPU & \ding{55} & \ding{55} & \ding{55} & \ding{55} \\
        \hline
        GPU & \ding{51} & \ding{55} & \ding{55} & \ding{55} \\
        \hline
        FPGA & \ding{55} & \ding{51} & \ding{55} & \ding{55} \\ 
        \hline
        ASIC & \ding{51} & \ding{51} & \ding{51} & \ding{55} \\ 
        \hline
        PIM/NDP & \ding{55} & \ding{51} & \ding{51} & \ding{51} \\ 
        \hline
    \end{tabular}
    \label{tab:operator}
\end{table}

\subsubsection{CPU}
\textbf{Storage Optimization.}
For the on-device inference of LLM with resource constraints, FlexInfer~\cite{flexinfer} proposes methods like asynchronous prefetching, balanced memory locking, and flexible tensor preservation to enhance memory efficiency and relieve bandwidth bottlenecks. For the Llama2-7B model, it achieves 12 tokens/s on an AMD 7995WX CPU.

\update{
\textbf{Fusion.}
V-Seek~\cite{rodrigo2025vseekacceleratingllmreasoning} optimizes LLM inference on the Sophon SG2042, the first commercially available many-core RISC-V CPU with vector processing capabilities. For DeepSeek R1 Distill Llama 8B and DeepSeek R1 Distill QWEN 14B, we achieve 4.32/2.29 token/s for token generation and 6.54/3.68 token/s for prompt processing.
}

\subsubsection{GPU}
\textbf{Fusion.}
To address the quadratic memory requirements in the attention computation, FlashAttention~\cite{flashattention, flashattention2} fuses the attention operation into a single operator by tiling input matrices (Q, K, V) and the attention matrix into blocks. Based on FlashAttention, FlashDecoding~\cite{flashdecoding} proposes additionally the parallel computation along the feature dimension, improving performance for small batch size during the decoding phase. FlashDecoding achieves 95.07 tokens/s and 54.19 tokens/s for Llama2-7B on a single NVIDIA A100 and RTX 3090 GPU respectively. Subsequently, FlashDecoding++~\cite{hong2024flashdecoding} optimizes the synchronization overhead in softmax computation by pre-determining a unified maximum based on statistical analysis in advance, enabling the parallel execution of subsequent operations and improving efficiency in typical LLMs like Llama2 and ChatGLM~\cite{chatglm}. FlashDecoding++ achieves 115.57 tokens/s and 61.66 tokens/s for Llama2-7B on a single NVIDIA A100 and RTX 3090 GPU respectively.
The linear operator is widely used in deep learning. In the common framework for deep learning (\textit{e.g.}, Pytorch), the backend of the linear operator is usually the General Matrix-Matrix Multiplication (GEMM) operation supported by NVIDIA. The naive implementation by HuggingFace achieves 44.60 tokens/s and 36.29 tokens/s for Llama2-7B on a single NVIDIA A100 and RTX 3090 GPU respectively. In the LLM framework (\textit{e.g.}, DeepSpeed~\cite{deepspeed}, vLLM~\cite{kwon2023vllm}, and OpenPPL~\cite{OpenPPL}), the GEMM implementation provided by cuBLAS~\cite{cuBLAS} APIs is generally optimized. For the Llama2-7B model on a single NVIDIA A100 GPU, they achieve 89.24 tokens/s, 88.45 tokens/s and  93.63 tokens/s. For the Llama2-7B model on a single NVIDIA RTX 3090 GPU, they achieve 51.28  tokens/s, 53.31 tokens/s and 55.99 tokens/s. To address the inefficiency of GEMM during decoding due to the reduced dimensions, TensorRT-LLM~\cite{tensorrt-llm} introduces a dedicated General Matrix-Vector Multiplication (GEMV) implementation to support the decoding phase of LLM when batch size equals 1. TensorRT-LLM achieves 98.19 tokens/s for Llama2-7B on a single NVIDIA A100 GPU.
For the smaller batch size during the decoding phase, FlashDecoding++~\cite{hong2024flashdecoding} introduces FlatGEMM to address the inefficiencies in cuBLAS~\cite{cuBLAS} and CUTLASS~\cite{CUTLASS} libraries and employs fine-grained tiling and double buffering techniques to improve parallelism and reduce the latency of memory access. Moreover, it adopts a heuristic selection mechanism to dynamically select the most efficient operator based on the input. The performance of FlashDecoding++ is shown above.
%
Operator fusion~\cite{flashattention,bytetransformer,deepspeed,hong2024flashdecoding} is a common and effective optimization to reduce the runtime memory access, eliminate kernel launching overhead and enhance parallelism. 
ByteTransformer~\cite{bytetransformer} and DeepSpeed~\cite{deepspeed} fuse the main operator (\textit{e.g.}, GEMM) and the lightweight operators(\textit{e.g.}, residual adding, layernorm and activation functions) into a single kernel to reduce the kernel launching overhead. 
FlashAttention~\cite{flashattention} mentioned above fuse the attention operator into one single kernel, reducing significantly the overhead of memory access and the memory requirements. 
FlashDecoding++~\cite{hong2024flashdecoding} also achieves the integration of seven fused kernels in a transformer block. 

\update{SpInfer~\cite{spinfer} is a high-performance framework tailored for sparsified LLM inference on GPUs. It introduces Tensor-Core-Aware Bitmap Encoding (TCA-BME) to minimize indexing overhead
and integrates an optimized SpMM kernel with Shared Memory Bitmap Decoding (SMBD) and an asynchronous pipeline. It achieves around 500 tokens/s with batch size 8 for OPT-13B model on a single RTX 4090 GPU. 
}

\update{
\textbf{Storage Optimization.}
FlashFormer~\cite{flashformer} introduces a proof-of-concept kernel for accelerating single-batch inference, which exploits increased
overlapping of memory movement and amortizes the launch overhead for the kernel over the entire forward pass. For Llama3.1-8B model, it achieves 181 tokens/s on a NVIDIA H100 GPU. To tackle  the memory bandwidth bottleneck during LLM inference, Dong et al.~\cite{dong2025accelerating} proposes L2 Cache-oriented asynchronous KV Cache prefetching method through computation-load overlap, which achieves around 3000 tokens/s for the Llama2-7B model on a single NVIDIA H20 GPU with batch size 64.
}

\subsubsection{FPGA}
\update{
\textbf{Nonlinear Function Approximation.}
HAAN~\cite{peng2025haan} is a holistic normalization acceleration approach for LLMs. 
HAAN exploits the correlation in normalization statistics among adjacent layers to bypass normalization computation by estimating statistics from preceding layers. 
}

\subsubsection{ASIC}

\textbf{Fusion.}
In 2020, Groq company introduces the Tensor Streaming Processor (TSP) architecture~\cite{abts2020think}, a functionally-sliced microarchitecture with memory units interleaved with vector and matrix compute units in order to take advantage of dataflow locality. 
The first TSP implementation yields a computational density of more than 1 TOPS/$mm^2$ for its 25$\times$29 mm 14nm chip at 900 MHz. 
In 2022, Groq company introduces a novel software-defined communication approach for large-scale interconnection networks of TSP elements~\cite{abts2022software}. 
This scalable communication fabric is based on a software-defined dragonfly topology, ultimately yielding a parallel machine learning system with elasticity to support a variety of workloads. 
Each TSP contributes 220 MB to the global memory capacity, with the maximum capacity limited only by the network's scale. 
The large-scale parallel system achieves up to 10,440 TSPs and more than 2 TB of global memory accessible in less than 3ms of end-to-end system latency.
Based on two previous works, Groq Language Processing Unit (LPU)~\cite{groq_lpu} 
is fabricated in 14nm with 185W power consumption.
According to~\cite{groq_lpu_result}, LPU achieves 814 tokens/s for Gemma-7B model.
Another commercial company, HyperAccel, also proposes a LLM inference chip with dataflow architecture. 
Latency Processing Unit (LPU)~\cite{moon2024lpu} introduces streamlined hardware that maximizes the effective memory bandwidth usage during end-to-end inference regardless of the model size to achieve up to 90\% bandwidth utilization for high-speed text generation. 
It consists of expandable synchronization link (ESL) that hides bulk of the data synchronization latency in a multi-device system to achieve nearperfect scalability.
Its on-chip power is 0.284W sythesised by Samsung 4nm and the total system power is 86W with 96GB HBM3.
For OPT-1.3B/6.7B/13B/30B, LPU achieves 769.23 tokens/s, 217.39 tokens/s, 112.40 tokens/s and 49.26 tokens/s, respectively.
\update{
Based on Cerebras WSE2, WaferLLM~\cite{he2025waferllm} is the first wafer-scale LLM inference system. For full LLM inference, WaferLLM achieves 2700 tokens/s and 2039 tokens/s on Llama3-8B and Llama2-13B with 434W and 354W, respectively.
}
The Wafer-Scale Engine (WSE-3)~\cite{lie2024wafer}, which powers the Cerebras CS-3 system, is the largest chip ever built. The WSE-3 is 57$\times$ larger than the largest GPU, has 52$\times$ more compute cores, and 880$\times$ more high performance on-chip memory. The only wafer scale processor ever produced, it contains 4 trillion transistors, 900,000 AI-optimized cores, and 44 gigabytes of high performance on-wafer memory. 
Each wafer consists of 84 dies with 40GB on-chip memory and 15kW power consumption.
For Llama3.1-8B, it can achieve about 1800 tokens/s, which is 20$\times$ faster than hyperscale clouds.

\textbf{Nonlinear Function Approximation.}
Constant Softmax (ConSmax)~\cite{liu2024consmax} is a software-hardware co-design as an efficient Softmax alternative. 
ConSmax employs differentiable normalization parameters to remove the maximum searching and denominator summation in softmax. 
It allows for massive parallelization while performing the critical tasks of softmax. 
In addition, a scalable ConSmax hardware utilizing a bitwidth-split LUT can produce lossless non-linear operation and support mix-precision computing. It further facilitates efficient LLM inference. 
Experimental results show that ConSmax achieves a minuscule power consumption of 0.43mW and area of 0.001$mm^2$ at 1GHz working frequency and 22nm CMOS technology. 
MARCA~\cite{li2024marca} is the first accelerator with reconfigurable architecture tailored for Mamba model.
It proposes a reduction alternative PE array architecture for both linear and element-wise operations. 
For linear operations, the reduction tree connected to PE arrays is enabled and executes the reduction operation.
For element-wise operations, the reduction tree is disabled and the output bypasses.
And it proposes a reusable nonlinear function unit based on the reconfigurable PE and decomposes the exponential function and activation function (SiLU) into element-wise operations to reuse the reconfigurable PEs.
Extensive experiments show that MARCA can achieve 23.78 tokens/s with batch size 1 for Mamba-2.8B model with 10.33W power consumption (11.89 tokens/s for $\sim$ 7B model).
\update{
PICACHU~\cite{qin2025picachu}, a plug-in coarse-grained reconfigurable accelerator tailored to efficiently handle nonlinear operations by using custom algorithms and a dedicated compiler toolchain. 
Evaluation shows that PICACHU achieves 96-132 tokens/s (2.4-3.3$\times$ speedup than A100) and $\sim$ 10.7W power consumption for Llama2-7B.
}

\textbf{Coarse-grained Processing.}
Tensor Contraction Processor (TCP)~\cite{kim2024tcp}, is composed of coarse-grained PEs, which are designed to be flexible enough to be configured as a large-scale single unit or a set of small independent compute units. 
TCP chip is designed and fabricated in 5nm technology with 256MB SRAM and 1.5 TB/s 48GB HBM3 under 150W. 
For Llama2-7B model, TCP achieves about 125 tokens/s with batch size 1 and 1176 tokens/s with batch size 8.
Habana Gaudi~\cite{habana_gaudi} and Gaudi2~\cite{habana_gaudi2} consists of two main compute engines, Matrix Multiplication Engine (MME) and Tensor Processor Core (TPC) cluster.
The TPC unit is a SIMD processor tailor-made for general deep learning operations. 
The biggest difference between GPU and Gaudi architecture is the size of MME. 
Gaudi architecture can handle 256$\times$256 matrix multiplication, which requires 512 input elements per cycle while GPU architecture with 16$\times$16 Tensor Core requires 8K input elements.
Therefore, Gaudi architecture can save 16$\times$ less read bandwidth requirements.
Besides compute engine, Gaudi2 includes 96 GB of HBM2E memories delivering 2.45 TB/sec bandwidth, in addition to a 48 MB of local SRAM.
Gaudi is fabricated in 16nm and Gaudi2 is fabricated in 7nm technique node with 600W power consumption.
For Llama2-7B and Llama2-13B models, Gaudi2 achieves from 81.97 to 111.11 tokens/s and from 49.02 to 64.52 tokens/s with batch size 1, respectively.

\subsubsection{PIM/NDP}

\textbf{Fusion.}
Modern DRAMs have multiple banks to serve multiple memory requests in parallel. However, when two requests go to the same bank, they have to be served serially, exacerbating the high latency of off-chip memory. 
Therefore, Kim et al.~\cite{kim2012case} propose Subarray-Level Parallelism (SALP) to overlap the latencies of different requests that go to the same bank. 
Based on SALP, SAL-PIM~\cite{han2024sal} proposes a subarray-level processing-in-memory architecture includes three architectural features. 
Two types of ALUs (SALU and C-ALU) are integrated into the subarray-level and the channel-level, respectively. 
S-ALU utilizes higher bandwidth than bank-level PIM to compute memory bound operation, and C-ALU supports accumulation and reduce-sum operation for multiple banks.
The SAL-PIM architecture is implemented by HBM2 8GB in 28nm CMOS technology with 68.973W power consumption. 
As a result, when the input size is from 32 to 128 and the output size is from 1 to 256, SAL-PIM achieves average 1.83$\times$ inference speedup for the text generation based on the GPT-2 medium model (345M) compared to the NVIDIA Titan RTX GPU.
PipePIM~\cite{jeong2024pipepim} introduces pipelining and dual buffering to maximize CU utilization in a digital PIM. 
PipePIM consists of two primary schemes: subarray-level pipelining (SAPI) and a dual vector buffer. 
The key ideas are to process activation, computation, and precharging on different subarrays in a pipelined manner by SAPI and concurrently perform buffer writes and computation by a dual vector buffer.
It is simulated in 22nm CMOS technology with 88.82$mm^2$ and 86.36$mm^2$ area. 
For LLMs like LLaMA and Mistral, the results show 1.2$\times$ and 1.14$\times$ speedup in Newton while 1.21$\times$ and 1.15$\times$ speedup in HBM-PIM~\cite{lee2021hardware}.

\textbf{Nonlinear Function Approximation.}
AttentionLego~\cite{cong2024attentionlego} proposes a PIM-based matrix-vector multiplication and look-up table-based Softmax design.
PIM-GPT~\cite{wu2024pim}, which achieves end-to-end acceleration of GPT inference with high performance and high energy efficiency.
At the hardware level, PIM-GPT is a hybrid system that includes DRAM-based PIM chips to accelerate VMM near data and an application-specific integrated circuit (ASIC) to support other functions that are too expensive for PIM including necessary nonlinear functions, data communication, and instructions for the PIM chips. 
At the software level, the mapping scheme is optimized to efficiently support the GPT dataflow.
Overall, PIM-GPT achieves 89$\times$ speedup and 221$\times$ energy efficiency over NVIDIA T4 GPU on 8 GPT models with up to 1.4 billion parameters.
SAL-PIM~\cite{han2024sal} also adopts LUT-based linear interpolation to perform complex nonlinear functions in PIM. 
\update{
SoftmAP~\cite{rakka2025softmap} deploys integer only low-precision Softmax on In-Memory Compute hardware to accelerate integer-only LLM inference. It achieves up to 1000$\times$ improvement in the energy-delay product compared to A100 and RTX3090 GPUs.
}

\begin{figure*}[!t]
  \centering
  \includegraphics[width=0.98\textwidth]{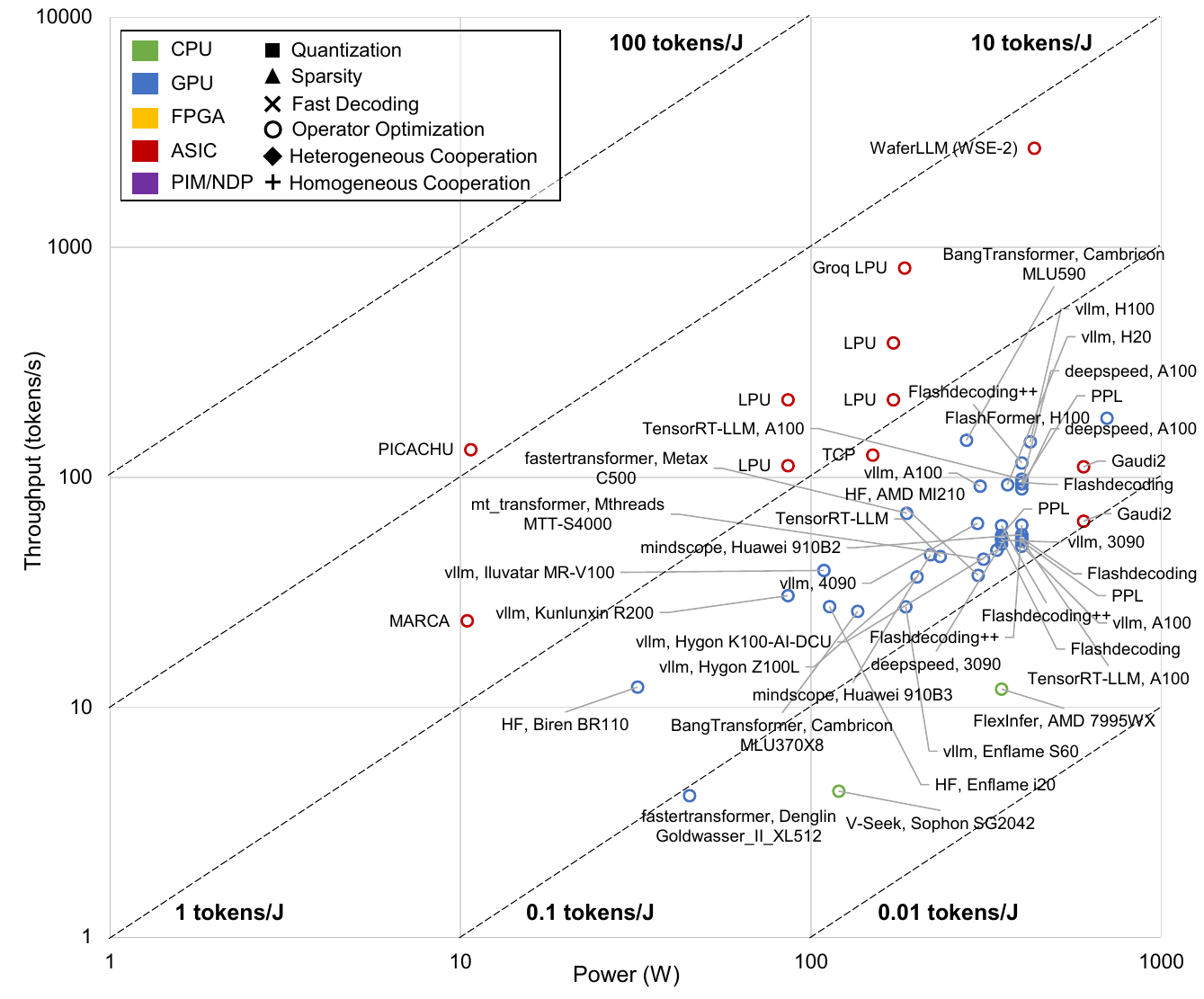}
  \vspace{-10pt}
  \caption{LLM ($\sim$ 7 billion parameters) decode stage throughput (batch size 1) vs power on different platforms with operator optimization.}
  \vspace{-10pt}
  \label{fig:sandian-operator}
\end{figure*}

\textbf{Storage Optimization.}
By observing that a key impediment to truly harness PIM acceleration is deducing optimal data-placement to place the matrix in memory banks, PIMnast~\cite{ibrahim2024balanced} proposes matrix tiling/ordering algorithms to tackle these factors and identify orchestration knobs that impact PIM acceleration.
For OPT-6.7B model, compared to the SoC baseline (AMD Ryzen PRO 7040 Series processors comprising eight CPU cores, 12 compute units of GPU cores, 16 AIE tiles, and eight channels of LPDDR5 memory for a peak memory bandwidth of 120 GB/s), PIMnast achieves 4.5$\times$ speedup for per-token latency.

\subsubsection{Comparison}

\update{
In Figure~\ref{fig:sandian-operator}, for operator optimization, power consumption ranges from 10.44W to 700W, with inference speeds between 4.13 tokens/s and 2700 tokens/s.
The energy efficiency ranges from 0.034 tokens/J to 12.34 tokens/J.
MARCA~\cite{li2024marca} (ASIC) achieves the lowest power consumption with 10.44W.
WaferLLM~\cite{he2025waferllm} (ASIC) achieves the highest throughput with batch size 1 (post-silicon results) and the highest energy efficiency.
\begin{itemize}
    \item 
    For CPUs, power consumption ranges from 120W to 350W, with inference speeds between 4.32 tokens/s and 12 tokens/s. The energy efficiency ranges from 0.034 tokens/J to 0.036 tokens/J. 
    \item 
    For GPUs, power consumption ranges from 32W to 700W, with inference speeds between 4.13 tokens/s and 181 tokens/s. The energy efficiency ranges from 0.092 tokens/J to 0.522 tokens/J. 
    \item 
    For ASICs, power consumption ranges from 10.44W to 600W, with inference speeds between 23.78 tokens/s and 2700 tokens/s, found in the upper left section of the figure. The energy efficiency ranges from 0.108 tokens/J to over 12.34 tokens/J, outperforming the GPU platform.
\end{itemize}
}

\subsection{Heterogeneous Cooperation}
\subsubsection{Overview}
Heterogeneous cooperation combines different types of computing platforms, such as CPUs, GPUs, FPGAs, and PIM/NDPs, to optimize system performance, energy efficiency, and flexibility. 
Each computing unit has unique strengths for specific tasks; for example, GPUs excel at parallel processing, FPGAs offer customizable hardware acceleration, and PIM/NDPs are specialized for memory-bound operations. 
By distributing tasks to the most suitable hardware, heterogeneous cooperation enhances computing efficiency, reduces power consumption, and lowers latency. 
It can be broadly categorized into two types, computing enhancement and memory enhancement.

\begin{figure*}[!thbp]
  \centering
  \vspace{-5pt}
  \includegraphics[width=0.88\textwidth]{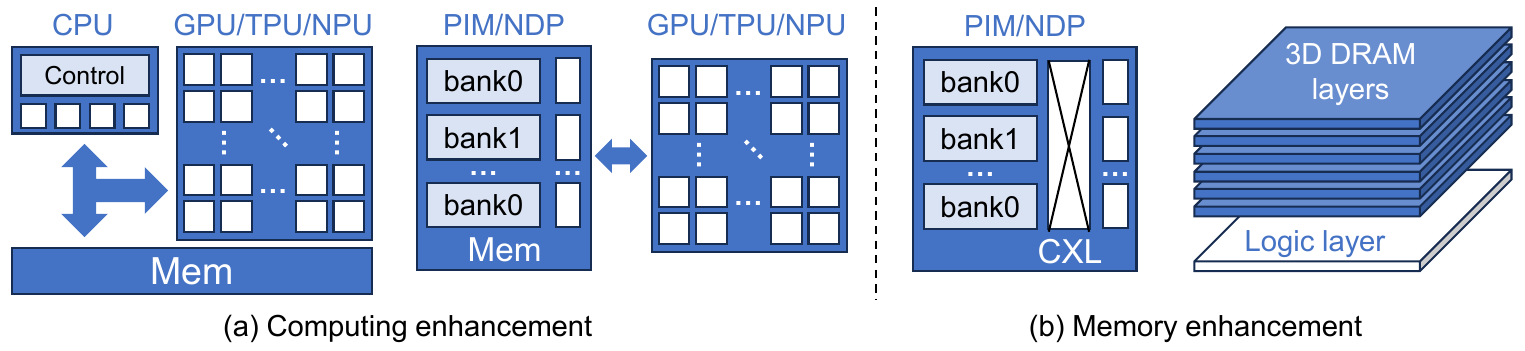}
  \vspace{-10pt}
  \caption{Heterogeneous cooperation.}
  \vspace{-5pt}
  \label{fig:heter}
\end{figure*}

\textbf{Computing Enhancement.}
Commonly, CPUs and PIMs may struggle when handling large-scale computation tasks. 
For instance, while CPUs offer versatility and flexibility, they are not well-suited for highly parallel processing. 
PIM reduces data transfer by performing computations within memory, but its computational power is limited. 
In such scenarios, powerful parallel computing hardware such as GPUs, NPUs, or TPUs are introduced to assist. 
In LLM inference, computation-intensive computations such as attention calculations are placed on GPUs, NPUs, or TPUs, while other computations can be handled by CPUs or PIMs.

\textbf{Memory Enhancement.} focuses on two key aspects: storage capacity enhancement and bandwidth enhancement. As we move into the LLM eras, computational tasks are becoming increasingly complex, leading to greater demands on memory capacity and bandwidth. From the perspective of storage capacity, memory enhancement is achieved by integrating more on-chip memory by using 3D storage stacks on the same chip area. This allows more data and models to be totally stored in a limited on-chip space, reducing the need to access external storage and improving overall efficiency.
Another memory enhancement is to accelerate memory-bound computations by increasing data transfer speeds. For heterogeneous cooperation, bandwidth often restricts computational performance. To address this issue, much methods are proposed to support high-speed on-chip~\cite{li20204gbps,li2023method,li2021gain} and chip-to-chip interconnect, such as Compute Express Link (CXL)~\cite{sharma2022compute} and NVLink~\cite{foley2017ultra}. These techniques enable low-latency, high-bandwidth data transmission between different chips, significantly speeding up data exchange processes.

\begin{table}[htbp]
    \centering
    \caption{Heterogeneous cooperation on CPU, GPU, FPGA, ASIC, and PIM/NDP}
    \begin{tabular} {|c|c|c|}
    
        \hline
        Hardware & Computing Enhancement & Memory Enhancement  \\
        \hline
        CPU & \ding{51} & \ding{55} \\
        \hline
        GPU & \ding{55} & \ding{51} \\
        \hline
        FPGA & \ding{51} & \ding{55} \\ 
        \hline
        ASIC & \ding{55} & \ding{55} \\ 
        \hline
        PIM/NDP & \ding{51} & \ding{51}\\ 
        \hline
    \end{tabular}
    \label{tab:heter}
\end{table}

\subsubsection{CPU}
\textbf{Computing Enhancement.}
\textbf{(1) CPU-GPU.}
Only using the CPU may result in slower performance, so many methods employ a combination of CPU and GPU to enhance LLM inference speed.
For personal computers, PowerInfer~\cite{song2023powerinfer} proposes that the hot-activated neurons should be preloaded onto the GPU for fast access, while cold-activated neurons are computed on the CPU, thus significantly reducing GPU memory demands and CPU-GPU data transfers.
PowerInfer further integrates adaptive predictors and neuron-aware sparse operators. 
For various models (OPT-30B/66B, Falcon-40B, and Llama-70B) on a Intel i9-13900K processor equipped with an NVIDIA RTX 4090, it achieves 8.32 tokens/s on average.
On a Intel i7-12700K processor equipped with an NVIDIA RTX 2080Ti, it achieves 5.77 tokens/s on average.
For smartphones, PowerInfer-2~\cite{xue2024powerinfer} further leverages the highly heterogeneous XPUs present in smartphone SoCs, such as asymmetric big.LITTLE CPU cores, GPU, and NPU.
On OnePlus 12 equipped with SnapDragon 8 Gen 3, PowerInfer-2 achieves 11.7 tokens/s and 10.5 tokens/s on Llama-7B and Mistral-7B, respectively.
On OnePlus Ace 2 equipped with SnapDragon 8+ Gen 1, PowerInfer-2 achieves 6.5 tokens/s and 6.3 tokens/s on Llama-7B and Mistral-7B, respectively.
For servers, Kim et al.~\cite{kim2024exploiting} propose an adaptive model to determine the LLM layers to be run on CPU and GPU, respectively, based on the memory capacity requirement and arithmetic intensity. 
They then propose CPU-GPU cooperative computing that exploits the AMX extensions of the latest Intel CPU. 
The evaluation demonstrates that for OPT-30B model, CPU-GPU cooperative computing achieves 336 tokens/s with batch size 1 and input tokens 2016 in prefill stage.
In decode stage, it achieves about 25 tokens/s with batch size 90 and input tokens 2016, and about 300 tokens/s with batch size 1400 and input tokens 64.

\update{
Dovetail~\cite{zhang2024dovetail}, Twinpilots~\cite{yu2024twinpilots} and Vellaisamy et al.~\cite{vellaisamy2025characterizing} accommodate GPU-CPU and the hierarchical memory architecture (GPU and CPU memory) to maximize CPU and GPU utilization and fully hide the data loading time behind the CPU compute time. 
On a server with an NVIDIA A10G GPU with 24GB memory and an AMD EPYC 7R32 CPU with 256GB memory, Twinpilots~\cite{yu2024twinpilots} can achieve $\sim$5 tokens/s and 12 tokens/s for Llama-13B with batch size 1 and 8.
PRIMA.CPP~\cite{li2025prima} proposes a piped-ring parallelism with prefetching and a scheduler to split model layers to heterogeneous devices. 
It can achieve 18.52 tokens/s and 22.73 tokens/s for Llama3-8B and Qwen2.5-7B on edge SoCs like Apple M1 ($\sim$14W) and Kirin 9000 ($\sim$7.9W).
HeteroLLM~\cite{chen2025heterollm} is a fast LLM inference engine with W4A16 quantization in mobile devices which supports both layer-level and tensor-level heterogeneous execution. 
Evaluation results show that HeteroLLM achieves 13-15 tokens/s for Llama2-7B on SnapDragon 8 Gen3 platform.
}

\subsubsection{GPU}
\update{
\textbf{Memory Enhancement.}
\textbf{(1) CPU-GPU.}
TightLLM~\cite{tightllm} proposes and implements an adaptive offloading-based inference system to maximize inference throughput with two key innovations. Its key-value (KV) distributor employs a trade-compute-for-transfer strategy to address growing transfer overhead by dynamically recomputing portions of the KV cache and its weight loader slices model weights and distributes the loading process across multiple batches, effectively overlapping data transfer with computation and amortizing the excessive weight loading overhead. For the OPT-13B and Llama2-13B models, it achieves 22 and 20 tokens/s with maximum batch size on a NVIDIA RTX 3090 GPU and a 24-core CPU.
}

\subsubsection{FPGA}
\update{
\textbf{Computing Enhancement.} 
\textbf{(1) FPGA-GPU.}
In latency-sensitive scenarios, a small batch or even one batch is usually required. 
This leads to the prefill and the decode stage of LLM inference being computational and memory bottlenecks, respectively.
Therefore, it is difficult for a homogeneous FPGA or GPU system to simultaneously address different computational bottlenecks in different stages of LLM inference, resulting in long prefill latency on FPGAs and low utilization during the decode stage on GPUs.
GLITCHES~\cite{yang2024glitches} is GPU-FPGA LLM inference through a collaborative heterogeneous system by employing GPUs for the prefill stage and FPGAs for the decode stage.
Experiments demonstrate that a GLITCHES heterogeneous LLM inference system with an A100/V100S GPU and 7xU280 FPGAs achieves 390-450 tokens/s for Llama2-7B.
HPU~\cite{rhee2025hpu} proposes the High-Bandwidth Processing Unit (HPU) and GPU-HPU heterogeneous system, and develops a real FPGA-based HPU prototype with HBM.
}

\subsubsection{PIM/NDP}

\textbf{Computing Enhancement.}
\textbf{(1) PIM-NPU.}
NeuPIMs~\cite{heo2024neupims} and IANUS~\cite{seo2024ianus} are heterogeneous PIM acceleration systems that jointly exploits a conventional GEMM-focused NPU and GEMV-optimized PIM devices. 
NeuPIMs~\cite{heo2024neupims} first proposes dual row buffers in each bank, facilitating the simultaneous management of memory read/write operations and PIM commands, to enable concurrent operations on both NPU and PIM platforms. 
Further, it employs a runtime sub-batch interleaving technique to maximize concurrent execution for the inherent dependencies between GEMM and GEMV in LLMs. 
Our evaluation demonstrates that NeuPIMs with $\sim$76W (0.6348W+75W) power consumption (32GB HBM) achieves $\sim$3k tokens/s with batch size 8 for GPT3-7B.
IANUS~\cite{seo2024ianus} proposes novel PIM access scheduling that manages not only the scheduling of normal memory accesses and PIM computations but also workload mapping across the PIM and the NPU. 
The evaluations show that for GPT-6.7B model, IANUS achieves 127.1 tokens/s with about 240W power consumption.
Cambricon-LLM~\cite{yu2024cambricon} proposes a chiplet-based hybrid architecture with NPU and a dedicated NAND flash chip to enable efficient LLM inference. 
It utilizes both the high computing capability of NPU and the data capacity of the NAND flash chip. 
The NAND flash chip is enhanced by in-flash computing to perform precise lightweight on-die processing, and the NPU performs matrix operations and handles special function computations. 
Experimental results show that Cambricon-LLM with $\sim$37W power consumption achieves 3.44 tokens/s ($\sim$) and 36.34 tokens/s for 70B and 7B LLMs, which is over 22$\times$ to 45$\times$ faster than existing flash-offloading technologies, respectively.
\textbf{(2) PIM-GPU.}
MoNDE~\cite{kim2024monde} is a near-data computing solution that efficiently enables Mixture-of-Experts (MoE) LLM inference on heterogeneous hardwares. 
MoNDE reduces the volume of MoE parameter movement by transferring only the hot experts to the GPU, while computing the remaining cold experts inside the host memory device. 
MoNDE can achieve inference latency comparable to an ideal GPU system with infinite memory.
AttAcc~\cite{choi2023unleashing,park2024attacc} is also an heterogeneous system equipped with GPUs to accelerate the attention layers. 
Compared to the monolithic state-of-the-art GPU system, AttAcc achieves significantly higher throughput (up to 2.81$\times$) and energy efficiency (up to 2.67$\times$) for GPT3-175B model.
In 2024, SK Hynix proposes LPDDR6-based heterogeneous LLM system called AiMX-xPU\cite{kim2024sk}, which consists 1 NVIDIA H100 GPU and 3 AiMX, achieving 167 tokens/s, 220 tokens/s, and 900 tokens/s with batch size 1, 8, and 32 for OPT-30B.
\update{
PAISE~\cite{lee2025paise} employs GPU-PIM heterogeneous computing resources to optimize LLM inference. 
For Llama2-7B model, the AMD MI100 GPU with HBM-PIM devices achives $\sim$400 tokens/s with batch size 1 and the power consumption is similar to the GPU-only system.
Pyramid~\cite{yan2025pyramid} optimizes GPU-PIM system for LLM inference by strategically allocating cross-level computational resources within PIM to meet diverse needs and leveraging the strengths of both technologies. 
Evaluation results demonstrate that Pyramid outperforms existing systems like NeuPIM, AiM, and AttAcc by factors of 2.31×, 1.91×, and1.72×, respectively.
The inherent activation sparsity in LLMs naturally divides weight parameters into two categories, termed “hot” and “cold” neurons, respectively. 
Hot neurons consist of only $\sim$20\% of all weight parameters but account for 80\% of the total computational load. In contrast, cold neurons account for just 20\% of the computational workload.
Hermes~\cite{liu2025make} leverages NDP units within DRAM DIMMs to enhance the performance of GPU by mapping hot neurons to GPU, while offloading cold neurons to NDP-DIMMs. 
Furthermore, Hermes introduce a lightweight predictor that ensures optimal real-time neuron partition and adjustment between GPU and NDP-DIMMs. 
Hermes facilitates the deployment of LLaMA2-70B on consumer-grade hardware at a rate of 13.75 tokens/s.
}
\textbf{(3) PIM-FPGA.}
Kang et al.~\cite{kang2023era} put the memory-bound GEMV calculations including projections, feed-forward-network layer, and the last fully-connected layer with pre-trained weight on HBM2-PIM and evaluate system with PIM-powered Xilinx Alveo U280 board. 
For GPT-1.3B model, after scaling the bandwidth by 8$\times$ for PIM technology, it can achieve about 347.83 tokens/s, which is 1.6$\times$ faster than the NVIDIA A100 GPU.
From 2022 to 2023, SK Hynix proposes GDDR6-based accelerator-in-memory named AiM~\cite{lee20221ynm,kwon2022system} and integrates it with 2 Xilinx XCVU9P FPGA chips for control and communication as a prototype called AiMX~\cite{AiMX} to achieve 330 tokens/s for OPT-6.7B with batch size 1.
\textbf{(4) PIM-TPU.}
H3D-Transformer~\cite{luo2024h3d} proposes a heterogeneous 3D-based accelerator design for transformer models, which adopts an interposer substrate with multiple 3D memory/logic hybrid cubes optimized for accelerating different MatMul workloads. 
An approximate computing scheme is proposed to take advantage of heterogeneous computing paradigms of mixed-signal compute-in-memory (CIM) and digital tensor processing units (TPU). 
From the system-level evaluation results, 10 TOPS/W energy efficiency is achieved for the GPT2 model, which is about 2.6$\times$-3.1$\times$ higher than the baseline with 7nm TPU and stacked FeFET memory.
3D-HI~\cite{sharma2023heterogeneous} leverage chiplet-based heterogeneous integration to design a Network-on-Interposer (NoI) architecture to accelerate LLM inference. 
3D-HI uses the streaming multiprocessors along with the associated memory controllers (SM-MCs) for multi-head attention and ReRAM cores for the feed-forward network, which optimize both achievable energy efficiency and throughput. 
Further, vertical integration on top of a 2.5D interposer helps to enhance overall system performance and alleviates the issue of memory bottlenecks. 
Experimental results demonstrate that 3D-HI lowers the latency and energy consumption by up to 22.8$\times$ and 5.36$\times$ with respect to an equivalent state-of-the-art chiplet-based platform.
\update{
PIM-LLM~\cite{malekar2025pim} leverages analog processing-in-memory (PIM) architectures and digital systolic arrays to accelerate 1-bit MatMul operations in projection layers and high-precision MatMul operations in attention, respectively. 
It can achieve 42.82 tokens/s and 170 tokens/J for Llama-7B with 128 sequence length.
}

\begin{figure*}[!t]
  \centering
  \includegraphics[width=0.98\textwidth]{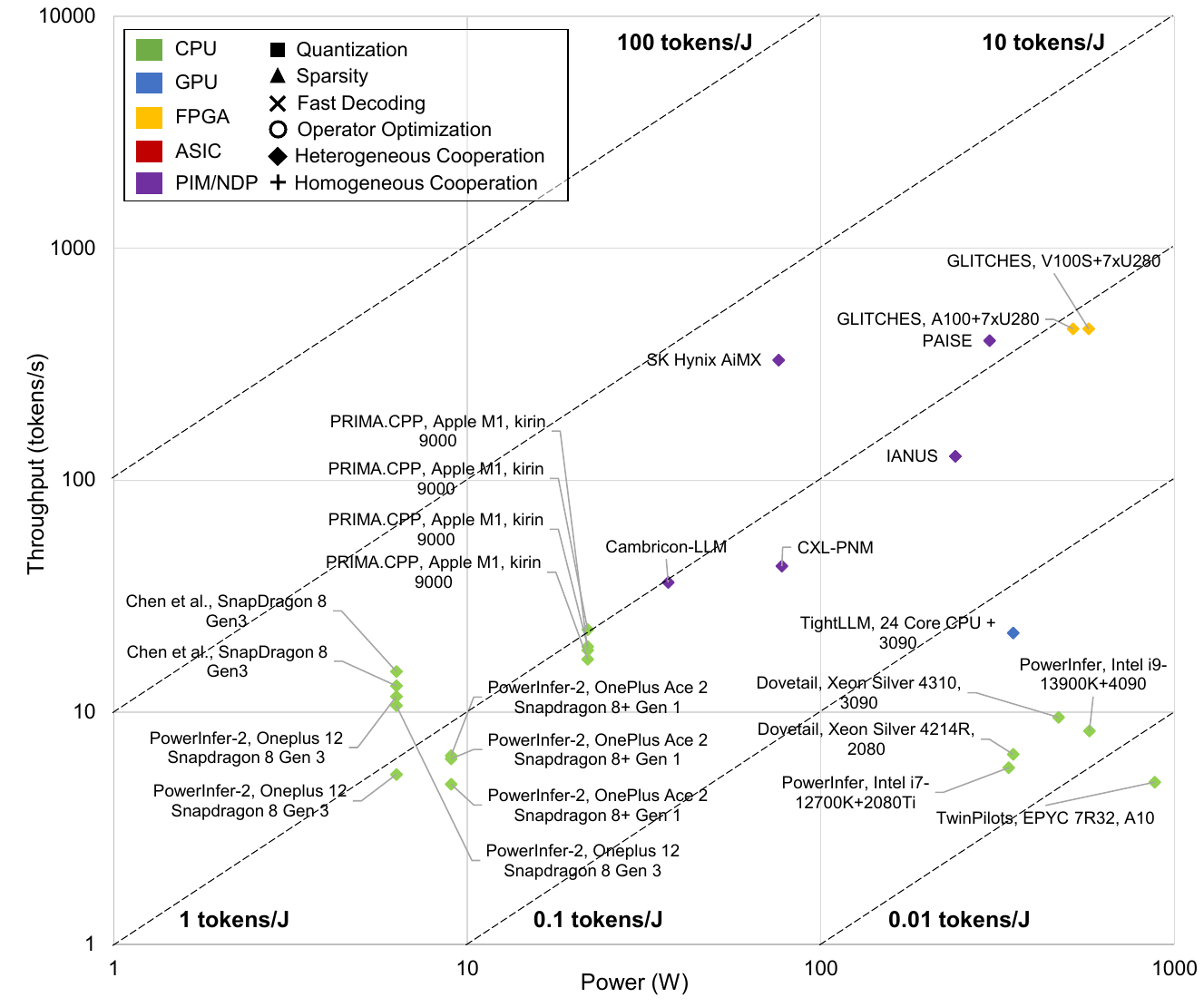}
  \vspace{-10pt}
  \caption{LLM ($\sim$ 7 billion parameters) decode stage throughput (batch size 1) vs power on different platforms with heterogeneous cooperation.}
  \vspace{-10pt}
  \label{fig:sandian-heter}
\end{figure*}

\textbf{Memory Enhancement.}
\textbf{(1) PIM-CXL.}
As the frequent transfers of these model parameters and intermediate values are performed over relatively slow device-to-device interconnects such as PCIe or NVLink, they become the key bottleneck for efficient acceleration of LLMs. 
Kim et al.~\cite{kim2024breakthrough} exploit PIM, which uses bank-level parallelization to provide higher internal memory bandwidth compared to traditional DRAM, resulting in a significant increase in on-DRAM compute bandwidth. 
In addition to achieving high capacity through Compute eXpress Link (CXL) memory expansion, CXL-PNM demonstrates performance improvements by integrating computational logic into memory products, consequently increasing internal memory bandwidth.
Evaluation results show that the performance of memory-bounded LLMs was significantly improved with PIM and PNM by up to 4.5$\times$ and 4.4$\times$, respectively.
CXL-PNM~\cite{park2024lpddr} first devises an LPDDR5X-based CXL memory architecture with 512GB of capacity and 1.1TB/s of bandwidth, which boasts 16$\times$ larger capacity and 10$\times$ higher bandwidth than GDDR6 and DDR5-based CXL memory architectures, respectively, under a module form-factor constraint. 
Second, it designs a CXL-PNM controller architecture integrated with an LLM inference accelerator, exploiting the unique capabilities of such CXL memory to overcome the disadvantages of competing technologies such as HBM-PIM and AxDIMM. 
The evaluation shows that for OPT-13B model, CXL-PNM achieves 42.68 tokens/s with 77.6W power consumption.
\textbf{(2) PIM-3D Stack.}
Sharda et al.~\cite{sharda2024accelerator} propose to use the capacitorless 3D stackable DRAM to store much larger LLMs compared to conventional DRAM at higher density. 
To reduce the intermediate data size, they propose to use a layer-wise sparsity-quantization hybrid (LSQH) algorithm, which induces sparsity based on calculations performed using low-bit quantization to reduce both the energy consumption and the data storage requirements. 
Finally, a 3D heterogeneously integrated accelerator is designed by stacking a 3D DRAM with logic dies designed in the 3nm technology node at 1GHz. 
The evaluation shows that for Llama2-13B, it achieves 163k tokens/s in prefill stage with 193W power consumption. 
\update{
PIM-AI~\cite{ortega2024pim} stacks computational units on DDR5/LPDDR5 memory chips to improve performance and energy efficiency. 
Simulations show that PIM-AI achieves 30 tokens/s and 35.97 tokens/J for Llama2-7B.
Lue et al.~\cite{lue2024prospects} propose a Flash-based Computing-in-Memory architecture using 3D NOR for accelerating LLM inference. 
For Llama-7B with W4A8, they can achieve $>$200 tokens/s inference speed.
}

\subsubsection{Comparison}

\update{
In Figure~\ref{fig:sandian-heter}, for heterogeneous cooperation, power consumption ranges from 6.3W to 880W, with inference speeds between 4.9 tokens/s and 1998 tokens/s.
The energy efficiency ranges from 0.0057 tokens/J to 46.66 tokens/J.
PowerInfer-2~\cite{xue2024powerinfer} with Snapdragon 8 Gen 3 (CPU) achieves the lowest power consumption with 6.3W.
Guo et al.~\cite{guo2024towards} (PIM/NDP) achieves the highest throughput with batch size 1 and the highest energy efficiency.
\begin{itemize}
    \item 
    For CPUs, power consumption ranges from 6.3W to 880W, with inference speeds between 4.9 tokens/s and 22.73 tokens/s, situated in the bottom part of the figure. The energy efficiency ranges from 0.0057 tokens/J to 2.381 tokens/J. In the end-side scenario, the CPU can achieve higher energy efficiency.
    \item 
    For GPUs, TightLLM is the only one using heterogeneous cooperation, which is similar to CPU platforms.
    \item
    For FPGAs, GLITCHES is the only one using heterogeneous cooperation, which achieves higher throughput than CPU/GPU platforms.
    \item 
    For PIM/NDPs, power consumption ranges from 11.52W to 300W, with inference speeds between 36.34 tokens/s and 1998 tokens/s, found in the upper part of the graph. The energy efficiency ranges from 0.787 tokens/J to 46.66 tokens/J, outperforming other platforms.
\end{itemize}
}
\subsection{Homogeneous Cooperation}

\subsubsection{Overview}
Due to the large size and high computational demands of LLMs, homogeneous cooperation can also enhance LLM inference.
Distributed computing like model parallelism is aimed at addressing memory limitations associated with LLMs. As model sizes continue to grow, a single hardware unit may not be able to accommodate the entire model. 
Model parallelism splits the model into multiple parts, with different hardware units processing different segments of the model concurrently.

\subsubsection{CPU}
Distributed computing emerges as a prevalent strategy to mitigate single-node memory constraints and expedite LLM inference performance. 
He et al.~\cite{he2024distributed} propose an efficient distributed inference optimization solution for LLMs on CPUs. 
On four 5th Gen Intel Xeon Scalable Processors the result shows the generation speed on Qwen-72B is 7.14 tokens/s.
He et al.~\cite{he2024inference} also propose new attention flow, SlimAttention, to reduce the KV cache size and ensure precision for efficient LLM inference on CPUs. 
The experimental results on Llama2-70b shows the latency of token generation is 4 tokens/s with 2 sockets and 11.4 tokens/s with 8 sockets on Intel Xeon 8563C CPUs.

\subsubsection{FPGA}
DFX~\cite{hong2022dfx} is a multi-FPGA acceleration which uses model parallelism and optimized dataflow to improve LLM inference speed in both prefilling and decoding phases. 
With the implementation on four Xilinx Alveo U280 FPGAs, DFX achieves about 120 tokens/s for GPT2-1.5B model. 
In this paper, we propose LoopLynx, a scalable dataflow architecture for efficient LLM inference that optimizes FPGA usage through a hybrid spatial-temporal design. The design of LoopLynx incorporates a hybrid temporal-spatial architecture, where computationally intensive operators are implemented as large dataflow kernels. This achieves high throughput similar to spatial architecture, and organizing and reusing these kernels in a temporal way together enhances FPGA peak performance. Furthermore, to overcome the resource limitations of a single device, we provide a multi-FPGA distributed architecture that overlaps and hides all data transfers so that the distributed accelerators are fully utilized. By doing so, LoopLynx can be effectively scaled to multiple devices to further explore model parallelism for large-scale LLM inference. Evaluation of GPT2 model demonstrates that LoopLynx can achieve comparable performance to state-of-the-art single FPGA-based accelerations. In addition, compared to Nvidia A100, our accelerator with a dual-FPGA configuration delivers a 2.52x speed-up in inference latency while consuming only 48.1\% of the energy.

\subsubsection{ASIC}
\update{
Chen et al.~\cite{chen2024agile} proposes an agile framework to support two homogeneous binary and floating-point PE arrays.
Experimental results illustrate it achieves $\sim$80 tokens/s for Llama-6.7B with 5.72W.
}
\section{Further Comparison}\label{sec:comparison}
\subsection{Setup}

\begin{figure*}[!t]
  \centering
  \includegraphics[width=0.95\textwidth]{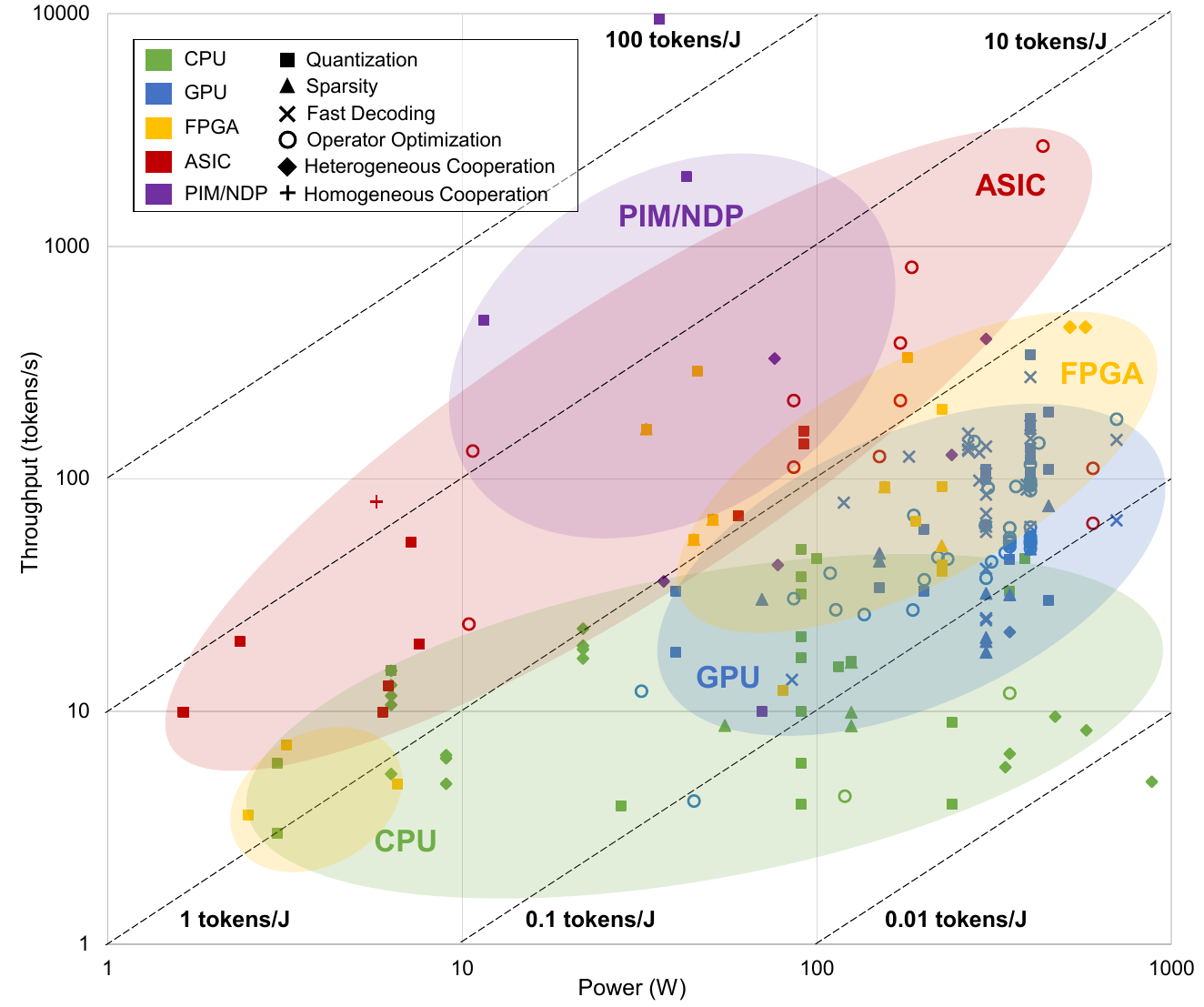}
  \vspace{-10pt}
  \caption{LLM ($\sim$ 7 billion parameters) decode stage throughput (batch size 1) vs power on different platforms with different optimization methods.}
  \vspace{-10pt}
  \label{fig:all_sandian}
\end{figure*}

In this section, we compare the performance of state-of-the-art optimization methods on hardware accelerators including CPUs, GPUs, FPGAs, ASICs and PIM/NDPs for generative LLM.
In traditional hardware capability comparisons, the focus is usually on peak computational performance and power consumption. 
However, in the context of generative large models, the emphasis shifts to inference speed (\textbf{tokens per second, tokens/s}) and hardware power consumption (\textbf{Watt, W}). 
The slope of the curve representing inference speed on the Y-axis versus hardware power consumption on the X-axis indicates the hardware efficiency in terms of tokens per joule (\textbf{tokens/J}).
Therefore, we have collected data on each hardware platform along with all optimization methods, highlighting their inference speed (mentioned explicitly or estimated) and actual power consumption while running generative large models ($\sim$ 7 billion parameters) with batch size 1 and 8, as shown in Figure~\ref{fig:all_sandian} and Figure~\ref{fig:bs8_all_sandian}, respectively.


In collecting the data, the sources for absolute inference speed are categorized into five types: 
\begin{itemize}
    \item For most CPU and GPU hardware platforms, the absolute inference speed is typically reported in the literature alongside the specific model being run.
    \item For GPU platforms with open-source and reproducible code, we directly measure the absolute inference speed through actual execution.
    \item There is limited work on FPGA hardware platforms. Some papers provide verified absolute inference data through physical testing, while others offer estimates without hardware validation. Consequently, we also include estimated absolute inference speeds for these cases.
    \item For most ASIC platforms, we first define a post-silicon work~\cite{10609625} with an absolute inference speed as the baseline and scale the results by the peak computational performance (TOPS or GOPS). \update{We use the open-source cycle-based accurate hardware simulator for BitMoD~\cite{chen2025bitmod} to simulate the performance of the baselines like OliVe~\cite{guo2023olive}.}
    \item Some works provide acceleration ratios relative to GPUs or other hardware platforms. In these cases, we scale existing data according to the reported acceleration ratios to estimate the absolute inference speed.
\end{itemize}

And the sources of power consumption are categorized into three types:
\begin{itemize}
\item For most CPU, and some GPU and FPGA hardware platforms, the papers typically specify the hardware model and quantity used. We use these specifications to determine the actual power consumption.
\item For GPU platforms with open-source and reproducible code, we directly measure the actual power consumption through real-world execution.
\item For most ASIC and PIM/NDP platforms, we consider both on-chip and off-chip memory access power consumption. On-chip power consumption is often reported in the literature. Off-chip power consumption is calculated by multiplying the energy per bit of the off-chip memory type by the model parameter count and then by the absolute inference speed.
\end{itemize}

\subsection{Hardware Comparison}
\subsubsection{Small batch size (bs=1)}

In Figure~\ref{fig:all_sandian}, for \textbf{CPU} platforms, power consumption ranges from 3W to 880W, with inference speeds between 3 tokens/s and 50 tokens/s, located in the bottom part of the figure. The energy efficiency ranges from 0.0057 tokens/J to 2.38 token/J. 
Additionally, we observe edge-side CPUs (including CPU SoCs) with 3W to 6W power consumption exhibit higher energy efficiency (0.544 tokens/J to 2.38 tokens/J) compared to GPUs. 
For \textbf{GPU} platforms, power consumption ranges from 40W to 700W, with inference speeds between 18 tokens/s and 343.4 tokens/s, situated in the middle right part of the figure. The energy efficiency ranges from 0.06 tokens/J to 0.8586 token/J.
For \textbf{FPGA} platforms, power consumption ranges from 2.49W to 572W, with inference speeds between 3.61 tokens/s and 450 tokens/s, also in the middle part of the figure. The energy efficiency ranges from over 0.1537 tokens/J to 6.304 tokens/J, which is higher than GPU platforms and similar with edge-side CPUs.
For \textbf{ASIC} platforms, power consumption ranges from 1.632W to 600W, with inference speeds between 9.95 tokens/s and 2700 tokens/s, found in the upper left part of the figure. The energy efficiency ranges from 0.1075 tokens/J to over 13.99 tokens/J, outperforming CPU, GPU and FPGA platforms.
For \textbf{PIM/NDP} platforms, power consumption ranges from 11.516W to 300W, with inference speeds between 10 tokens/s and 1998 tokens/s, located in the upper left part of the figure. The energy efficiency ranges from 0.1429 tokens/J to 46.66 tokens/J, outperforming other hardware platforms.

\begin{figure*}[!t]
  \centering
  \includegraphics[width=0.95\textwidth]{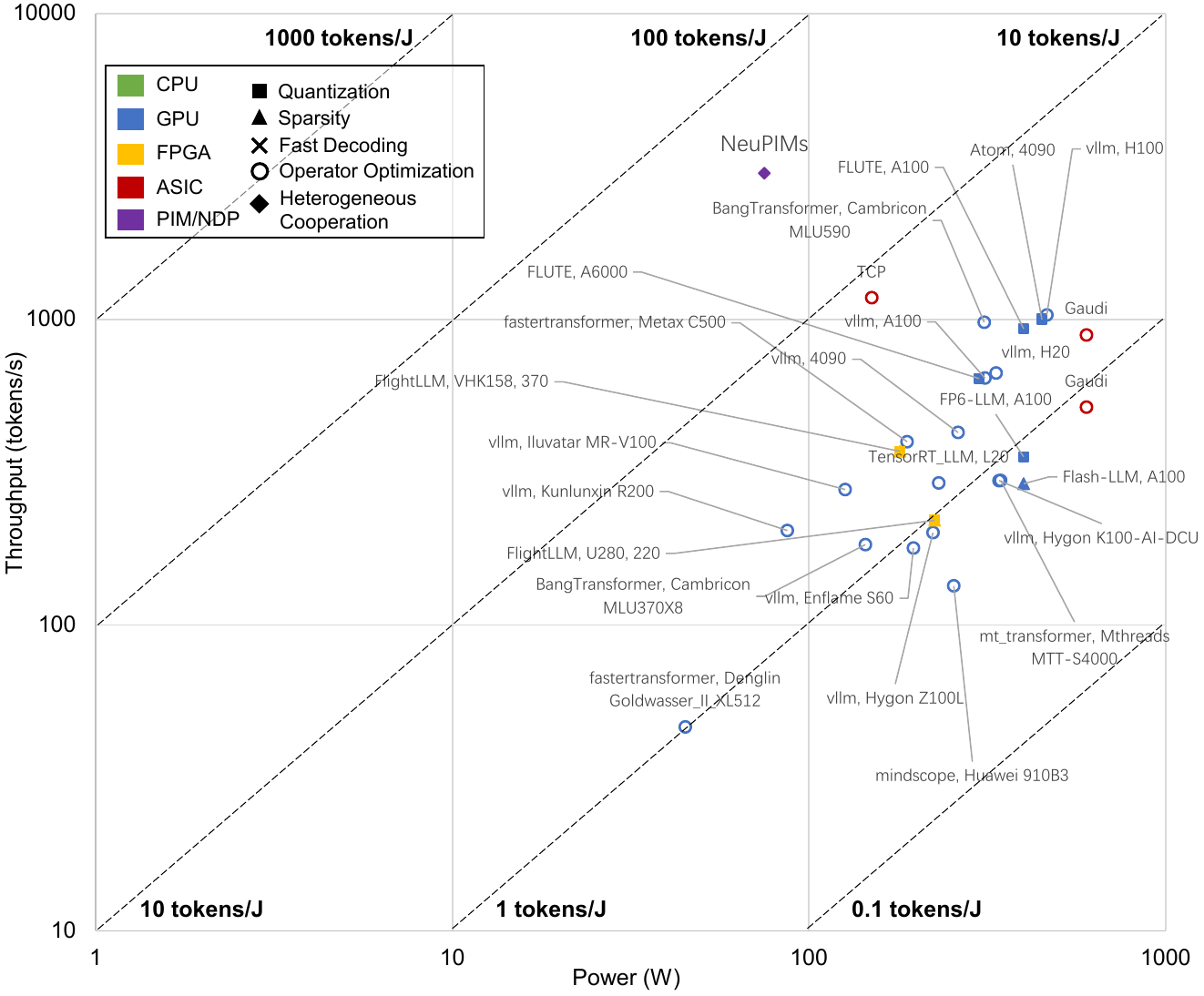}
  \vspace{-10pt}
  \caption{LLM ($\sim$ 7 billion parameters) decode stage throughput (batch size 8) vs power on different platforms with different optimization methods.}
  \vspace{-10pt}
  \label{fig:bs8_all_sandian}
\end{figure*}

\subsubsection{Large batch size (bs=8)}
Compared to small batch size, the results of inference throughput on large batch size are limited. We can only collect seldom results in Figure~\ref{fig:bs8_all_sandian}.
For \textbf{GPU} platforms, power consumption ranges from 45W to 464W, with inference speeds between 46 tokens/s and 1033 tokens/s. The energy efficiency ranges from 0.527 tokens/J to 3.155 token/J.
For \textbf{FPGA} platforms, power consumption ranges from 180W to 255W, with inference speeds between 220 tokens/s and 370 tokens/s, also in the middle right part of the figure. The energy efficiency ranges from over 0.978 tokens/J to 2.056 tokens/J.
For \textbf{ASIC} platforms, power consumption ranges from 150W to 600W, with inference speeds between 516 tokens/s and 1176 tokens/s. The energy efficiency ranges from 0.86 tokens/J to over 7.84 tokens/J.
For \textbf{PIM/NDP} platforms, power consumption is about 75W, with inference speeds about 3000 tokens/s. The energy efficiency is up to 40 tokens/J.
Compared to small batch size 1, larger batch size 8 can achieve significantly higher throughput. For example, on a GPU, throughput increases from 18-194 tokens/s to 46-1033 tokens/s, representing an improvement of 2.56$\times$-5.32$\times$. Similarly, energy efficiency improves from 0.067-0.825 tokens/J to 0.527-3.155 tokens/J, an increase of 3.82$\times$-7.87$\times$.


\subsection{Optimization Method Comparison}
\subsubsection{Small batch size (bs=1)}
We then compare different optimization methods on the same platforms.
In Figure~\ref{fig:all_sandian}, the methods employed on \textbf{CPU} platforms include quantization, sparsity, and heterogeneous cooperation. Quantization can achieve higher absolute inference speeds, reaching approximately 50 tokens/s, while sparsity and heterogeneous cooperation achieve speeds of 16.3 tokens/s and 22.73 tokens/s, respectively. Quantization and heterogeneous cooperation can achieve higher energy efficiency with 2.38 tokens/J, compared to 0.16 tokens/J for sparsity.
On \textbf{GPU} platforms, the methods used include quantization, sparsity, fast decoding, and operator optimization. Quantization, fast decoding can achieve higher absolute inference speeds, reaching approximately 343.4 tokens/s and 274.4 tokens/s, respectively, while operator optimization and sparsity achieve only about 181 tokens/s and 76.69 tokens/s. Regarding efficiency, these three methods show relatively small differences in performance compared to sparsity (0.434 tokens/J), with efficiencies of 0.859 tokens/J, 0.686 tokens/J, and 0.522 tokens/J, respectively.
On \textbf{FPGA} platforms, the methods employed include quantization, sparsity and heterogeneous cooperation, which are often used together. The highest on-board deployed speed reaches 290 tokens/s, with an efficiency of up to 6.304 tokens/J. 
Some works achieve a performance of 450 tokens/s with an energy efficiency of 0.87 tokens/J through heterogeneous GPUs and FPGAs.
On \textbf{ASIC} platforms, the methods employed include quantization, and operator optimization. Operator optimization can achieve the highest absolute inference speeds, reaching approximately 2700 tokens/s. In comparison, quantization achieves speeds of 161.1 tokens/s. In terms of efficiency, quantization and operator optimization offer similar and high energy efficiency with 8.49 tokens/J and 12.34 tokens/J, respectively.
On \textbf{PIM/NDP} platforms, the main methods used are quantization, fast decoding, and heterogeneous cooperation. Quantization can achieve higher throughput with 1998 tokens/s while fast decoding and heterogeneous cooperation only achieve 174.018 tokens/s and 400 tokens/s, respectively. In terms of efficiency, quantization also achieves higher energy efficiency with 46.66 tokens/J while fast decoding and heterogeneous cooperation only achieve 9.94 tokens/J and 4.34 tokens/s, respectively.


\subsubsection{Large batch size (bs=8)}
In Figure~\ref{fig:bs8_all_sandian}, on \textbf{GPU} platforms, the methods used include quantization, sparsity, and operator optimization. Quantization and operator optimization achieve higher absolute inference speeds, reaching approximately 1000 tokens/s and 1033 tokens/s, respectively, while sparsity achieves about 290 tokens/s. Regarding efficiency, quantization, and operator optimization also achieve higher energy efficiency with 3.155 tokens/J and 2.328 tokens/s compared to sparsity (0.725 tokens/J).
On \textbf{FPGA} platforms, the methods employed include quantization and sparsity, which are often used together. The highest inference speed reaches 370 tokens/s, with an efficiency of up to 2.056 tokens/J.
On \textbf{ASIC} platforms, the methods employed include operator optimization. It can achieve higher absolute inference speeds, reaching approximately 1176.47 tokens/s with 7.843 tokens/J energy efficiency.
On \textbf{PIM/NDP} platforms, the main methods used are heterogeneous cooperation. Heterogeneous cooperation can achieve an inference speed of about 3000 tokens/s and 40 tokens/J.

\section{Further Discussion}\label{sec:discussion}


\begin{figure}[!th]
  \centering
  \includegraphics[width=0.8\textwidth]{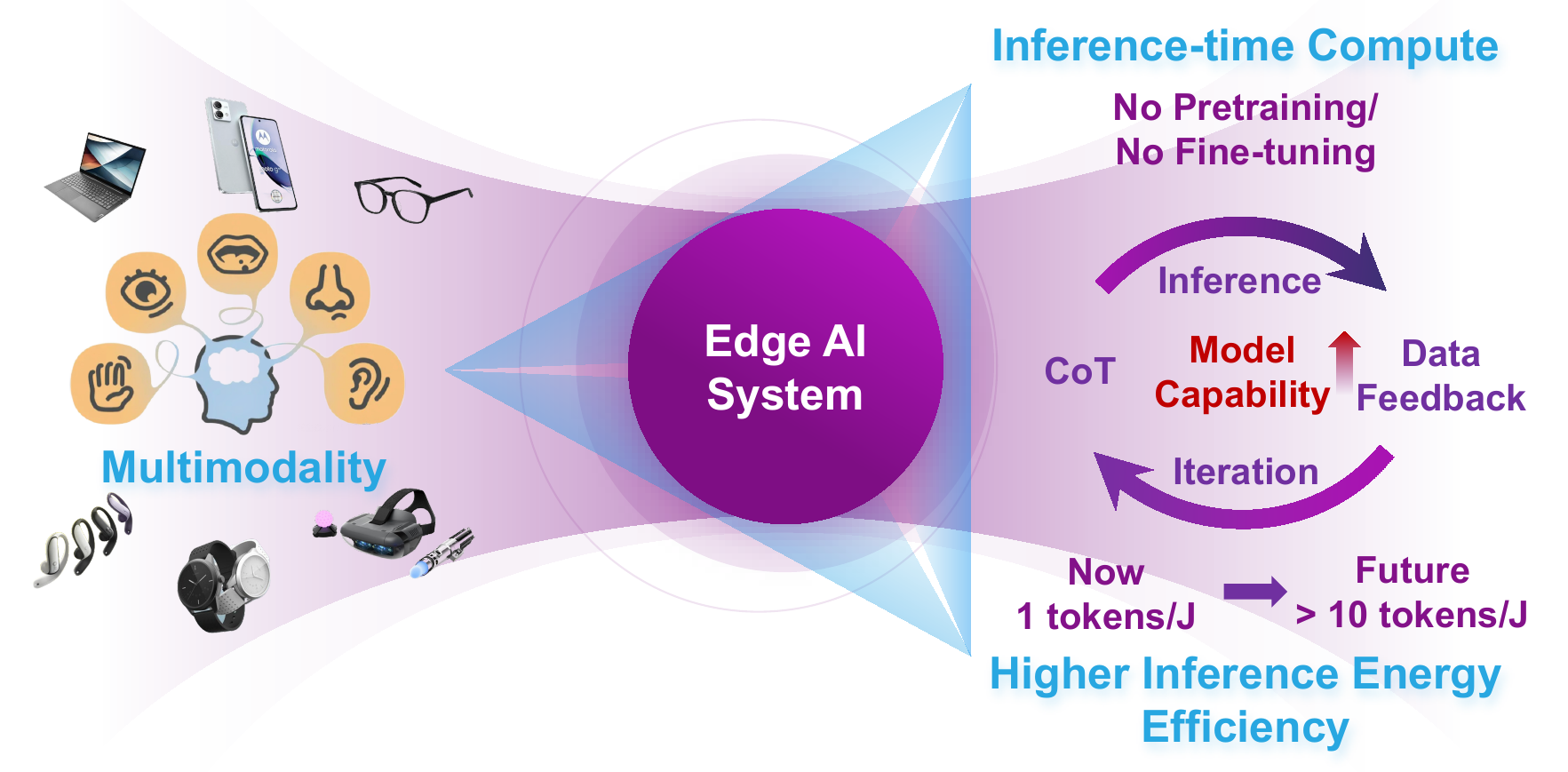}
  \caption{The trends of edge AI system: multimodality, inference-time computation, and higher inference energy efficiency.}
  \vspace{-10pt}
  \label{fig:trend}
\end{figure}

The development of edge intelligence has gained significant momentum, driven by the increasing capability of LLMs and the increasing demands of edge applications. 
As we look toward the future, three trends emerge: \textbf{multimodality}, \textbf{inference-time compute}, and \textbf{higher inference energy efficiency}, which is promising to redefine the capabilities of edge AI systems, as shown in Fig.~\ref{fig:trend}.
Multimodality focuses on integrating diverse data types, such as visual, auditory, textual, and sensory inputs, to enhance the system's ability equipped with LLMs to understand and adapt to complex scenarios, supporting a wide range of edge devices like PCs, smartphones, AR glasses, and smartwatches. 
Inference-time compute emphasizes the deep thinking during LLM inference leveraging iterative processes such as Chain-of-Thought (CoT) reasoning to continuously improve model capabilities. It is achieved by performing a post-training process.
Higher energy efficiency highlights the demand for increasing energy efficiency from 1 tokens/J to >10 tokens/J for high real-time scenarios, such as embodied intelligence represented by robotics control and autonomous driving.

\subsection{Multimodality}



\begin{figure}[!b]
  \centering
  \includegraphics[width=0.98\textwidth]{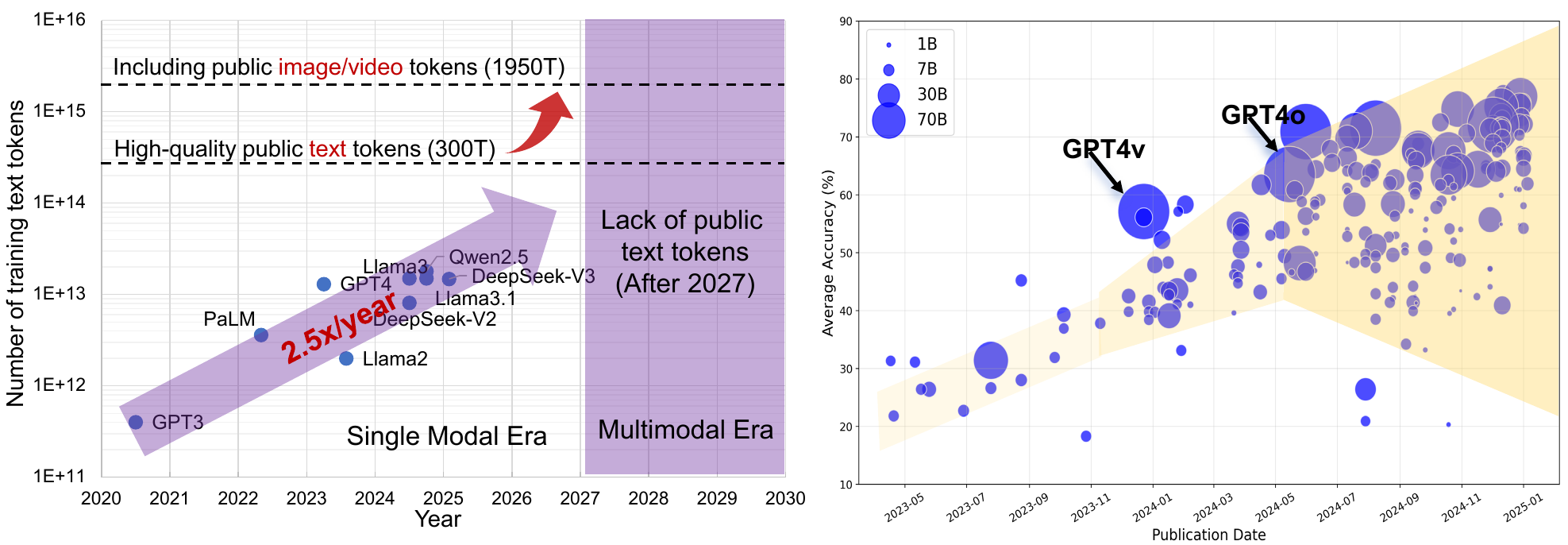}
  \caption{The timeline of multimodal (Vision-Language) LLMs: average accuracy, publication date, and model size. The results are collected from OpenCompass~\cite{OpenCompass}.}
  \label{fig:mllm_trend}
\end{figure}

As shown in Fig.~\ref{fig:mllm_trend} left, we summarize the release dates and the number of training text tokens for typical LLMs over the past five years. 
The number of training tokens has been growing at a rate of 2.5$\times$ per year. 
GPT-3 is trained using only 0.4 trillion text tokens, while Qwen2.5 and Llama3.1 use 18 trillion and 15 trillion tokens, respectively. 
According to the result of Epoch AI, the total number of human-generated public text tokens is $\sim$300 trillion~\cite{villalobos2024run}. 
At this growth rate, it is expected that by 2027, we will face a shortage of available public text tokens for training LLMs.
In addition to text tokens, there is a vast amount of other modalities of tokens on the Internet, such as image and video tokens (total $\sim$1650 trillion~\cite{villalobos2024run}). 
Therefore, from the perspective of token availability, by leveraging multiple modalities of tokens for training LLMs, we can continue the LLMs' scaling law~\cite{kaplan2020scaling}.
From the perspective of human interaction, humans rely on various senses as gateways to interact with the physical world, which is inherently multimodal. 
Recent research has also proven that multimodal learning provably performs better than single modal due to the more accurate estimate of the latent space representation~\cite{huang2021makes}.
These factors indicate that future LLMs should have the capability of multimodal interactions to perceive, process, and generate information across multiple modalities. 


Currently, more than 80\% multimodal LLMs focus on both vision and language tasks~\cite{zhang2024mm}.
As shown in Fig.~\ref{fig:mllm_trend}, their development can be roughly divided into three stages. 
First, from May to September 2023, the model size is small and the accuracy ranges from 20\% to 40\%. 
Then, from December 2023 to April 2024, with the release of GPT4v, the model accuracy significantly improves, reaching nearly 60\%.
After May 2024 with the release of GPT4o, more multimodal models are released and the accuracy reaches almost 80\%.
It should be noted that the development of multimodal LLMs (MLLMs) heavily relies on the construction of multimodal datasets. 
And another significant trend in MLLMs is the concept of native multimodality. 
By aligning diverse modalities during the training stage, native MLLMs enable end-to-end inputs and outputs. 

\subsection{Inference-time Compute}



\begin{figure}[!t]
  \centering
  \includegraphics[width=0.98\textwidth]{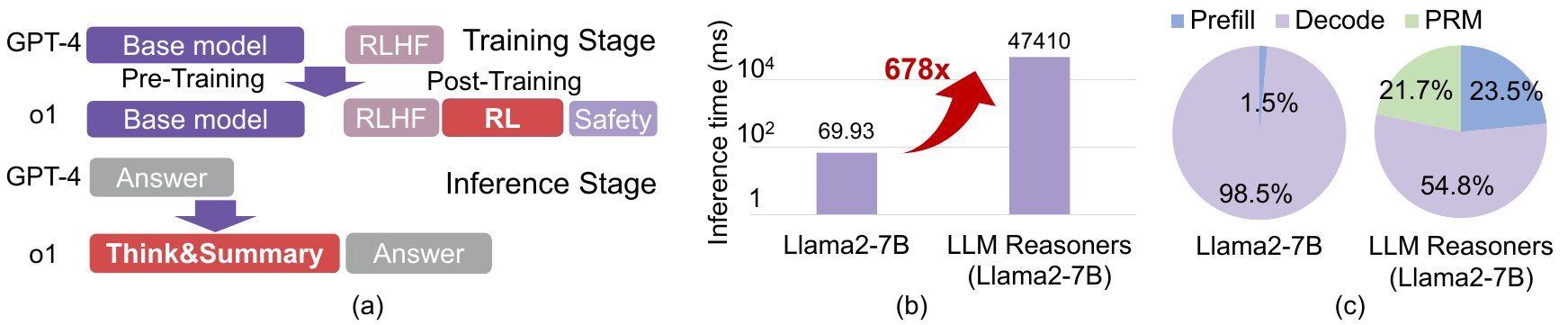}
  \vspace{-5pt}
  \caption{Inference-time Compute: (a) the difference of training and inference stages, (b) the increment of inference time, and (c) runtime breakdown of inference time.}
  \vspace{-15pt}
  \label{fig:o1}
\end{figure}

In the LLM era, data is the fossil fuel, while the age of pre-training will end because the data is not growing~\cite{llyanips}. 
Inference-time compute, as represented by OpenAI's o1 paradigm~\cite{openai_o1}, is now seen as the critical pathway to enhance model capabilities in the next LLM era. 
The principles of o1 can be illustrated through its distinction from GPT4, as shown in Fig.~\ref{fig:o1}(a). During the training stage, GPT4 follows the conventional workflow: pre-training a base model followed by fine-tuning through reinforcement learning from human feedback (RLHF). 
In contrast, o1 introduces several innovations in the post-training stage. Beyond incorporating reinforcement learning (RL), o1 places particular emphasis on safety optimization to ensure that o1 is equipped for more nuanced tasks and safe decision-making during deployment.
The difference between GPT4 and o1 becomes more pronounced in the inference stage. GPT4 generates answers directly without explicitly involving iterative reasoning. By comparison, o1 introduces a ``Think\&Summary'' phase before producing answers. This step leverages advanced techniques such as Chain-of-Thought (CoT) reasoning and tree search to perform more intricate logical reasoning and step-by-step problem solving. 
Compared to GPT4o, the o1 paradigm achieves a 50\% improvement in accuracy on the complex logical reasoning tasks like AI Math Evaluation (AIME) test dataset. However, the inference cost is increased by 10-100$\times$~\cite{o1inferencecost}.

Inference-time compute can enhance the model capabilities but also substantially increase inference time, particularly for complex reasoning tasks. 
As shown in Fig.~\ref{fig:o1}(b), in the GSM8K benchmark, we employ Llama2-7B as the backbone LLM, combined with the open-source LLM-Reasoners~\cite{hao2024llm}. 
The inference time is increased by 678$\times$ compared to using Llama2-7B directly.
This substantial time overhead primarily stems from the iterative multi-step process inherent in inference-time compute, which is necessary to satisfy the requirements of more complex reasoning task. 
Therefore, choosing the appropriate scenarios and conditions for employing inference-time compute remains a topic worthy of further research.
Additionally, inference-time compute also changes the runtime breakdown of the LLM inference. 
As shown in Fig.~\ref{fig:o1}(c), the proportion of the prefill stage significantly increases from 1.5\% to 23.5\%, while the decode stage decreases from 98.5\% to 54.8\%. 
The increase in the prefill stage is partly due to the need to merge the human input tokens with the template input tokens as the whole LLM input, which effectively extends the input sequence length. 
Moreover, in the iterative process, intermediate output tokens are cumulatively fed back as inputs, leading to a gradual increase in the sequence length for prefill.
Meanwhile, the introduction of the Process Reward Model (PRM) accounts for 21.7\% of the time. 
This shift of runtime breakdown highlights that future optimization methods for inference-time compute should consider all prefill stage, decode stage, and PRM simultaneously.

Inference-time compute is poised to become the foundational LLM inference paradigm of next-generation AI systems. For edge applications like autonomous driving and robotics, the advanced model capabilities are essential while the exponential growth of inference cost presents a significant challenge for both hardware and software. 
Real-time responsiveness, higher throughput, and efficient resource utilization are becoming essential attributes for supporting these edge applications.

\subsection{Higher Inference Energy Efficiency}



\begin{figure}[!t]
  \centering
  \includegraphics[width=0.8\textwidth]{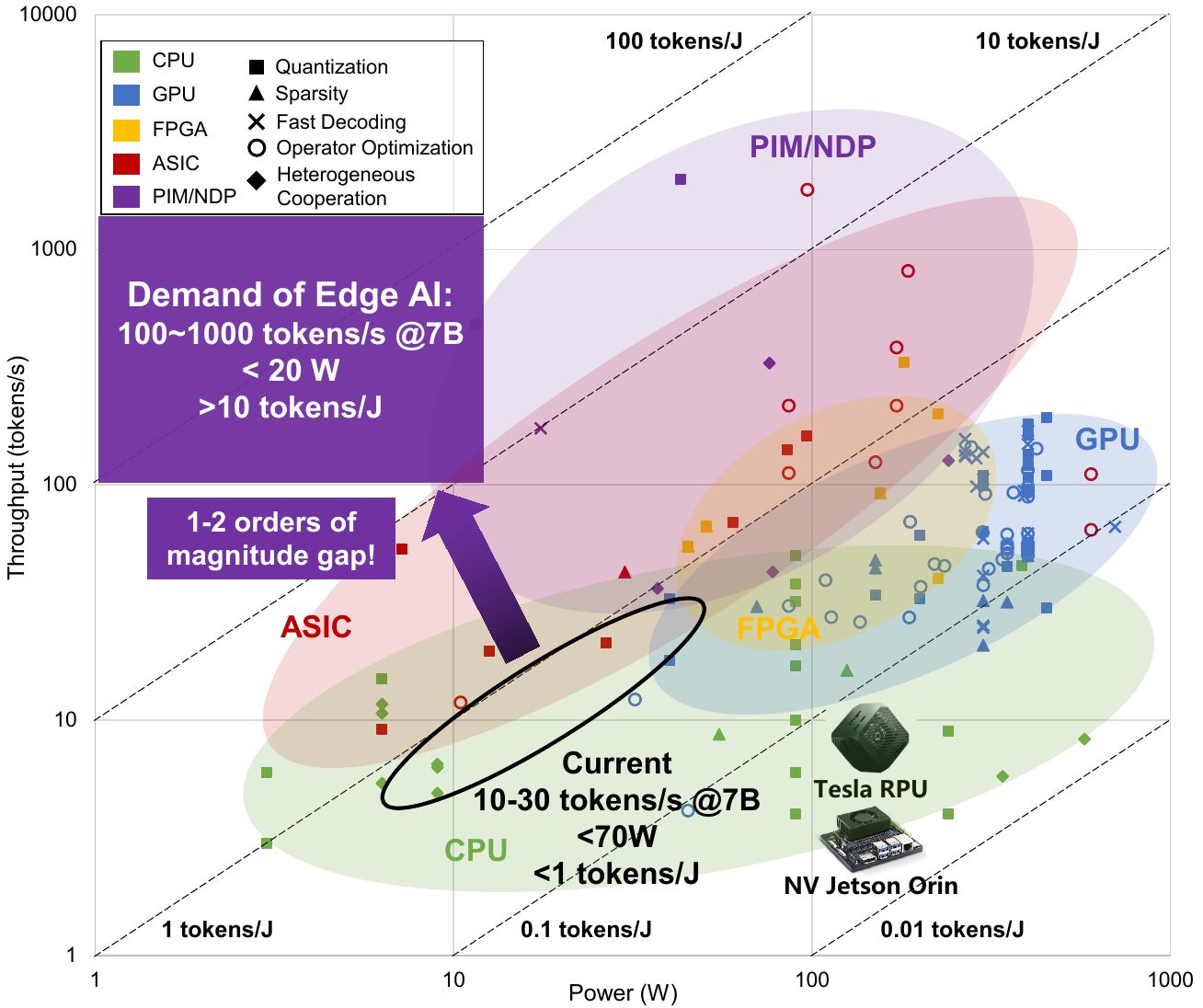}
  \vspace{-10pt}
  \caption{The performance gap between the future and the current edge AI systems.}
  \vspace{-10pt}
  \label{fig:edge_gap}
\end{figure}

In future edge applications, particularly in high real-time scenarios such as embodied intelligence represented by robotics control and autonomous driving, the system's control frequency needs to reach 100-1000 Hz~\cite{stewart1990implementing,bena2023hybrid,mafi2009parallel,atkins2023mintnet}. This implies that systems using LLMs as the decision-making center for actions or instructions should achieve an inference speed of at least 100-1000 tokens/s. Since edge devices need to consider energy supply, the power of edge chips should be <20W (the power consumption of human brain~\cite{markram2012human,magistretti2015cellular}) and then the inference efficiency should exceed 10 tokens/J. 
However, as shown in the Fig.~\ref{fig:edge_gap}, current commercial edge chips such as Tesla RPU~\cite{teslarpu} and NVIDIA Jetson Orin~\cite{nvidiaorin}, even with software optimization methods, only achieve an actual inference speed of 10-30 tokens/s and an efficiency of <1 tokens/J. 
This presents a gap of 1-2 orders of magnitude in both absolute inference speed and efficiency compared to future edge application demands. 

To achieve this goal, we should continue the algorithm-hardware co-design approach. 
On the algorithm side, more aggressive model compression methods can be employed with high-quality datasets for retraining to minimize accuracy loss. 
On the hardware side, in addition to designing specific hardware units to support algorithm optimizations, it is necessary to design more innovative chip architectures, such as 3D DRAM-stacked integrated architectures~\cite{li2025fine} that significantly reduce data transfer latency and energy consumption, dataflow architectures that optimize data transfer paths to improve computational efficiency, and wafer-scale architectures that provide higher computational and storage capabilities through ultra-large-scale integration. 
These co-design approaches can effectively bridge the current gaps in inference speed and energy efficiency of edge-side chips, meeting the demands of future applications.


\section{Conclusion}\label{sec:conclusion}

Generative LLMs like GPT series and Llama series are currently the main focus due to their high algorithm performance.
The advancements in generative LLMs are closely intertwined with the development of hardware capabilities. 
This paper presents a comprehensive survey of efficient generative LLM inference on different hardware platforms.
We provide an overview of the algorithm architecture of mainstream generative LLMs and summarize different optimization methods for different platforms such as CPU, GPU, FPGA, ASIC, and PIM/NDP. 
Furthermore, we perform a qualitative and quantitative comparison of inference performance with batch sizes 1 and 8 on different hardware platforms by considering hardware power consumption, absolute inference speed (tokens/s), and energy efficiency (tokens/J).
We point out that the development of edge intelligence has gained significant momentum, and three trends (multimodality, inference-time compute, and higher inference energy efficiency) are promising to redefine the capabilities of edge AI systems. 
Based on the software and hardware optimization, future edge chips are expected to achieve high throughput and low energy consumption, eliminating the 1-2 orders of magnitude gap in edge AI inference.



\bibliographystyle{unsrt}  
\bibliography{main}

\end{document}